\documentclass[final, 5p, times,authoryear, twocolumn]{elsarticle}
\usepackage{float}
\usepackage{amssymb}
\usepackage{amsmath}
\usepackage{amsthm}
\usepackage[colorlinks=true,allcolors=blue]{hyperref}
\hypersetup{
pdftitle={CATNUS: Coordinate-Aware Thalamic Nuclei Segmentation Using T1-Weighted MRI},
pdfauthor={Anqi Feng; Zhangxing Bian; Samuel W. Remedios; Savannah P. Hays; Blake E. Dewey; Alexa Colinco; Jiachen Zhuo; Reha S. Erzurumlu; Dan Benjamini; Jerry L. Prince; for the Alzheimer's Disease Neuroimaging Initiative}
}
\usepackage{hypernat}
\usepackage{booktabs}
\usepackage{subcaption}
\usepackage{threeparttable} 
\usepackage{array}
\usepackage{multirow}
\usepackage{booktabs}
\usepackage{makecell}
\journal{NeuroImage}
\begin{document}
\begin{frontmatter}

\title{
CATNUS: Coordinate‑Aware Thalamic Nuclei Segmentation\\ Using T1‑Weighted MRI
}

\author[aff1,aff2]{Anqi Feng\corref{cor1}}
\cortext[cor1]{Corresponding author.}
\ead{afeng11@jhu.edu}
\author[aff1]{Zhangxing Bian}
\author[aff3]{Samuel W. Remedios}
\author[aff1]{Savannah P. Hays}
\author[aff4]{Blake E. Dewey}
\author[aff5]{\\Alexa Colinco}
\author[aff5]{Jiachen Zhuo}
\author[aff6]{Reha S. Erzurumlu}
\author[aff2]{Dan Benjamini}
\author[aff1]{Jerry L. Prince}
\author{\\for the Alzheimer’s Disease Neuroimaging Initiative%
\textsuperscript{\ensuremath{\dagger}}}

\renewcommand{\thefnote}{\ensuremath{\dagger}}
\fntext[adni]{%
Data used in preparation of this article were obtained from the
Alzheimer’s Disease Neuroimaging Initiative (ADNI) database
(\url{https://adni.loni.usc.edu}). As such, the investigators within
the ADNI contributed to the design and implementation of ADNI and/or
provided data but did not participate in analysis or writing of this
report. A complete listing of ADNI investigators can be found at:
\url{http://adni.loni.usc.edu/wp-content/uploads/how_to_apply/ADNI_Acknowledgement_List.pdf}
}

\affiliation[aff1]{
  organization={Department of Electrical and Computer Engineering, Johns Hopkins University},
  city={Baltimore},
  state={MD},
  country={USA}
  }
\affiliation[aff2]{
  organization={Laboratory of Behavioral Neuroscience, National Institute on Aging, National Institutes of Health},
  city={Baltimore},
  state={MD},
  country={USA}
  }
\affiliation[aff3]{
  organization={Department of Computer Science, Johns Hopkins University},
  city={Baltimore},
  state={MD},
  country={USA}
  }
\affiliation[aff4]{
  organization={Department of Neurology, Johns Hopkins University School of Medicine},
  city={Baltimore},
  state={MD},
  country={USA}
  }
\affiliation[aff5]{
  organization={Department of Diagnostic Radiology and Nuclear Medicine, University of Maryland School of Medicine},
  city={Baltimore},
  state={MD},
  country={USA}
}
\affiliation[aff6]{
  organization={Department of Neurobiology, University of Maryland School of Medicine},
  city={Baltimore},
  state={MD},
  country={USA}
}
\begin{abstract}
Accurate segmentation of thalamic nuclei from magnetic resonance images is important due to the distinct roles of these nuclei in overall brain function and to their differential involvement in neurological and psychiatric disorders. 
However, segmentation remains challenging given the small size of many nuclei, limited intrathalamic contrast and image resolution, and inter-subject anatomical variability. 
In this work, we present CATNUS (Coordinate-Aware Thalamic NUclei Segmentation), which segments 13 thalamic nuclei (or nuclear groups) using a 3D U-Net architecture enhanced with coordinate convolution, to improve localization of the smallest labeled nucleus.
To support broad clinical applicability, we provide pre-trained model variants that can operate on quantitative T1 maps as well as on widely used magnetization-prepared rapid gradient echo (MPRAGE) and fast gray matter acquisition T1 inversion recovery (FGATIR) sequences.
We benchmarked CATNUS against established methods, including FreeSurfer, THOMAS, and HIPS-THOMAS, demonstrating improved segmentation performance and robust test-retest reliability across multiple nuclei.
We compared CATNUS segmentations with the Morel atlas, showing good consistency in the major thalamic nuclear organization.
CATNUS further demonstrated strong out-of-distribution generalization on traveling-subject datasets spanning multiple scanners, field strengths, and vendors, producing reliable and anatomically coherent segmentations across diverse acquisition conditions.
Moreover, CATNUS showed strong downstream utility in distinguishing Alzheimer’s disease from cognitively normal controls using thalamic volumetric features.
Overall, CATNUS provides a robust and generalizable solution for thalamic nuclei segmentation, with strong potential to facilitate large-scale neuroimaging studies and support real-world clinical assessment.
\end{abstract}



\begin{keyword}
Thalamic nuclei segmentation \sep Quantitative T1 mapping \sep Coordinate convolution
\end{keyword}
\end{frontmatter}

\section{Introduction}
\label{sec1}
The thalamus is a pair of ovoid, gray matter structures located deep within the brain. 
It is a major component of the diencephalon, sitting above the brainstem, flanking the third ventricle, and bordered by the cerebral hemispheres~\citep{jones2012thalamus,torrico2019neuroanatomy}. 
The thalamus not only serves as a central relay for directing sensory and motor signals to their target cortical areas, but also actively contributes to motor control, consciousness, alertness, and sleep regulation~\citep{sommer2003role,sherman2006exploring,sherman2007thalamus,sherman2016thalamus}.

The thalamus consists of multiple nuclei, each connected to specific cortical and subcortical regions~\citep{jones2012thalamus}. 
For example, the ventral lateral (VL) nucleus relays cerebellar signals to the motor and premotor cortex to support movement coordination, while the ventral anterior (VA) nucleus passes signals from the basal ganglia to the motor and premotor cortex to help with movement planning~\citep{keun2021structural}. 
The ventral posterior lateral nucleus (VPL) and ventral posterior medial nucleus (VPM) process somatosensory information from the body and face, and project to the somatosensory cortex and insular cortex~\citep{price2002central,keun2021structural}. 
The mediodorsal (MD) nucleus supports higher cognitive functions like working memory and emotion regulation~\citep{mitchell2015mediodorsal}, while the pulvinar nucleus helps control visual attention and sensory integration~\citep{grieve2000primate}. 
Additionally, disruptions in thalamic nuclei are implicated in multiple neurological and psychiatric disorders. For example, alterations in the MD nucleus are associated with schizophrenia and major depressive disorder~\citep{alelu2008mediodorsal}; atrophy of the anterior, MD and pulvinar nuclei has been linked to cognitive impairment in Alzheimer's disease~\citep{forno2023thalamic}; degeneration of the central median (CM) nucleus contributes to motor and cognitive dysfunction in Parkinson's disease~\citep{henderson2000degeneration}; and structural damage to the pulvinar nucleus is commonly observed in multiple sclerosis~\citep{minagar2013thalamus}.
Moreover, thalamic nuclei are important targets for treatment. Deep brain stimulation (DBS) of the anterior nucleus is well established for drug-resistant focal epilepsy~\citep{fisher2010electrical}; the CM nucleus is an emerging target for Lennox–Gastaut syndrome~\citep{dalic2022dbs}; and stimulation of the central lateral (CL) nucleus is being explored to improve the recovery of executive function after traumatic brain injury~\citep{schiff2023thalamic}. Given these roles, accurate thalamic nuclei segmentation is vital for both neuroscience research and clinical applications.

In vivo thalamic nuclei segmentation primarily relies on magnetic resonance imaging (MRI).
However, the small size of nuclei and low intrathalamic contrast make manual delineation time-consuming and highly dependent on expert knowledge, even with advanced imaging techniques developed to enhance contrast~\citep{tourdias2014visualization}. 
These challenges have motivated the development of automatic segmentation methods for efficient and reliable thalamic mapping. 
Existing approaches use different imaging features and parcellation principles, leading to inconsistent nucleus definitions and boundary delineation that complicate direct comparison across methods~\citep{segobin2024roadmap}. 
These methods span structural MRI, diffusion MRI, functional MRI, and multimodal approaches.

Recent segmentation efforts have widely adopted structural MRI (sMRI) due to its broad availability and high spatial resolution. 
\cite{deoni2007segmentation} introduced an unsupervised genetic algorithm to cluster intrathalamic regions using quantitative T1 and T2 maps, though the lack of spatial constraints led to fragmented, anatomically inconsistent clusters.
\cite{traynor2011segmentation} addressed this by incorporating spatial priors into T1/T2 map-derived segmentations, enabling reproducible division into six subregions. However, reliance on synthetic data with idealized contrasts, hard boundaries, and a single anatomical atlas limited its generalizability to real-world data.
To improve anatomical accuracy, \cite{iglesias2018probabilistic} introduced the FreeSurfer method, constructing a probabilistic atlas of 26 thalamic nuclei from ex vivo 7T MRI and histology, applied to in vivo T1-weighted Magnetization-Prepared Rapid Gradient Echo (MPRAGE) images through a Bayesian inference framework. While accurate, this approach is computationally intensive (often requiring several hours), and is prone to oversegmentation into adjacent subcortical regions.
\cite{su2019thalamus} proposed THOMAS, a fast multi-atlas segmentation using 7T white-matter-nulled (WMn) MPRAGE, also known as Fast Gray Matter Acquisition T1 Inversion Recovery (FGATIR), but this sequence is not commonly used in standard imaging protocols. 
\cite{liu2020generation} introduced a multi-contrast 7T atlas evaluated on 3T T1-weighted MRI using statistical shape models, which performed well but assumed a pre-segmented thalamic boundary and ignored intrathalamic intensity variation, limiting its sensitivity to disease-related structural changes. 
\cite{majdi2020automated} used cascaded multi-planar residual U-Nets to segment thalamic nuclei from WMn and conventional MPRAGE images.   However, the method relies on 2D networks that do not fully exploit 3D anatomical information.
\cite{datta2021fast} extended THOMAS to MP2RAGE sequences, achieving comparable performance, though MP2RAGE remains less commonly used in routine clinical practice. 
\cite{umapathy2022convolutional} proposed a CNN-based segmentation method using standard T1-weighted MRI, with optional synthesized WMn-like contrast to enhance small nuclei detection and group difference sensitivity. However, it used THOMAS-derived labels as ground truth and was sensitive to the quality of synthesized contrast.
Most recently, HIPS-THOMAS~\citep{vidal2024robust} addressed accessibility issues by generating WMn-like contrast from standard T1-weighted scans using polynomial synthesis, achieving near WMn-MPRAGE performance. However, it was evaluated by comparing against segmentations derived from WMn-MPRAGE, rather than against manual annotations. Additionally, the simple polynomial model could amplify noise, and segmentation remained unreliable for small nuclei such as the habenula and VLA.

Diffusion MRI (dMRI) offers unique insights into thalamic microstructure by capturing local fiber orientations, enabling differentiation of nuclei beyond intensity-based sMRI. 
\cite{wiegell2003automatic} performed k-means clustering on voxels within the thalamus based on the principal diffusion orientation and diffusion anisotropy. However, the method was limited by the angular resolution of early Diffusion Tensor Imaging (DTI), so small nuclei could not be distinguished, and it required the number of clusters to be pre-specified.
\cite{ziyan2006segmentation} proposed a spectral clustering method, constructing a voxel affinity matrix based on tensor similarity and applying a normalized cut algorithm with Markov relaxation to encourage spatial coherence. Still, this method remained sensitive to parameter choices and was  computationally intensive. 
\cite{jonasson2007level} introduced a level set method, assigning voxels based on similarity to representative tensors within evolving surfaces, and incorporating fractional anisotropy (FA) and mean diffusivity (MD) to enhance segmentation. Nevertheless, the method required careful initialization, and was sensitive to noise and errors in tensor estimation.
\cite{rittner2010segmentation} computed gradients of diffusion tensor fields to highlight boundary transitions, followed by segmentation using watershed or region growing on the resulting gradient map. However, this approach was sensitive to noise and lacked priors on nuclear shape or location, leading to uncertain boundaries in regions with gradual diffusion changes.
\cite{mang2012thalamus} clustered thalamic voxels based purely on local principal diffusion directions, assuming each nucleus has a dominant fiber orientation. While effective for large nuclei, it may merge nuclei with mixed or crossing fibers, failing to resolve small nuclei.
\cite{ye2013parcellation} developed a multi-object geometric deformable model based on DTI, where multiple deformable surfaces (each corresponding to a nucleus) evolved simultaneously, guided by local diffusion features and shape priors, to achieve joint thalamic segmentation. However, this method required an initial thalamus mask and approximate surface initialization, and was computationally intensive.
\cite{battistella2017robust} proposed a k-means clustering method using orientation distribution functions (ODFs) computed from high-angular-resolution diffusion imaging (HARDI), combining ODF shape and spatial proximity to segment seven thalamic groups. While accurate, the method demanded advanced acquisition protocols and careful parameter tuning.
Despite the methodological advances, dMRI-based approaches generally face several limitations. First, most acquisitions rely on echo planar imaging (EPI)~\citep{stehling1991echo}, which provides relatively low spatial resolution (typically 2 to 3 mm isotropic), leading to partial volume effects that obscure small and adjacent nuclei. Moreover, the thalamus is primarily composed of gray matter, which exhibits low diffusion anisotropy and thus offers limited contrast for nuclei differentiation. 

Beyond local microstructural features, diffusion MRI can also distinguish thalamic nuclei based on their unique patterns of structural connectivity with cortical and subcortical regions.
\cite{behrens2003non} pioneered a probabilistic tractography method that parcellated the thalamus based on voxel-wise connectivity to predefined cortical regions, enabling segmentation aligned with functional cortico-thalamic circuits. However, the method depended on arbitrary cortical region-of-interest (ROI) definitions and might not align with true anatomical boundaries.
\cite{johansen2005functional} validated this approach by demonstrating correspondence between connectivity-defined segments and both histological nuclei and functional MRI (fMRI) activations. However, this validation did not address the method’s original limitations, and the approach was not evaluated for small nuclei.
\cite{o2011clustering} applied Independent Component Analysis (ICA) to thalamocortical tractography data to identify independent connectivity patterns, enabling thalamic parcellation without predefined cortical ROIs. Still, it did not guarantee one-to-one alignment  with traditional nuclei, often spanning multiple cytoarchitectonic nuclei or splitting one nucleus into pieces with different connectivity. 
Recently, \cite{he2025deepnuparc} proposed DeepNuParc, which uses tractography-derived connectivity features of thalamic voxels, an autoencoder for dimensionality reduction, and k-means clustering for thalamic parcellation.
However, the method was developed and evaluated using high-quality HCP diffusion MRI, leaving its performance on routine clinical acquisitions unclear.
Overall, connectivity-based thalamic parcellation faces several limitations. 
First, it often fails to match the true anatomical boundaries of thalamic nuclei. This is partly because many neighboring nuclei project to similar cortical regions, leading multiple distinct nuclei to be grouped together. Also, some individual nuclei exhibit heterogeneous projection patterns, which can lead to a single anatomical nucleus being split into multiple fragments.
Moreover, crossing fibers make tractography results less reliable. In such regions, the inferred pathways can be inaccurate or incomplete, which leads to incorrect connectivity assignments for thalamic voxels.

To better leverage the complementary strengths of structural and diffusion MRI, several studies have explored hybrid approaches for thalamic nuclei segmentation.
\cite{stough2014automatic} introduced a hierarchical random forest that first segmented the whole thalamus and then parcellated it into six nuclei, using T1-weighted MRI, diffusion-derived features, and tractography-based connectivity.
\cite{glaister2016thalamus} extended this method by incorporating atlas-derived nuclei priors to better localize small structures like the lateral and medial geniculates. However, both approaches shared limitations, such as dependence on accurate whole-thalamus segmentation, limited generalizability due to small training sets, and sensitivity to tractography quality, especially in low-resolution or noisy diffusion data.
\cite{tregidgo2023accurate} extended the MPRAGE-based FreeSurfer method by introducing a Bayesian framework that integrates sMRI and dMRI through joint likelihood modeling, combining an improved probabilistic atlas as a prior to estimate voxel-wise posterior probabilities, with final labels assigned via maximum a posteriori (MAP) estimation. Yet, the method did not model partial volume effects and imposed a reflective symmetry constraint on dMRI distributions, which might limit accuracy in low-resolution data or anatomically asymmetric cases.
\cite{yan2023segmenting} adopted Uniform Manifold Approximation and Projection (UMAP) to project multimodal MRI features, including T1-weighted, T2-weighted, multiple T1-weighted images at different inversion times (multi-TI), and diffusion-derived features, into a low-dimensional space, and performed thalamic nuclei segmentation using k-nearest neighbor classification.
\cite{feng2023label} and \cite{feng2024ratnus} both adopted a two-stage 3D U-Net architecture that first segmented the whole thalamus and then further parcellated the thalamus into 13 nuclei, leveraging multimodal data including multi-TI images, quantitative T1 and proton density (PD) maps and diffusion-derived features.
\cite{gao2025iterative} proposed iterative prior-guided parcellation (iPGP), which first registers the Morel atlas to each subject’s space using structural MRI and then iteratively reassigns boundary voxels based on spatially constrained local diffusion features.
Yet, validation relied on reproducibility, diffusion-feature separability, and behavioral associations rather than subject-specific histological references, leaving the anatomical accuracy of the resulting boundaries uncertain.
Despite these efforts, the use of multimodal features remains subject to several practical constraints. 
First, different modalities capture distinct structural properties: sMRI provides anatomical boundaries, dMRI encodes local microstructural information, and tractography-derived connectivity reflects long-range projections. These representations often misalign spatially or functionally, leading to conflicting cues at nucleus boundaries, where anatomically distinct nuclei may share similar connectivity profiles, and a single nucleus may show heterogeneous diffusion patterns. 
Second, most current multimodal methods rely on supervised learning, yet the available ground truth labels are typically derived from sMRI or dMRI alone and may not correspond well across modalities, reducing supervision reliability when conflicts arise.
Moreover, the resolution gap between high-resolution sMRI and typically lower-resolution dMRI limits the visibility of small nuclei in diffusion space, further complicating voxel-level fusion across modalities.

Resting-state fMRI has been used to identify functional subdivisions of the thalamus. \cite{zhang2008intrinsic} used partial correlations and a winner-take-all strategy to assign each thalamic voxel to the predefined cortical region showing the strongest correlation. However, this approach depended on predefined cortical regions and thus produced relatively coarse functional subdivisions. To reduce this dependence, \cite{fan2015functional} applied k-means clustering to the whole-brain functional connectivity profiles of thalamic voxels. However, the reported parcellation was group-level, and stable subject-specific results were not established. 
\cite{kumar2017functional} applied group ICA to instantaneous connectivity time courses and used split-half reproducibility analysis to identify stable thalamic functional parcels, which were compared with the Morel atlas and structural and functional connectivity patterns.
However, these parcels did not show a one-to-one correspondence with fine cytoarchitectonic nuclei. \cite{van2018functional} further developed a top-down instantaneous connectivity parcellation framework using group ICA, achieving greater correspondence with the Morel atlas than standard ICA and Craddock clustering. However, the method required a biologically meaningful initial region of interest, and the resulting functional parcels did not fully reflect cytoarchitectonic boundaries. 
Overall, these approaches characterize group-level functional organization rather than subject-specific anatomical nuclei. Moreover, functional connectivity reflects physiological associations that may not align with histological boundaries, and the resulting parcellations are sensitive to fMRI data quality and the selected parcellation scale.

In this work, we present CATNUS (Coordinate‐Aware Thalamic NUclei Segmentation), a deep learning-based framework for thalamic nuclei segmentation built upon a coordinate-aware 3D U-Net~\citep{cciccek20163d}, targeting 13 nuclei or nuclear groups. 
CATNUS accepts quantitative T1 maps as primary input, which offer favorable contrast for thalamic nuclei segmentation~\citep{feng2025segmenting},  and can also operate on MPRAGE or FGATIR for broader clinical applicability.
This work extends our earlier studies~\citep{feng2024ratnus,feng2025segmenting} by 
incorporating coordinate convolution, 
expanding benchmark comparisons, evaluating consistency with the Morel atlas, 
testing out-of-distribution generalization on traveling-subject datasets,
and assessing discrimination between Alzheimer’s disease and cognitively normal controls using thalamic volumetric features.

\begin{figure*}[!ht]
\centering
\includegraphics[width=0.96\textwidth]{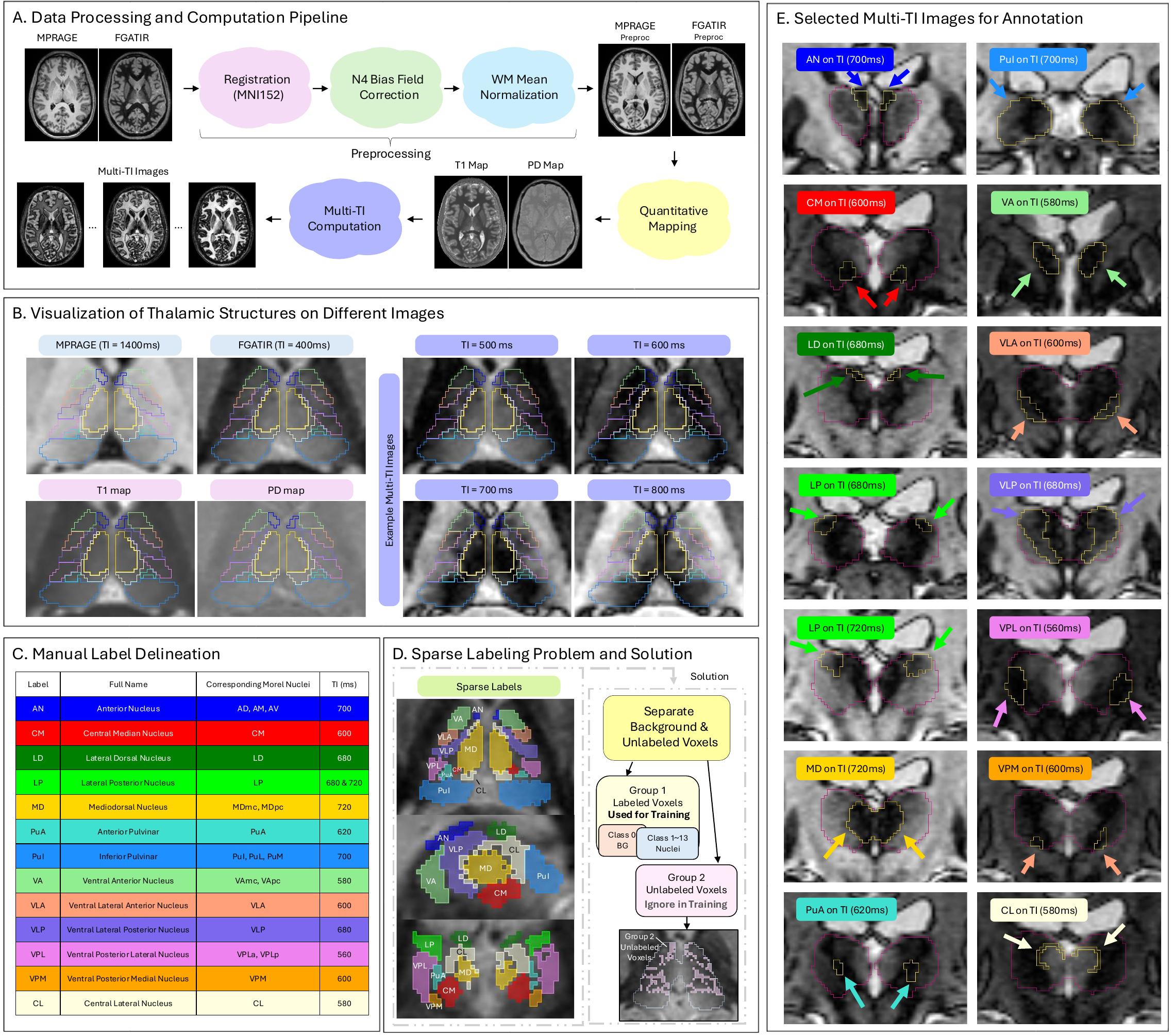}
\caption{Overview of the CATNUS preprocessing and annotation workflow.
A. Data preprocessing and computation pipeline. MPRAGE and FGATIR images are co-registered to the MNI152 space, followed by joint N4 bias field correction and white matter (WM) mean normalization. Quantitative T1 and PD maps are computed, and multi-TI images are generated using the estimated T1 and PD maps. 
B. Visualization of thalamic structures on different images. Representative MPRAGE, FGATIR, T1 and PD maps, and multi-TI images illustrate how varying inversion times enhance intra-thalamic visibility and facilitate nuclear delineation. 
C. Manual label delineation. Summary of the 13 CATNUS labels, including their abbreviations, full names, corresponding Morel nuclei, selected TI values used for annotation, and the color-coding scheme. 
D. Sparse-labeling problem and solution. Annotations were restricted to high-confidence voxels (left). Unlabeled voxels were explicitly separated from background and excluded from model training (right). 
E. Selected multi-TI images for annotation. Examples of nuclei annotated on specific TI images, illustrating how different inversion times highlight distinct nuclear boundaries. Red and yellow outlines indicate the thalamic boundary and the corresponding nucleus boundary, respectively. When only yellow is visible, the nucleus boundary coincides with the thalamic boundary on this slice.}
\label{fig:s2-data_and_labels_visualization}
\end{figure*}

\section{Methods}
\label{sec2}
\subsection{Overview}\label{sec:2.1}
This section describes the overall methodology of CATNUS. 
Section ~\ref{sec:2.2} introduces the study cohorts.
Section~\ref{sec:2.3} presents the MRI acquisition protocol and imaging parameters.
Section~\ref{sec:2.4} details the preprocessing, quantitative mapping, and multi‐TI computation pipeline.
Section~\ref{sec:2.5} describes the manual delineation procedure.
Section~\ref{sec:2.6} presents the coordinate‐aware 3D U‐Net segmentation framework, including network architecture, data augmentation, loss function, and post‐processing refinement.
Finally, Section~\ref{sec:2.7} explains the unified labeling protocol used to align our nuclei definitions with those of existing methods for comparison.

\subsection{Study Cohorts}\label{sec:2.2}
The primary labeled cohort was acquired as part of a mild traumatic brain injury (mTBI) study, approved by the University of Maryland's ethics review board. 
It included 24 participants: 14 controls and 10 participants with mTBI. 
The control group consisted of one healthy control and 13 orthopedic controls, who had no prior head injury, had a negative head CT, and were admitted for non-brain-related injuries.
The mTBI participants met the criteria for mild head injury, including a Glasgow Coma Scale (GCS) score of 13--15 at admission, together with documented loss of consciousness, a positive head CT, or facial trauma and wounds.
The controls were 34.9 $\pm$ 14.4 years old (range: 18--59 years; 6F/8M), and the mTBI participants were 42.7 $\pm$ 13.4 years old (range: 20--59 years; 4F/6M). 
For the mTBI participants, the injury-to-scan interval was 9.0 $\pm$ 5.4 days (range: 1--20 days), and the admission GCS score was 14.6 $\pm$ 0.7 (range: 13--15). 
Injury mechanisms included assault, fall, motor vehicle collision, and stab injury. Clinical review did not identify direct thalamic lesions in these mTBI participants.

In addition to the primary labeled cohort, we used a separate test--retest cohort for segmentation reproducibility evaluation.
The test--retest cohort included 11 controls, each scanned in two MRI sessions approximately six months apart. 
For nine of these participants, the first-session scan was included in the primary labeled cohort, while the other two participants were included only in the test--retest cohort. All second-session scans were excluded from the primary labeled cohort.

\subsection{Acquisition Protocol}\label{sec:2.3}
Each participant underwent an MRI session with MPRAGE and FGATIR sequences, both acquired using a 64-channel head-and-neck coil with a 256 $\times$ 256 mm$^{2}$ field of view, a 256 $\times$ 256 acquisition matrix, 176 sagittal slices, 1 mm isotropic resolution, a repetition time (TR) of 4000 ms, an echo time (TE) of 3.37 ms, a flip angle of 6$^\circ$, GRAPPA acceleration of 2, and partial Fourier sampling. 
The two sequences differed only in the inversion time (TI): 1400 ms for MPRAGE and 400 ms for FGATIR. 
All scans were performed on a 3T Siemens Prisma Fit scanner.

\subsection{Data Preprocessing and Computation}\label{sec:2.4}
We processed MPRAGE and FGATIR images to compute quantitative T1 and PD maps, which were then used to calculate multi-TI images. 
The T1 maps served as inputs for segmentation models, while the multi-TI images were used for thalamic nuclei delineation and visualization.
The preprocessing and computation pipeline is described in detail below and illustrated in Figure~\ref{fig:s2-data_and_labels_visualization}A. Example MPRAGE and FGATIR images, T1 and PD maps, and selected multi-TI images are shown in Figure~\ref{fig:s2-data_and_labels_visualization}B.

We selected T1 maps as the primary input modality because they encode the core tissue-specific T1 contrast that underlies T1-weighted acquisitions. As demonstrated in our previous work~\citep{feng2025segmenting}, T1 maps provide better segmentation performance compared to individual T1-weighted sequences such as MPRAGE or FGATIR. However, not all clinical environments can acquire the dual inversion time scans needed to compute T1 maps. Thus, we also trained models on MPRAGE and FGATIR separately to maximize clinical accessibility.

\subsubsection{Data Preprocessing}
Image preprocessing consisted of three sequential steps:
(1) Registration. MPRAGE and FGATIR images were co-registered to the MNI ICBM152 nonlinear symmetric 2009c template using a coarse-to-fine strategy with ANTs ~\citep{avants2009advanced}, where MPRAGE was first rigidly aligned to the MNI template, followed by FGATIR alignment to MPRAGE.
(2) N4 bias field correction using N4ITK~\citep{tustison2010n4itk}.
(3) White matter mean normalization~\citep{reinhold2019evaluating}.
Because MPRAGE and FGATIR were acquired with identical parameters except for the inversion time (TI), their quantitative relationship must be carefully preserved for reliable signal modeling and parameter estimation. Therefore, these steps were applied jointly to ensure that both modalities remain on a consistent intensity scale.

\textit{N4 bias field correction.}
Let $I_{\text{M}}(x)$ and $I_{\text{F}}(x)$ denote the MPRAGE and FGATIR images after registration to the MNI space. We first estimated the bias fields independently for each modality using N4:
\begin{equation}
    B_{\text{M}}(x) = \text{N4}(I_{\text{M}}(x)), \quad B_{\text{F}}(x) = \text{N4}(I_{\text{F}}(x)).
\end{equation}
To harmonize correction across modalities, we computed the geometric mean of the two bias fields:
\begin{equation}
    B_{\text{harm}}(x) = \sqrt{B_{\text{M}}(x) \cdot B_{\text{F}}(x)}.
\end{equation}
This harmonized bias field was then applied to both images:
\begin{equation}
    \hat{I}_{\text{M}}(x) = \frac{I_{\text{M}}(x)}{B_{\text{harm}}(x)}, \quad
    \hat{I}_{\text{F}}(x) = \frac{I_{\text{F}}(x)}{B_{\text{harm}}(x)}.
\end{equation}

\textit{White matter mean normalization.}
We generated a white matter (WM) mask $\mathcal{M}_{\text{WM}}$ from the MPRAGE image using fuzzy C-means clustering, and computed the mean WM intensity:
\begin{equation}
    \mu_{\text{WM}} = \frac{1}{|\mathcal{M}_{\text{WM}}|} \sum_{x \in \mathcal{M}_{\text{WM}}} \hat{I}_{\text{M}}(x),
\end{equation}
where \( |\mathcal{M}_{\text{WM}}| \) is the total number of voxels in the white matter mask.\\
Both bias-corrected images were then scaled by this mean value:
\begin{equation}
    \tilde{I}_{\text{M}}(x) = \frac{\hat{I}_{\text{M}}(x)}{\mu_{\text{WM}}}, \quad
    \tilde{I}_{\text{F}}(x) = \frac{\hat{I}_{\text{F}}(x)}{\mu_{\text{WM}}}.
\end{equation}
With these preprocessing steps, the images are standardized for accurate quantitative estimation.

\subsubsection{Quantitative Mapping and Image Computation}
For a given voxel $v$ within a T1-weighted inversion recovery image, the signal intensity $I(v)$ can be modeled as:
\begin{equation}
I(v) = PD(v) \left[ 1 - 2\exp\left(-\frac{\text{TI}}{T_1(v)}\right) + \exp\left(-\frac{\text{TR}}{T_1(v)}\right) \right],
\label{eq:inversion}
\end{equation}
where $PD(v)$ and $T_1(v)$ denote the proton density and T1 relaxation value at voxel $v$, while TI and TR are the inversion and repetition times. Given two input images (MPRAGE and FGATIR) and their known acquisition parameters, we obtained two measurements per voxel, and fitted two instances of Equation~\ref{eq:inversion} to estimate $PD(v)$ and $T_1(v)$ at each voxel using nonlinear least squares fitting.

FGATIR was acquired with a short inversion time (TI = 400 ms), placing many tissues in a state of negative longitudinal magnetization. However, MR scanners store magnitude images, which discard signal polarity by taking the absolute value. To preserve the signed signal properties required for accurate fitting of Equation~\ref{eq:inversion}, we negated the FGATIR magnitude image prior to parameter estimation, thereby recovering the true signal distribution expected by the model. This negation assumption holds for thalamic tissues with relatively long T1 values, but may not apply to tissues with much shorter T1 (e.g., fat).

After estimating $PD(v)$ and $T_1(v)$,  we computed multi-TI images to support the delineation of thalamic nuclei. Specifically, for each voxel, we used Equation~\ref{eq:inversion} to calculate signal intensity by fixing TR and plugging in the estimated $PD(v)$ and $T_1(v)$ values, while varying the TI. We took the absolute value of the simulated signal to match the magnitude format of clinical MR images. These multi-TI images enhance thalamic contrast by modulating tissue visibility across different inversion times. Depending on the TI, specific tissues are nulled and inter-compartment differences are amplified, enabling visual identification of some thalamic nuclei boundaries that are not discernible in MPRAGE or FGATIR images. Example multi-TI images in Figure~\ref{fig:s2-data_and_labels_visualization}B show how TI variations affect tissue contrast and highlight different features. In our implementation, TI is sampled from 400~ms to 1400~ms in 20~ms increments. This range spans meaningful contrast variation across thalamic tissue, and the 20~ms step is the minimum visually detectable change in our images. 

\subsection{Manual Label Delineation}\label{sec:2.5}
Manual delineation was performed for 13 thalamic nuclei or nuclear groups using the Morel atlas as the reference~\citep{morel1997multiarchitectonic}. 
The digital Morel atlas in the MNI152 space~\citep{krauth2010mean,jakab2012generation} was first warped to each subject’s space, followed by manual correction based on multi-TI images.
The annotation protocol was developed by A.C. (a medical student) and J.Z. (a clinical MRI expert). Following this protocol, A.C. performed the manual delineations under the guidance of J.Z.
R.E. (an expert neuroanatomist) reviewed the 13 label definitions and the overall protocol.

We used multi-TI images to support manual annotation because different TI values highlight different thalamic nuclei and boundaries. 
For each CATNUS label, we evaluated the local contrast between the corresponding registered Morel label and its surrounding tissue across TI images. 
This contrast profile was used to identify candidate TI values. 
The final TI images were selected by the annotator based on boundary visibility and subject-specific anatomy across the coronal, axial, and sagittal views. 
The selected TI values are listed in Figure~\ref{fig:s2-data_and_labels_visualization}C, representative annotation examples are shown in Figure~\ref{fig:s2-data_and_labels_visualization}E, and the contrast analysis is provided in Supplementary Note~S1.

For each subject, the Morel labels were first registered to the subject image space and used as initial references. Morel nuclei that could not be reliably separated on the multi-TI images were combined into a single CATNUS label, while structures outside the current annotation protocol were not assigned CATNUS labels. This resulted in 13 CATNUS labels. The correspondence between CATNUS labels and Morel nuclei is summarized in Figure~\ref{fig:s2-data_and_labels_visualization}C. The detailed protocol, including label definitions, starting points, and anatomical landmarks, is provided in Supplementary Table S1.
Manual refinement of the registered Morel labels was performed primarily on the selected TI images, with neighboring TI images and the quantitative T1 map used to confirm the boundaries. 
During refinement, voxels outside the intended nuclear regions were removed, and boundary voxels that could not be confidently assigned because of weak contrast or partial volume effects were left unlabeled. This resulted in sparse annotations containing only high-confidence voxels. 
To handle these sparse labels during training, we separated true background from unlabeled intrathalamic voxels. 
The 13 CATNUS labels were assigned class indices 1--13. 
Voxels outside the manually localized thalamic region were assigned background class 0 and included in training, while unlabeled voxels within the thalamic region were assigned an ignore label 100 and excluded from loss computation. 
Across the 24 annotated subjects, labeled voxels accounted for $68.3\% \pm 4.0\%$ of the manually localized thalamic region. This sparse-label formulation is illustrated in Figure~\ref{fig:s2-data_and_labels_visualization}D.
Manual correction examples and consistency checks are provided in Supplementary Note S3.

\begin{figure*}[!ht]
\centering
\includegraphics[width=0.97\textwidth]{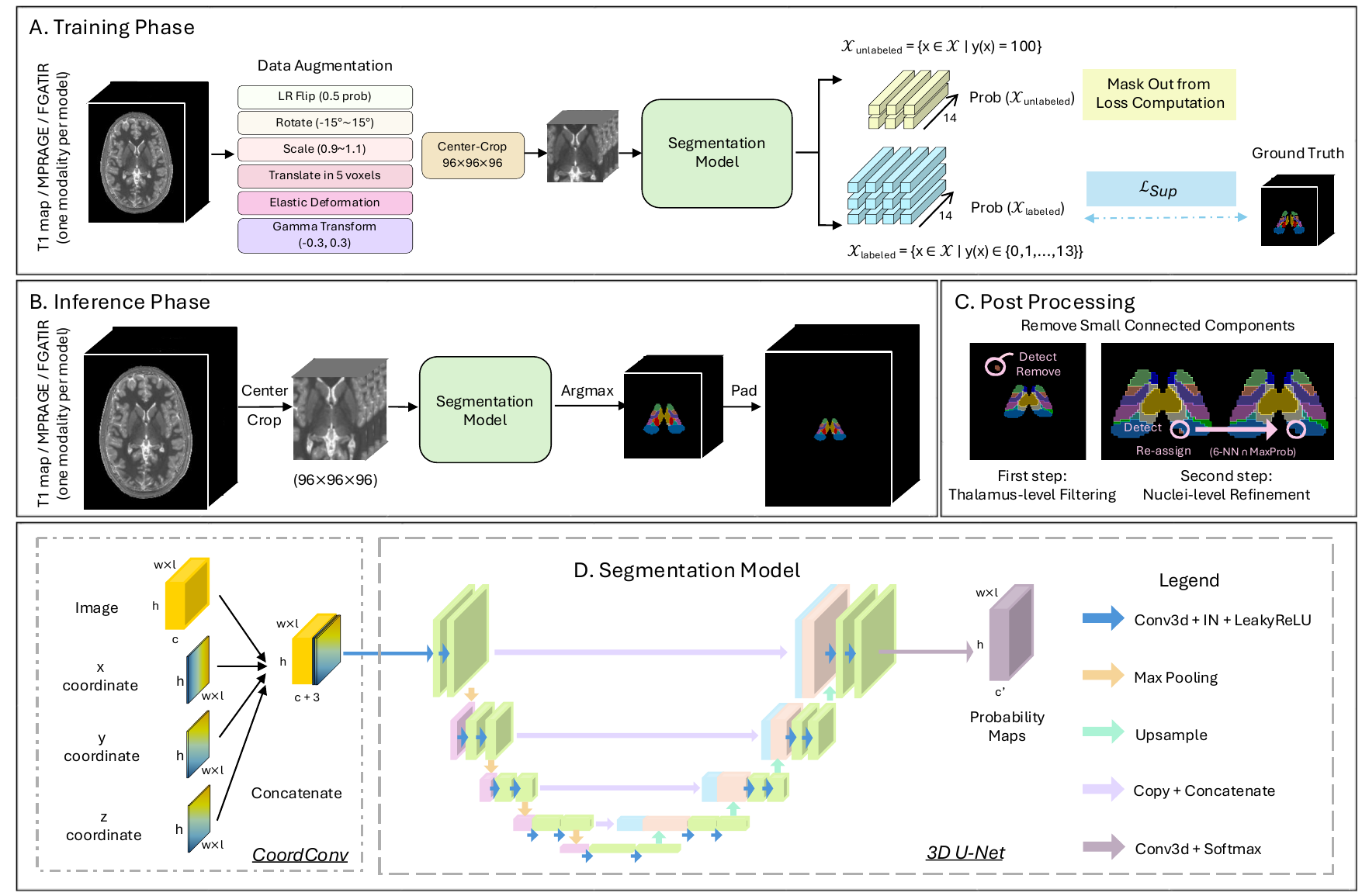}
\caption{Overview of the CATNUS segmentation framework. 
CATNUS includes three separately trained model variants for T1 maps, MPRAGE, and FGATIR, respectively. All variants share the same model architecture and training, inference, and post-processing procedures.
A. Training Phase. Input images undergo spatial and intensity augmentations, followed by center-cropping to a fixed size. The model is supervised using Dice loss computed only on voxels with manual labels, while unlabeled voxels are masked out from loss computation.
B. Inference Phase. A cropped image is passed through the model, and each voxel is assigned the class with the highest predicted probability. The output is then padded back to the original volume size.
C. Post Processing. Predicted segmentations are refined sequentially, first by thalamus-level filtering, then by nucleus-level refinement. 
D. Segmentation Model. The model takes an image and its voxel-wise spatial coordinates (x, y, z) as input, processes them via coordinate convolution, and outputs probability maps using a 4-level 3D U-Net.}
\label{fig:s2-segmentation_framework}
\end{figure*}

\subsection{Segmentation Framework}\label{sec:2.6}
CATNUS employs a segmentation framework based on a coordinate-aware 3D U-Net~\citep{cciccek20163d} to classify each voxel into 14 classes (0: background, 1–13: nuclei). 
It comprises three separately trained single-input model variants for quantitative T1 maps, MPRAGE, and FGATIR, respectively. All three variants share the same model architecture and training \& inference procedures.
Each model was trained using Dice loss computed exclusively on manually labeled voxels, while unlabeled regions were excluded from loss computation. The overall workflow is illustrated in Figure~\ref{fig:s2-segmentation_framework}, and the subsequent subsections detail the data augmentation, model architecture, loss formulation, and post-processing.

\subsubsection{Data Augmentation}
To improve model generalization and account for anatomical variability, we apply diverse data augmentation during training (illustrated in Figure~\ref{fig:s2-segmentation_framework}A). 
Spatial augmentations include left-right flipping (p=0.5), affine transformations with rotations within $\pm$ 15$^\circ$, isotropic scaling between 0.9 and 1.1, and translations within 5 voxels. 
We also apply elastic deformation using a 3D B-spline grid with seven control points per axis and a maximum displacement of seven voxels. 
Following augmentation, all volumes are center-cropped to 
a fixed size of 96$\times$96$\times$96, which fully covers the thalamus region across subjects in MNI space. 
Spatial transformations are applied jointly to images and labels using trilinear interpolation for images and nearest-neighbor interpolation for labels.
Intensity augmentation is applied only to images via random gamma correction, with log gamma values sampled uniformly from [$-$0.3, 0.3]. 
During validation and testing, only center cropping is applied.

\subsubsection{Model Architecture}
Our segmentation model (illustrated in Figure~\ref{fig:s2-segmentation_framework}D) uses a 4-level 3D U-Net~\citep{cciccek20163d} as the backbone, which has an encoder–decoder architecture with skip connections and a central bottleneck block. 

\textit{Encoder Path}: The encoder contains four downsampling stages. Each stage applies two consecutive 3D convolution blocks (3$\times$3$\times$3 kernels, instance normalization, LeakyReLU activation), followed by 2$\times$2$\times$2 max pooling to reduce spatial resolution while increasing feature channels.

\textit{Decoder Path}: The decoder contains four upsampling stages that mirror the encoder. Each upsampling stage follows a three-step process: (1) trilinear interpolation upsampling with a 2$\times$ scale factor, (2) 3D convolution with instance normalization and LeakyReLU activation, and (3) concatenation with corresponding encoder features via skip connections, followed by two 3D convolution blocks for feature integration.

\textit{Coordinate Convolution}:
In our previous study~\citep{feng2025segmenting}, we observed recurrent localization failures for PuA, the smallest nucleus in our label set, particularly in the MPRAGE and FGATIR models.
To address this issue, CATNUS incorporates coordinate convolution~\citep{liu2018intriguing} at the input layer. This adds three channels representing normalized voxel-wise $x$, $y$, and $z$ coordinates to the input image. The coordinates are computed for the center-cropped 96$\times$96$\times$96 volume after data augmentation and normalized to the range [$-1,1$], with 0 corresponding to the spatial center along each axis.

\textit{Padding Strategy}: We replace the default zero padding in all convolutional layers with replicate padding. In coordinate convolution, the value 0 represents the spatial center along each axis. If zero padding is used, voxels with this same value would be introduced at the image borders, creating ambiguity as the network may confuse padded border regions with true center locations. Replicate padding avoids this by copying edge values, preserving the correct spatial context at the boundaries.

\textit{Output}: The network outputs 14-channel voxel-wise probability maps through a final 3D convolution followed by the softmax activation.

\subsubsection{Loss Function}
We employ supervised Dice loss computed only on labeled voxels to handle the sparse annotation problem (illustrated in Figure~\ref{fig:s2-segmentation_framework}A). Let the labeled voxel set be:
\[
\mathcal{X}_{\text{labeled}} = \{ x \in \mathcal{X} \mid y(x) \in \{0, 1, \dots, 13\} \},
\]
where \( y(x) \) denotes the ground-truth label for voxel \( x \), and \( \mathcal{X} \) is the set of all voxels in the volume.
For \( x \in \mathcal{X}_{\text{labeled}} \), we compute the mean class-wise Dice loss between the predicted probabilities \( P_{\text{pred}}(c, x) \) and the one-hot ground-truth \( P_{\text{gt}}(c, x) \):
\[
\mathcal{L}_{\text{dice}} = 1 - \frac{1}{|C|} \sum_{c \in C}
\frac{
2 \sum\limits_{x \in \mathcal{X}_{\text{labeled}}} P_{\text{pred}}(c, x) \cdot P_{\text{gt}}(c, x)
}{
\sum\limits_{x \in \mathcal{X}_{\text{labeled}}} \left[ P_{\text{pred}}(c, x) + P_{\text{gt}}(c, x) \right] + \varepsilon
},
\]
where \( C = \{0, 1, \dots, 13\} \) includes 14 classes, with 0 as background and the rest corresponding to thalamic nuclei, and \( \epsilon \) is a small constant for numerical stability. This formulation encourages overlap between the predicted and ground-truth regions for each class, while ignoring unlabeled voxels that might introduce training noise.

\subsubsection{Post-processing}
To improve anatomical plausibility and reduce spurious predictions, we apply two sequential post-processing steps after inference (illustrated in Figure~\ref{fig:s2-segmentation_framework}C). First, to address false positive predictions in non-thalamic regions, we perform thalamus-level filtering.
We create a binary mask by combining all predicted thalamic nuclei (labels 1-13) versus background (label 0), then remove any connected components that are smaller than 200 voxels, as these typically represent spurious predictions far from the thalamic region.
Second, to eliminate fragmented predictions within each class, we apply nucleus-level refinement. 
For each class, disconnected components smaller than 4 voxels are removed from their original class and reassigned to the neighboring class with the highest predicted probability, excluding the original class, within their 6-connected neighborhood.
These two steps work together to promote spatial consistency in the final segmentation.

\begin{figure}[!ht]
\includegraphics[width=1\linewidth]{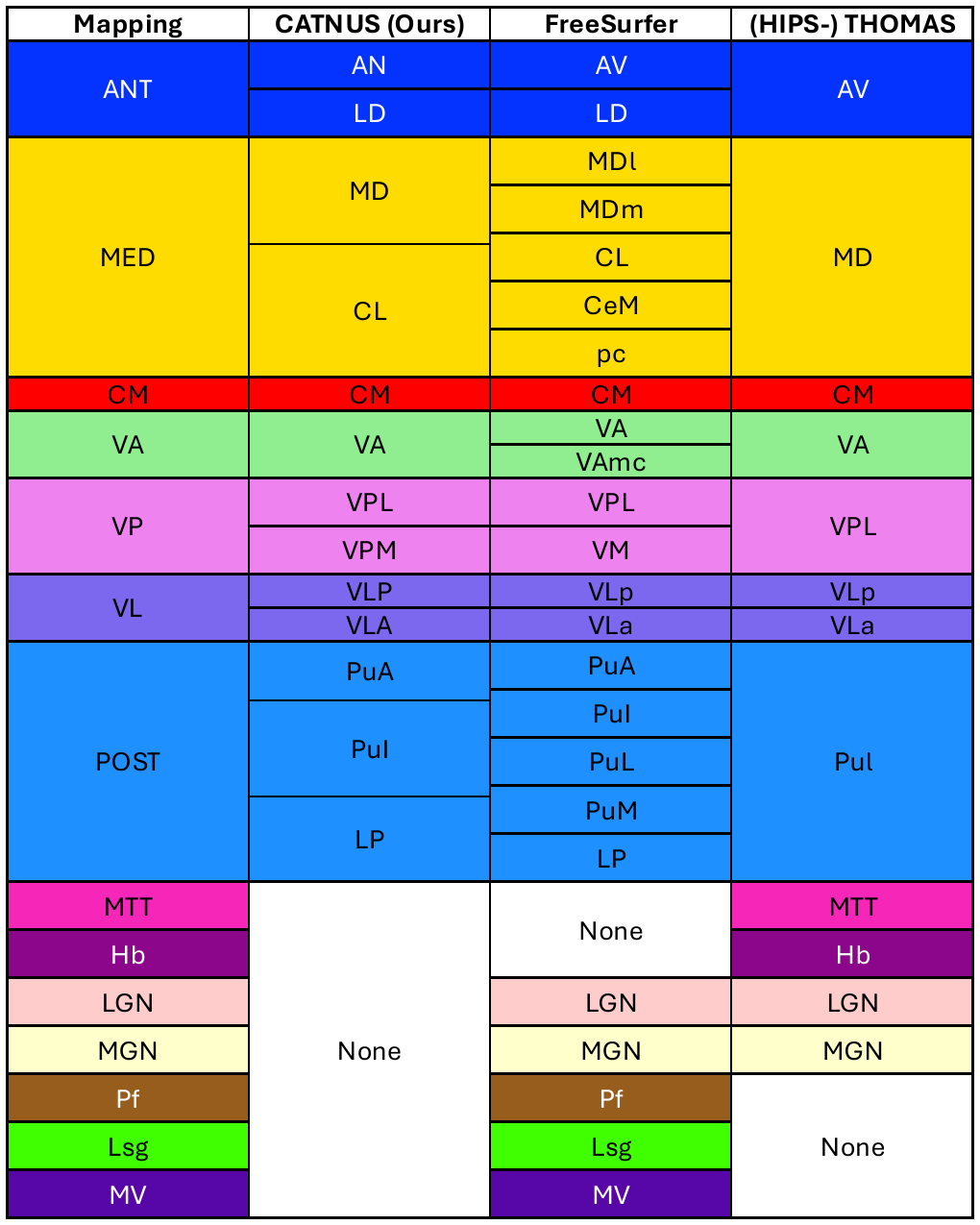}
\caption{A comprehensive mapping of thalamic nuclei categories between CATNUS, FreeSurfer, THOMAS and HIPS-THOMAS. The first column shows the harmonized comparison
labels, and the second, third, and fourth columns list the corresponding method-specific labels. 
All comparison labels and individual nuclei are assigned distinct colors to help with visual differentiation. THOMAS and HIPS-THOMAS share the same labeling and color conventions. Abbreviations for nuclei in FreeSurfer, THOMAS, and HIPS-THOMAS are adopted from ~\cite{iglesias2018probabilistic}, 
~\cite{su2019thalamus}
and ~\cite{vidal2024robust}.}
\label{fig:s2-label_unifying_protocols}
\end{figure}

\subsection{Labeling Scheme Harmonization}
\label{sec:2.7}
We compared CATNUS against three widely used thalamic nuclei segmentation tools: FreeSurfer, THOMAS, and HIPS-THOMAS. FreeSurfer uses a Bayesian inference framework and operates on standard MPRAGE scans~\citep{iglesias2018probabilistic}. 
THOMAS is a multi-atlas segmentation method designed for white-matter-nulled (WMn) MPRAGE ~\citep{su2019thalamus}. 
Although THOMAS can also be applied to MPRAGE, its performance is much lower than with FGATIR. Because FGATIR is available in our dataset and provides WMn-like contrast, we evaluated THOMAS only on FGATIR scans.
HIPS-THOMAS extends THOMAS by synthesizing WMn-like contrast from conventional T1-weighted images~\citep{vidal2024robust}.
Therefore, HIPS-THOMAS was applied to MPRAGE scans.
As each method follows a different labeling protocol, we used the Morel atlas~\citep{morel1997multiarchitectonic} as a structural reference to map method-specific labels into a harmonized comparison space.
Based on anatomical correspondences, we mapped nuclei into seven comparison labels: ANT, MED, CM, VA, VP, VL, and POST.
Nuclei without clear correspondence, such as Pf, Lsg, and MV (present only in FreeSurfer); MTT and Hb (present in THOMAS and HIPS-THOMAS); and MGN and LGN (absent from our labels), were retained as distinct classes.
The mapping and color-coding scheme is illustrated in Figure~\ref{fig:s2-label_unifying_protocols}.
Since THOMAS and HIPS-THOMAS share the same labeling definitions, we present them as a single column.
Note that this mapping was used only for cross-method comparison, not to define a new anatomical naming scheme.

\section{Experiments and Results}\label{sec:3}

\subsection{Experimental Setup}\label{sec:3.1}
\textit{Experiments Overview}: We evaluated CATNUS through five experiments: 
(1) within-domain segmentation across input modalities and coordinate convolution settings (Section~\ref{sec:3.2}); 
(2) comparisons against established methods (Section~\ref{sec:3.3});
(3) consistency analysis with the Morel atlas (Section~\ref{sec:3.4});
(4) out-of-distribution generalization on traveling-subject datasets (Section~\ref{sec:3.5});
and (5) downstream Alzheimer’s disease discrimination
(Section~\ref{sec:3.6}).
These together assess segmentation accuracy, robustness, generalizability, and downstream utility.

\textit{General Training Configuration}: Models were trained on single-channel volumetric data, where the input was either a T1 map, MPRAGE, or FGATIR. We used the Adam optimizer with an initial learning rate of 1e--3 and a weight decay of 1e--4. The batch size was set to 1 due to GPU memory constraints with 3D volumetric inputs. To ensure reproducibility and fair comparisons, we fixed the random seed to 1234.

\textit{Cross-validation Experiments}:
For experiments requiring ground truth evaluation, we performed 8-fold cross-validation using our 24 manually labeled subjects. In each fold, 19 subjects were used for training, 2 for validation, and 3 for testing, ensuring each subject appeared in the test set exactly once. The learning rate was reduced by 10\% if the validation loss did not improve for 5 consecutive epochs. Training was run for up to 200 epochs, with early stopping applied if no improvement was observed over 15 consecutive epochs. These cross-validation models were used to evaluate different input modalities and with/without coordinate convolution (Section~\ref{sec:3.2}), and to compare against benchmark methods (Section~\ref{sec:3.3}), with final performance reported as mean and standard deviation across all 24 test subjects. 
We also evaluated model sensitivity to reductions in the number of labeled training voxels (Supplementary Note S4).
Since our ground truth labels are sparse, where the rater only annotated voxels with high confidence, many thalamic voxels remain unlabeled. 
In such cases, overlap-based metrics like the Dice score can be misleading, as they may penalize predictions in potentially valid but unlabeled regions.
To avoid this issue, we adopted the true positive rate (TPR) as our primary evaluation metric, which directly measures recovery of confidently labeled voxels under partial supervision.

\textit{Full-dataset-training Experiment}: For the remaining analyses, we trained CATNUS variants with all 24 labeled subjects to maximize model performance. Without a validation set, the learning rate was scheduled to decay by 10\% every 10 epochs, and training was run for exactly 200 epochs without early stopping. The resulting models were then used for the consistency analysis with the Morel atlas (Section~\ref{sec:3.4}), generalization testing on external datasets (Section~\ref{sec:3.5}), and downstream Alzheimer’s disease differentiation (Section~\ref{sec:3.6}).

\begin{figure*}[!ht]
\centering
\includegraphics[width=0.97\textwidth]{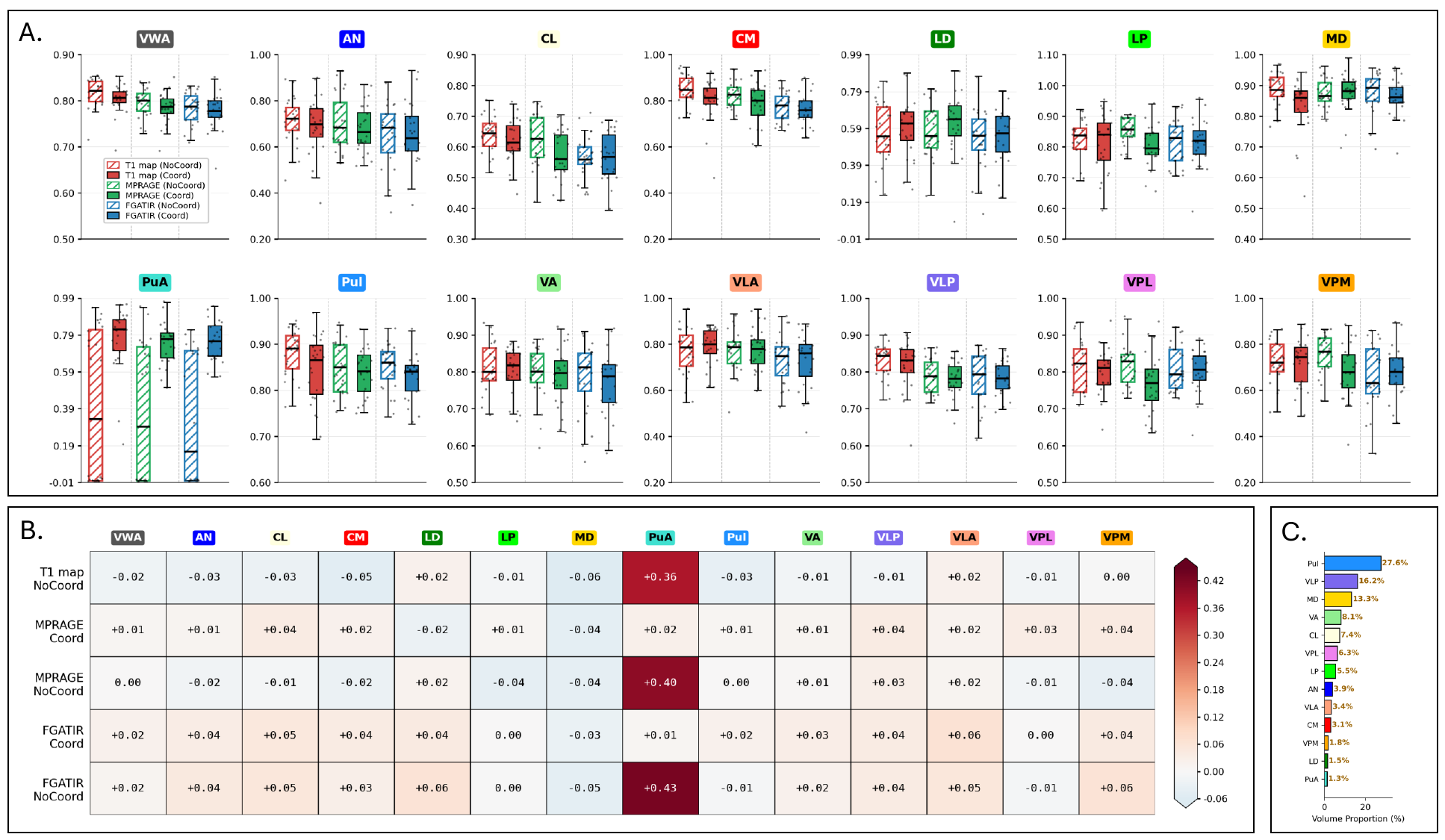}
\caption{Within-domain segmentation performance across three input modalities (T1 map, MPRAGE, and FGATIR) with or without coordinate convolution (CoordConv). Performance was evaluated across 24 manually labeled subjects using 8-fold cross-validation.
A. Boxplots show the distribution of true positive rates (TPR) for each thalamic nucleus and the volume-weighted average (VWA), grouped by modality and CoordConv setting.
B. Heatmap shows the TPR difference between T1 map with CoordConv and each other configuration, computed as $\mathrm{TPR}(\mathrm{T1~map~with~CoordConv}) - \mathrm{TPR}(\mathrm{model})$. Positive values indicate higher TPR for T1 map with CoordConv, while negative values indicate higher TPR for the compared model. The T1 map with CoordConv reference row is omitted because all differences are zero.
C. Bar plot shows the volume fraction of each nucleus within the labeled thalamic nuclei. }
\label{fig:s3-quantitative_results}
\end{figure*}

\begin{figure*}[!ht]
\centering
\includegraphics[width=0.97\textwidth]{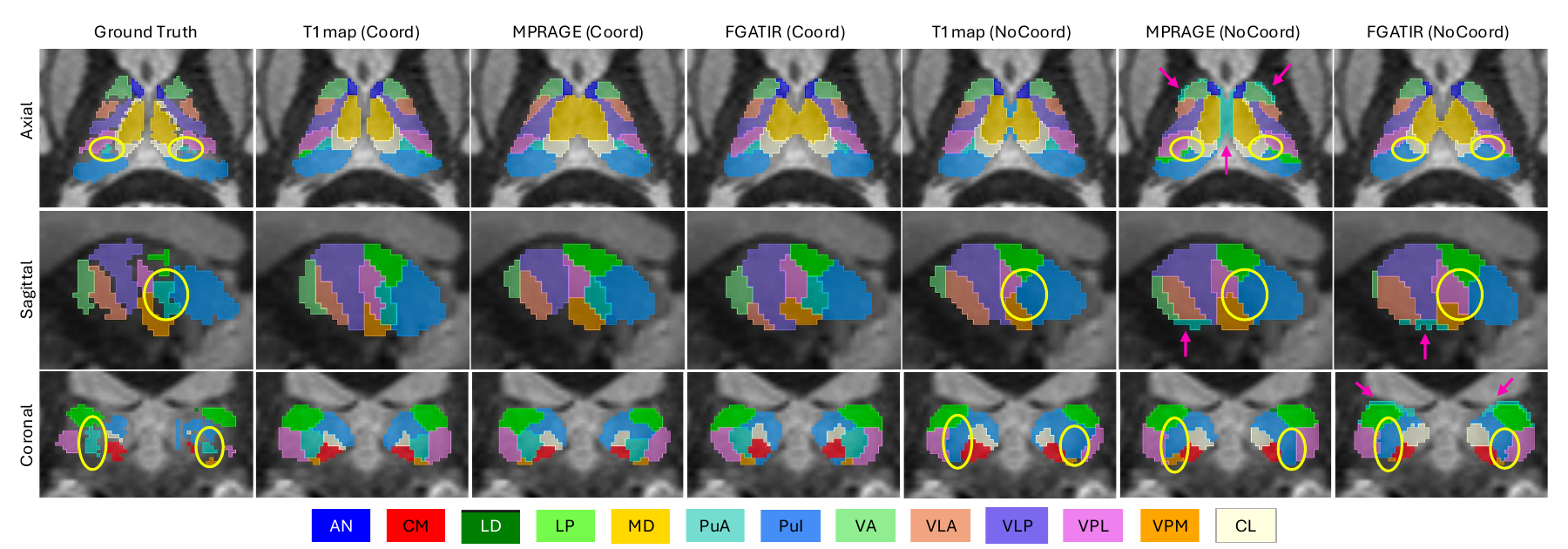}
\caption{Qualitative comparisons of segmentation results across input modalities (T1 map, MPRAGE, FGATIR) and coordinate convolution (CoordConv) settings, overlaid on FGATIR images. 
Each row shows a representative slice in a different anatomical plane (axial, sagittal, coronal), with each slice taken from a different subject. 
Ground-truth annotations (leftmost column) are compared against model predictions from the six configurations.
Yellow circles highlight common failure cases, especially for the small PuA nucleus, which is frequently missed when CoordConv is not used. 
Pink arrows indicate additional mislocalizations of the PuA nucleus along thalamic boundaries.}
\label{fig:s3-qualitative_results}
\end{figure*}

\subsection{Within-domain Segmentation Across Modalities and Coordinate Convolution Settings }\label{sec:3.2}

\textit{Experimental Design:}
We systematically evaluated the effects of input modality and coordinate convolution (CoordConv) on segmentation performance. 
Specifically, we trained six models, where the input was either a T1 map, MPRAGE, or FGATIR scan, each with or without CoordConv. 
All models shared the same architecture and training protocol to ensure controlled comparison. To isolate the effects of modality and coordinate encoding, no post-processing was applied during this evaluation.
Performance was measured using the true positive rate (TPR) for each of the 13 thalamic nuclei. 
Additionally, we computed a volume-weighted average (VWA) TPR per subject to assess overall segmentation accuracy. 
Final results are reported as the mean and standard deviation of per-nucleus and VWA TPRs across all 24 subjects.

\textit{Quantitative Results:}
Figure~\ref{fig:s3-quantitative_results} summarizes the quantitative performance of the six model configurations.
Overall, all three input modalities achieved reasonable segmentation performance. 
T1 map generally performed better than MPRAGE and FGATIR, with the trend most evident in VWA and several nuclei, such as CL, CM, PuI, VLP, and VLA. 
The modality differences were often small and varied across nuclei, suggesting that although T1 map showed the best overall performance, MPRAGE and FGATIR also provided useful segmentation information.
The effect of CoordConv varied across nuclei. 
For VWA and several individual nuclei, models without CoordConv performed slightly better than their corresponding CoordConv models, indicating that adding coordinate information did not lead to a uniform overall improvement. 
However, this pattern was different for very small nuclei, especially PuA. 
PuA was the smallest nucleus in our annotation protocol, accounting for only 1.3\% of the labeled nucleus volume. Without CoordConv, PuA predictions frequently failed, while adding CoordConv increased PuA TPR by approximately 36--42 percentage points across all three modalities.
A smaller improvement was also observed for LD, the second smallest nucleus, though the effect was less pronounced than for PuA.
Taken together, T1 map showed more favorable overall modality performance, while CoordConv greatly benefited the smallest nucleus, PuA, despite limited effects in some other regions.

\textit{Statistical Analysis:}
We first performed a two-way repeated-measures ANOVA separately for each thalamic nucleus and the VWA TPR. Input modality and CoordConv setting were treated as within-subject factors, with TPR as the dependent variable. Greenhouse--Geisser correction was applied to account for potential violations of sphericity, and generalized eta-squared ($\eta_G^2$) was reported as the effect size. The ANOVA showed a significant main effect of input modality in VWA and in eight nuclei, including AN, CL, CM, LD, PuI, VLP, VLA, and VPM, indicating that segmentation performance varied across input modalities. However, CoordConv showed no significant main effect, and no significant interaction between input modality and CoordConv was observed, indicating that CoordConv did not produce a consistent global effect.
We then performed paired Wilcoxon signed-rank post-hoc tests with Holm correction for multiple comparisons. All significant results were defined by Holm-corrected $p < 0.05$. Modality comparisons were performed separately under the CoordConv and no-CoordConv settings.
Under the CoordConv setting, T1 map was significantly higher than MPRAGE in VLP, while MPRAGE was significantly higher than T1 map in MD. T1 map was significantly higher than FGATIR in CL, CM, and VLA. MPRAGE and FGATIR showed no significant differences. No other significant differences were observed.
Under the no-CoordConv setting, the advantage of T1 map was more pronounced. T1 map was significantly higher than MPRAGE in VWA, CM, PuI, and VLP, and significantly higher than FGATIR in VWA, AN, CL, CM, PuI, VLP, and VPM. MPRAGE was significantly higher than FGATIR in CL, CM, and VPM. No other significant differences were observed.
We also compared CoordConv and no-CoordConv within each input modality. CoordConv significantly improved PuA across all three input modalities. However, several significant differences favored models without CoordConv, including MD for T1 map, LP and VPM for MPRAGE, and PuI for FGATIR. No other significant differences were observed. 
Overall, the statistical analysis supports the quantitative findings. T1 map generally showed the strongest performance. CoordConv did not improve all nuclei, but showed a consistent and significant benefit for PuA across all input modalities.

\textit{Qualitative Results:}
Figure~\ref{fig:s3-qualitative_results} shows segmentation outputs across input modalities and CoordConv settings. 
Overall, the six configurations captured the major thalamic structures and produced segmentations broadly consistent with the ground-truth labels.
However, the main qualitative differences were observed for the smallest nucleus, PuA. 
Models without CoordConv often fail to predict PuA at the correct positions, highlighted with yellow circles, where PuA is visible in the ground-truth but absent in model predictions lacking coordinate encoding. 
In addition, we observe systematic mislocalizations of PuA in the MPRAGE and FGATIR models without CoordConv, where false positives appear along the outer boundaries of the thalamus, indicated by pink arrows.
These qualitative patterns are consistent with the quantitative and statistical findings, supporting improved PuA localization with CoordConv.

Overall, T1 map showed the strongest performance, while MPRAGE and FGATIR still provided useful segmentation results. 
CoordConv did not improve every nucleus, and no-CoordConv models were slightly better in some cases. However, these differences were generally small. The main exception was PuA. Without CoordConv, PuA was often missed or misplaced, representing an important failure mode for this very small nucleus. CoordConv helped prevent this failure across input modalities and improved PuA localization. 
Therefore, we selected T1 map with CoordConv as the primary CATNUS configuration to balance strong overall performance with robust PuA localization, while keeping the MPRAGE and FGATIR CoordConv models as alternatives when T1 maps are unavailable. 
A supplementary sensitivity analysis of all six configurations with reduced training-label availability is provided in Supplementary Note S4.

\begin{figure*}[!ht]
\centering
\includegraphics[width=0.9\textwidth]{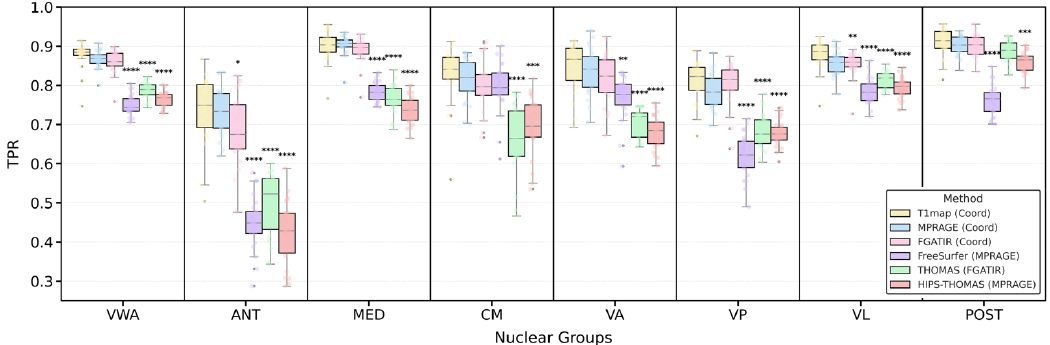}
\caption{
Quantitative comparison of thalamic nuclei segmentation performance against benchmark methods.
True positive rates (TPRs) were averaged over 24 subjects from 8-fold cross-validation, evaluated for seven harmonized nuclear groups and their volume-weighted average (VWA).
The first three methods, T1map (Coord), MPRAGE (Coord), FGATIR (Coord), are coordinate-aware CATNUS variants.
The latter three are literature baselines: FreeSurfer (MPRAGE), THOMAS (FGATIR), and HIPS‐THOMAS (MPRAGE).
Each box shows the distribution of TPRs for a nuclear group and method, with overlaid dots indicating subject-level values.
Asterisks indicate statistically significant differences from T1map (Coord) after Bonferroni correction (*p<0.05, **p<0.01, ***p<0.001, ****p<0.0001).}
\label{fig:s3-benchmark_quantitative_results}
\end{figure*}

\begin{figure*}[!ht]
\centering
\includegraphics[width=0.9\textwidth]{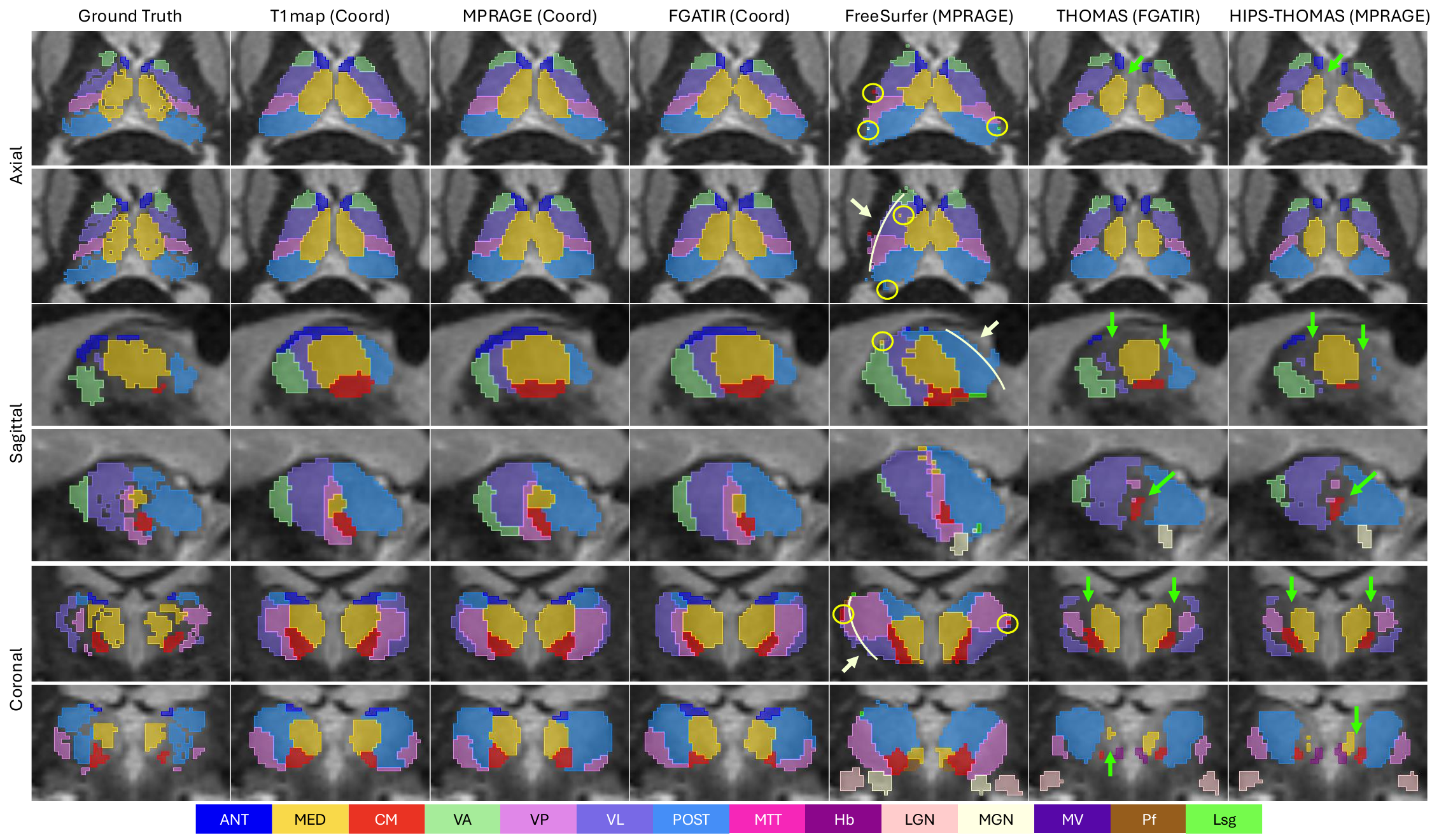}
\caption{
Qualitative comparison of thalamic nuclei segmentation performance against benchmark methods. 
Columns represent the manual ground truth, our three coordinate-aware CATNUS variants (T1 map, MPRAGE, and FGATIR), and three established methods including FreeSurfer (MPRAGE), THOMAS (FGATIR), and HIPS-THOMAS (MPRAGE).
Rows show representative axial, sagittal, and coronal slices, each from a different subject.
All predictions are overlaid on FGATIR images for visualization.
The visualized segmentations include the seven thalamic nuclear groups evaluated in this study, as well as additional nuclei (e.g., MGN, LGN, MTT, Hb, Pf, Lsg, MV) that appear in some benchmark outputs but are not labeled in the ground truth. 
Our CATNUS variants closely match manual annotations, producing smooth boundaries and anatomically coherent shapes. 
In contrast, FreeSurfer frequently generates fragmented predictions (yellow circles) and over-segments into adjacent structures beyond thalamic boundaries (white arrows). 
THOMAS and HIPS-THOMAS generate similar segmentations, yielding smoother results with fewer scattered errors than FreeSurfer, but tend to under-segment the thalamus, leaving some regions unlabeled (green arrows).}
\label{fig:s3-benchmark_qualitative_results}
\end{figure*}

\begin{table*}[htbp]
\centering
\caption{Test-retest Dice scores for CATNUS and benchmark methods. Results are computed on 11 control participants using the harmonized seven-group scheme shown in Figure~\ref{fig:s2-label_unifying_protocols} and the volume-weighted average (VWA), reported as mean $\pm$ standard deviation. The highest value in each row is shown in bold.}
\small
\begin{tabular}{lcccccc}
\toprule
Group
& \makecell{T1 map\\(Coord)} 
& \makecell{MPRAGE\\(Coord)} 
& \makecell{FGATIR\\(Coord)} 
& \makecell{FreeSurfer\\(MPRAGE)} 
& \makecell{THOMAS\\(FGATIR)} 
& \makecell{HIPS-THOMAS\\(MPRAGE)} \\
\midrule
ANT & \textbf{0.8736$\pm$0.0788} & 0.8623$\pm$0.0743 & 0.8054$\pm$0.2166 & 0.7642$\pm$0.0523 & 0.7982$\pm$0.1597 & 0.7809$\pm$0.1024\\ 
MED & \textbf{0.9554$\pm$0.0216} & 0.9478$\pm$0.0327 & 0.9195$\pm$0.1026 & 0.8949$\pm$0.0249 & 0.9081$\pm$0.0592 & 0.9244$\pm$0.0255\\ 
CM & 0.9212$\pm$0.0388 & 0.9083$\pm$0.0549 & 0.8728$\pm$0.1079 & 0.8359$\pm$0.0446 & 0.8416$\pm$0.1045 & \textbf{0.9274$\pm$0.0201} \\
VA & \textbf{0.9213$\pm$0.0461} & 0.8983$\pm$0.0680 & 0.8948$\pm$0.0674 & 0.8305$\pm$0.0361 & 0.8588$\pm$0.0813 & 0.8521$\pm$0.0588\\ 
VP & \textbf{0.9102$\pm$0.0590} & 0.8930$\pm$0.0716 & 0.8920$\pm$0.0720 & 0.8551$\pm$0.0467 & 0.8650$\pm$0.0792 & 0.9068$\pm$0.0386\\ 
VL & \textbf{0.9410$\pm$0.0300} & 0.9263$\pm$0.0476 & 0.9150$\pm$0.0648 & 0.8767$\pm$0.0398 & 0.8984$\pm$0.0714 & 0.8210$\pm$0.0994\\ 
POST & \textbf{0.9504$\pm$0.0309} & 0.9422$\pm$0.0347 & 0.9344$\pm$0.0481 & 0.8966$\pm$0.0304 & 0.9103$\pm$0.0605 & 0.8964$\pm$0.0414\\ 
\midrule
VWA & \textbf{0.9391$\pm$0.0331} & 0.9275$\pm$0.0431 & 0.9124$\pm$0.0714 & 0.8752$\pm$0.0328 & 0.8944$\pm$0.0688 & 0.9068$\pm$0.0309 \\
\bottomrule
\end{tabular}
\label{tab:test-retest_comparisons}
\end{table*}

\subsection{Comparisons Against Established Methods}\label{sec:3.3}
\textit{Experimental Design:}
We benchmarked our three coordinate‑aware models (trained separately on T1 map, MPRAGE, and FGATIR)  against three established segmentation methods, FreeSurfer~\citep{iglesias2018probabilistic}, THOMAS~\citep{su2019thalamus} and HIPS-THOMAS~\citep{vidal2024robust}.
These methods were run with fixed software versions: FreeSurfer 7.4.1, THOMAS v2.1, and HIPS-THOMAS Release\_2.0.
As described in Section~\ref{sec:2.7}, we ran FreeSurfer with MPRAGE, THOMAS with FGATIR, and HIPS-THOMAS with MPRAGE.
All input images underwent the identical preprocessing pipeline described in Section~\ref{sec:2.4}, including registration to the MNI152 template, N4 bias field correction, and white matter mean normalization.
Our model predictions, the benchmark segmentation results, and the ground‑truth labels were mapped to the harmonized seven-group scheme shown in Figure~\ref{fig:s2-label_unifying_protocols} for comparisons.  
For quantitative analysis, we focused only on the seven common nuclear groups, excluding the background and nuclei not shared across all methods.
We computed the true positive rate (TPR) for each of the seven groups by comparing model outputs with the ground truth, and derived a volume-weighted average (VWA) TPR across groups. 
For each method, we reported the mean value and standard deviation of these per‑subject metrics and visualized the full distributions in Figure~\ref{fig:s3-benchmark_quantitative_results}. 
Statistical comparisons were performed using paired Wilcoxon signed-rank tests against our coordinate‐aware model trained with T1 maps, selected for its strong overall performance.
Bonferroni correction was applied for multiple comparisons, and significance levels are indicated as follows: *p<0.05, **p<0.01, ***p<0.001, ****p<0.0001.  

\textit{Quantitative Results:}
In Figure~\ref{fig:s3-benchmark_quantitative_results}, all three coordinate-aware models showed higher overall TPRs than the benchmark methods, both in volume-weighted average (VWA) and across individual nuclear groups.
Among the three variants, the T1 map with CoordConv model achieved the best performance, reaching a VWA TPR of 0.88 $\pm$ 0.03, significantly higher than FreeSurfer (0.75 $\pm$ 0.03; p < 0.0001), THOMAS (0.79 $\pm$ 0.02; p < 0.0001), and HIPS-THOMAS (0.76 $\pm$ 0.02; p < 0.0001). 
At the nuclear group level, it showed significant improvements over FreeSurfer in ANT, MED, VA, VP, VL, and POST; over THOMAS in ANT, MED, CM, VA, VP, and VL; and over HIPS-THOMAS across all seven groups.
The T1 map with CoordConv model also displayed a consistent trend of higher TPRs than its MPRAGE and FGATIR counterparts, though most differences are not statistically significant.

\textit{Qualitative Results:}
In Figure~\ref{fig:s3-benchmark_qualitative_results}, we show segmentation outputs from CATNUS compared with FreeSurfer, THOMAS, and HIPS-THOMAS.
All our coordinate-aware models produced segmentations that closely align with the manual ground-truth, exhibiting smooth boundaries, accurate localization for nuclear groups, and clear delineation of thalamic borders across all views. 
These segmentations successfully preserve internal thalamic structure and avoid spillage into surrounding non-thalamic regions, reflecting high spatial precision. 
In contrast, FreeSurfer frequently produced fragmented or noisy outputs, with scattered false positives marked by yellow circles.
It also tends to over-segment the thalamus, spilling into adjacent subcortical regions beyond the thalamus boundary, as indicated by white arrows. 
THOMAS and HIPS-THOMAS generally produce very similar results. 
Compared with FreeSurfer, both methods yield visually smoother and more anatomically coherent segmentations, with improved nuclear localization and fewer isolated errors.
However, they systematically under-segment the thalamus, leaving many regions unlabeled, as highlighted by green arrows.
Overall, these qualitative observations reinforce our quantitative results, demonstrating that CATNUS produces anatomically coherent and reliable segmentations that outperform existing benchmark methods.

\textit{Test-retest Analysis:}
To evaluate segmentation reproducibility, we performed test--retest analysis using the 11 control participants described in Section~\ref{sec:2.2}. 
For each participant, the v2 scan was rigidly co-registered to the corresponding v1 scan and processed using the pipeline described in Section~\ref{sec:2.4}. 
For CATNUS, the nine participants whose v1 scans were included in the primary labeled cohort were segmented at both sessions using the corresponding held-out cross-validation model, ensuring that the model had not been trained on that participant.
For the remaining two participants, both sessions were segmented using the fold-0 cross-validation model. 
Benchmark methods were applied directly to both sessions. All segmentation results were then mapped into the seven harmonized nuclear groups.
For each group, let $A$ denote the segmentation mask in the v1 scan, and $B$ denote the corresponding segmentation mask in the v2 scan. 
The Dice coefficient was defined as:
\begin{equation}
\mathrm{Dice}(A,B)=\frac{2|A\cap B|}{|A|+|B|}.
\label{eq:dice}
\end{equation}
We computed Dice scores separately for each group and derived a VWA Dice score for each participant. 
The final results in Table~\ref{tab:test-retest_comparisons} show the mean and standard deviation across subjects. 
The coordinate-aware models achieved high test-retest Dice scores overall. Among them, the T1 map with CoordConv model achieved the highest VWA Dice score and the highest mean Dice in six of seven nuclear groups, outperforming FreeSurfer, THOMAS, and HIPS-THOMAS in nearly all groups. 
Statistical testing was performed on the VWA Dice score. The T1 map with CoordConv model was compared with each other method using two-sided paired Wilcoxon signed-rank tests, with Holm--Bonferroni correction across the five comparisons. 
The T1 map with CoordConv model achieved significantly higher VWA Dice than the FGATIR with CoordConv model (p = 0.0488), FreeSurfer (p = 0.0049), THOMAS (p = 0.0049), and HIPS--THOMAS (p = 0.0049). The comparison with the MPRAGE with CoordConv model did not reach statistical significance after correction (p = 0.0537).
These findings indicate that CATNUS produces robust and reproducible segmentations across repeated scans.

\subsection{Consistency With the Morel Atlas}\label{sec:3.4}
\begin{figure*}[ht]
  \centering
  \includegraphics[width=0.9\linewidth]{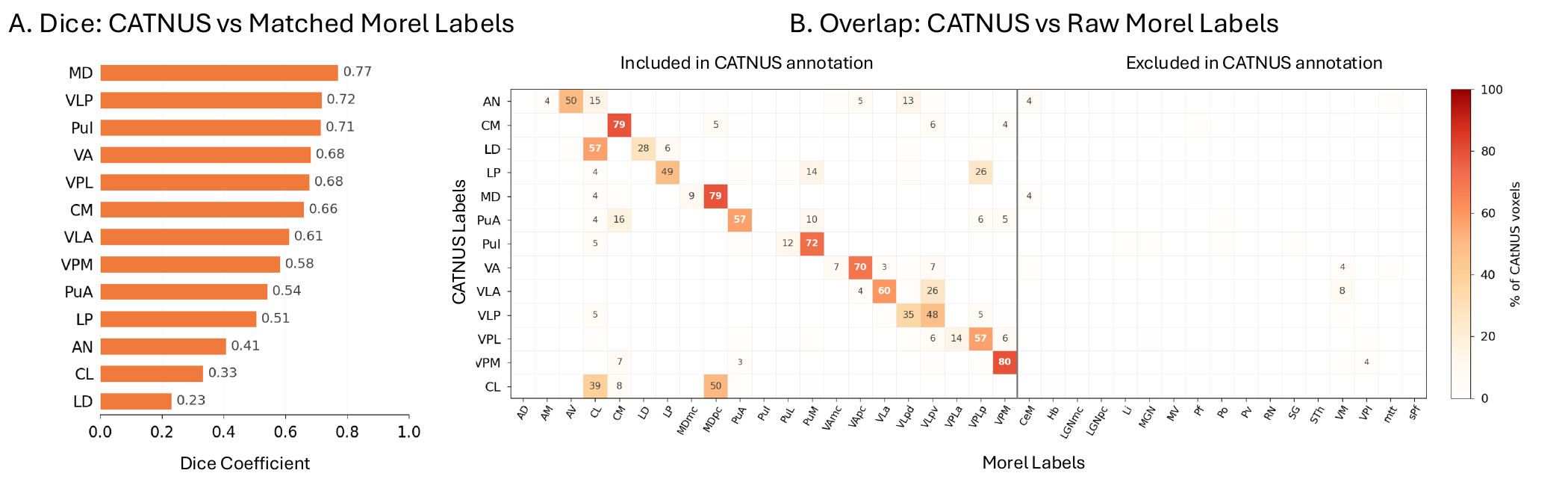}
  \caption{Quantitative comparison between CATNUS and the Morel atlas in template space.
  A. Dice overlap between CATNUS predictions and matched Morel labels. 
  B. Voxel-level overlap between CATNUS predictions and raw Morel labels. Each cell shows the percentage of voxels from a CATNUS prediction that fall within a raw Morel label. Raw Morel labels are grouped by whether they were included in the CATNUS annotation protocol. Only overlaps $\geq 3\%$ are annotated.
 }
  \label{fig:atlas-quantitative}
\end{figure*}

\begin{figure}[ht]
  \centering
  \includegraphics[width=\linewidth]{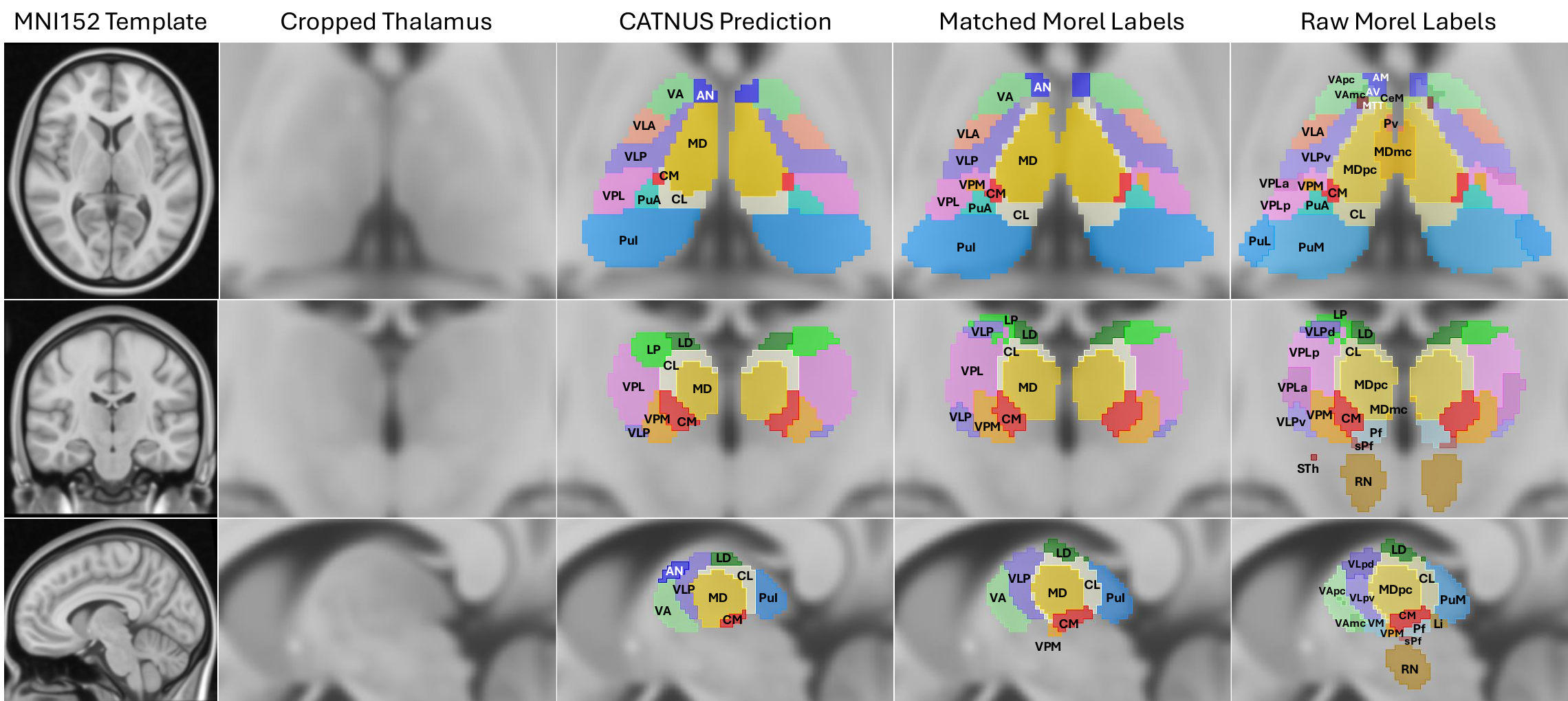}
  \caption{Qualitative comparison between CATNUS predictions and the Morel atlas in template space. Columns show the MNI152 T1w template, cropped thalamic region, CATNUS predictions, matched Morel labels, and raw Morel labels in axial, coronal, and sagittal views. Matched Morel labels were generated by merging Morel subnuclei according to the CATNUS 13-label definitions and removing structures not included in the CATNUS annotation protocol. Raw Morel labels show the original atlas labels before merging or exclusion.
 }
  \label{fig:atlas-qualitative}
\end{figure}

\textit{Experimental Design:}
To examine whether CATNUS predictions reflect histology-defined thalamic nuclei organization, we compared CATNUS predictions with the Morel atlas.
Specifically, we first registered the Morel T1w template from its original FSL standard space to the MNI152 space used by CATNUS, and then ran the CATNUS MPRAGE model on the processed template image.
We then applied the same registration to the Morel labels with nearest-neighbor interpolation.
For comparison, we used two Morel label representations.
Matched Morel labels were generated by merging Morel nuclei according to the CATNUS 13-label definitions and removing structures outside the CATNUS annotation protocol. Raw Morel labels kept the original Morel nuclei.
Agreement between CATNUS predictions and the matched Morel labels was quantified using the Dice coefficient.
 Correspondence with the raw Morel labels was evaluated using voxel-level overlap, defined for a CATNUS label $C$ and a raw Morel label $M$ as $|C \cap M| / |C| \times 100\%$, representing the percentage of voxels in the CATNUS prediction that fell within the Morel label.

\textit{Quantitative Results:}
Figure~\ref{fig:atlas-quantitative} summarizes the quantitative comparison between CATNUS predictions and the Morel atlas.
CATNUS predictions showed good agreement with the matched Morel labels for major nuclei, with a volume-weighted average Dice of 0.65. 
The highest Dice values were observed for MD, VLP, PuI, VA, VPL, and CM, reaching 0.77, 0.72, 0.71, 0.68, 0.68, and 0.66, respectively. 
VLA, VPM, PuA, and LP showed moderate agreement, while AN, CL, and LD were lower, likely reflecting residual registration error, smoothing in the average template, and boundary sensitivity of small nuclei.
The overlap matrix showed that CATNUS predictions mainly overlapped with the expected Morel nuclei included in the CATNUS annotation protocol. 
For example, CATNUS MD, VA, VLP, VPL, and PuI primarily overlapped with Morel MDmc/MDpc, VAmc/VApc, VLpv/VLpd, VPLa/VPLp, and PuI/PuL/PuM, respectively.
In contrast, CATNUS predictions showed only limited overlap with Morel nuclei excluded from the annotation protocol, indicating that the predictions were largely restricted to the target nuclei.
Overall, these results support the consistency between CATNUS predictions and the major thalamic nuclei organization defined by the Morel atlas.

\textit{Qualitative Results:}
Figure~\ref{fig:atlas-qualitative} shows that CATNUS predictions had similar overall spatial organization and thalamic shape to the matched Morel labels, while showing limited overlap with additional structures in the raw Morel labels that were excluded from the CATNUS annotation protocol. These observations further support the spatial consistency between CATNUS predictions and the Morel atlas.

\subsection{Out-of-Distribution Testing}\label{sec:3.5}
\begin{figure}[ht]
  \centering
  \includegraphics[width=0.6\linewidth]{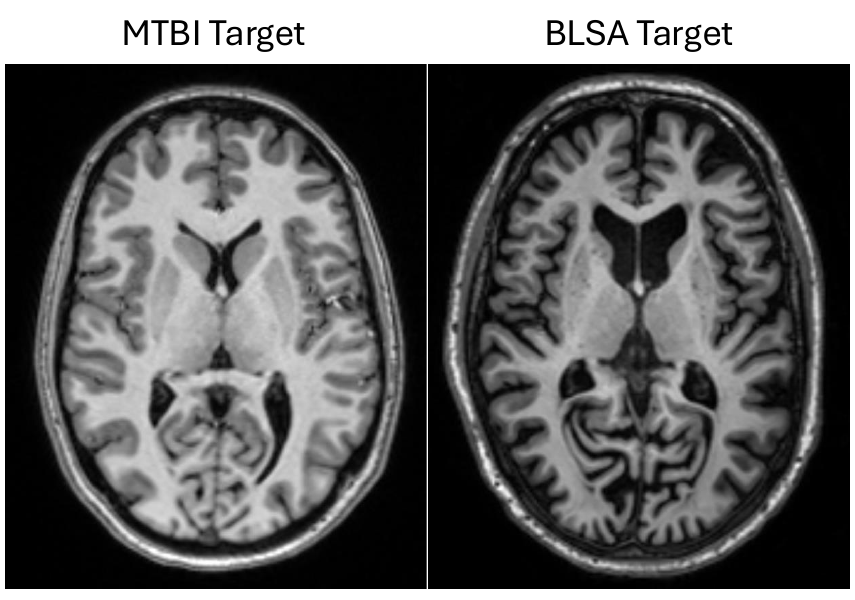}
  \caption{MTBI and BLSA harmonization targets used in OOD evaluation. Both are 3 T T1-weighted images shown on the same slice after identical preprocessing (registration to MNI152, N4 bias correction, and white-matter mean normalization). The MTBI target depicts a healthy young adult from the mild traumatic brain injury study (SIEMENS Prisma), whereas the BLSA target depicts a healthy aging participant from the Baltimore Longitudinal Study of Aging (Philips Achieva).
 }
  \label{fig:ood-Harm}
\end{figure}

\textit{Experimental Design:}
To evaluate model generalizability beyond the training domain, we performed out-of-distribution (OOD) testing using traveling-subject datasets.
These datasets scan the same individuals across multiple sites, allowing quantitative assessment of segmentation stability under scanner and protocol variation, eliminating inter-subject differences as a confound.
We used two publicly available traveling-subject datasets. 
The Frequently Traveling Human Phantom (FTHP) dataset~\citep{opfer2023automatic} provides T1-weighted scans of a single healthy male (about 50 years old) acquired on 116 scanners over 2.5 years.
For this analysis, 80 scans were used, consisting of four T1-weighted scans each from 20 different scanners (10 at 3.0T and 10 at 1.5T). 
The Multisite, Multiscanner, and Multisubject Acquisitions for Studying Variability (MASiVar) dataset~\citep{cai2021masivar} contains both diffusion and T1-weighted scans. 
We used only T1-weighted scans from cohort II, comprising five healthy adults (3 males, 2 females, ages 27–47), each scanned on four scanners with one to two sessions per scanner. 
Scanner details of both datasets are summarized in Table~\ref{tab:scanner_info}.
All scans were registered to MNI152 space, followed by independent N4 bias field correction and white-matter mean normalization, without joint processing with FGATIR.
To systematically examine the impact of scanner- and protocol-dependent intensity differences on segmentation stability, we tested three input conditions:
(1) Non-harmonized, preprocessed images reflecting inherent robustness to domain shift;
(2) Harmonized-to-MTBI, images harmonized to the training domain (MTBI dataset) using Harmonization with Attention-based Contrast, Anatomy, and Artifact Awareness (HACA3)~\citep{zuo2023haca3}, assessing whether stability can be recovered when input contrast matches the training distribution;
(3) Harmonized-to-BLSA, images harmonized to an independent domain from the Baltimore Longitudinal Study of Aging (BLSA)~\citep{resnick2000one},  reducing the possibility that improvements in (2) arise from the model recognizing familiar training patterns.
The MTBI and BLSA targets differ in scanner vendor, acquisition protocol, and participant demographics, shown in Figure~\ref{fig:ood-Harm}, avoiding confusion that stability improvements may stem from target similarity.

\begin{table}[t]
\centering
\small
\caption{Scanner Information, including vendor, model, and field strength (FS) for three datasets. The MTBI dataset was used for training, while the FTHP and MASiVar are traveling-subject datasets for out-of-distribution (OOD) testing. In MASiVar, Site V includes two Philips Achieva scanners, denoted as ‘Achieva’ and ‘Achieva*’.}
\label{tab:scanner_info}
\begin{tabular}{lllll}
\hline
Dataset & Site & Vendor & Model & FS \\
\hline
MTBI & Site A & SIEMENS & Prisma & 3.0T \\
\hline
\multirow{20}{*}{FTHP} & Site B & Philips & Intera & 1.5T \\
 & Site C & SIEMENS & Aera & 1.5T \\
 & Site D & SIEMENS & Avanto & 1.5T \\
 & Site E & Philips & Achieva & 1.5T \\
 & Site F & SIEMENS & Espree & 1.5T \\
 & Site G & SIEMENS & Symphony & 1.5T \\
 & Site H & GE & Optima  MR450 & 1.5T \\
 & Site I & Philips & Ingenia & 1.5T \\
 & Site J & GE & Optima MR360 & 1.5T \\
 & Site K & GE & Signa HDxt & 1.5T \\
 & Site L & GE & Discovery MR750 & 3.0T \\
 & Site M & SIEMENS & Skyra & 3.0T \\
 & Site N & Philips & Ingenia & 3.0T \\
 & Site O & SIEMENS & Verio & 3.0T \\
 & Site P & Philips & Achieva & 3.0T \\
 & Site Q & SIEMENS & MAGNETOM Vida & 3.0T \\
 & Site R & SIEMENS & Spectra & 3.0T \\
 & Site S & SIEMENS & MAGNETOM Lumina & 3.0T \\
 & Site T & GE & SIGNA Architect & 3.0T \\
 & Site U & SIEMENS & Prisma & 3.0T \\
\hline
\multirow{4}{*}{MASiVar} 
 & Site V & Philips & Achieva & 3.0T \\
 & Site V & Philips & Achieva* & 3.0T \\
 & Site W & GE & Discovery MR750 & 3.0T \\
 & Site X & SIEMENS & Skyra & 3.0T \\
\hline
\end{tabular}
\end{table}

\begin{table*}[htbp]
\centering
\caption{
Quantitative results of out-of-distribution (OOD) segmentation using traveling-subject datasets.
We assess CATNUS generalizability on the FTHP and MASiVar datasets under three harmonization conditions: non-harmonized, harmonized to MTBI, and harmonized to BLSA. Segmentation stability was quantified using the coefficient of variation (CV = standard deviation / mean volume), reported in percentages ($\%$). Lower CV values indicate greater segmentation stability and are highlighted in bold. “Overall” represents the combined thalamic volume across all nuclear groups.
}
\label{tab:stability}
\vspace{0.2cm}
(a) FTHP - Non-harmonized Data\\
\vspace{0.1cm}
\begin{tabular}{lccccccccc}
\toprule
\begin{tabular}[c]{@{}c@{}}CV\\(Std/Mean)\end{tabular}
& ANT & MED & CM & VA & VP & VL & POST & Overall \\
\midrule
CATNUS & 15.8\% & 10.3\% & 9.3\% & 14.5\% & 13.1\% & 7.8\% & 6.9\% & 6.2\% \\
HIPS-THOMAS & 17.9\% & 9.2\% & 10.2\% & 7.0\% & 8.2\% & 7.2\% & 7.4\% & 6.7\% \\
FreeSurfer & {\textbf{11.2\%}} & {\textbf{7.6\%}} & {\textbf{4.9\%}} & {\textbf{6.5\%}} & {\textbf{5.5\%} }& {\textbf{4.2\%}} & {\textbf{6.6\%}} & {\textbf{4.0\%}} \\
\bottomrule
\end{tabular}

\vspace{0.3cm}
(b) FTHP - Harmonized Data to MTBI\\
\vspace{0.1cm}
\begin{tabular}{lccccccccc}
\toprule
\begin{tabular}[c]{@{}c@{}}CV\\(Std/Mean)\end{tabular}
& ANT & MED & CM & VA & VP & VL & POST & Overall \\
\midrule
CATNUS & {\textbf{2.7\%}} & {\textbf{2.0\%}} & {\textbf{2.4\%}} & {\textbf{3.3\%}} & {\textbf{2.7\%}} & {\textbf{1.8\%}} & {\textbf{1.3\%}} & {\textbf{1.1\%}} \\
HIPS-THOMAS & 16.7\% & 8.7\% & 10.2\% & 6.8\% & 7.2\% & 7.0\% & 7.0\% & 6.5\% \\
FreeSurfer & 7.1\% & 3.8\% & 3.5\% & 3.8\% & 3.3\% & 2.2\% & 3.7\% & 1.9\% \\
\bottomrule
\end{tabular}

\vspace{0.3cm}
(c) FTHP - Harmonized Data to BLSA\\
\vspace{0.1cm}
\begin{tabular}{lccccccccc}
\toprule
\begin{tabular}[c]{@{}c@{}}CV\\(Std/Mean)\end{tabular}
& ANT & MED & CM & VA & VP & VL & POST & Overall \\
\midrule
CATNUS & {\textbf{3.0\%}} & 
{\textbf{2.0\%}} & {\textbf{2.0\%}} & 3.9\% & {\textbf{2.4\%}} & {\textbf{1.6\%}} & {\textbf{1.3\%}} & {\textbf{1.1\%}} \\
HIPS-THOMAS & 18.3\% & 8.2\% & 9.4\% & 6.6\% & 7.4\% & 6.9\% & 6.9\% & 6.5\% \\
FreeSurfer & 6.4\% & 3.6\% & 3.6\% & {\textbf{3.3\%}} & 3.3\% & 2.3\% & 3.4\% & 1.8\% \\
\bottomrule
\end{tabular}

\vspace{0.3cm}
(d) MASiVar - Non-harmonized Data\\
\vspace{0.1cm}
\begin{tabular}{lccccccccc}
\toprule
\begin{tabular}[c]{@{}c@{}}CV\\(Std/Mean)\end{tabular}
& ANT & MED & CM & VA & VP & VL & POST & Overall \\
\midrule
CATNUS & {\textbf{5.1\%}} & 7.2\% & 12.9\% & 8.2\% & 15.0\% & 11.6\% & 12.2\% & 5.8\% \\
HIPS-THOMAS & 7.6\% & {\textbf{2.9\%}} & {\textbf{5.3\%}} & {\textbf{2.4\%}} & {\textbf{5.7\%}} & {\textbf{2.5\%}} & {\textbf{1.6\%}} & {\textbf{1.8\%}} \\
FreeSurfer & 7.5\% & 6.6\% & 10.0\% & 5.0\% & 14.3\% & 9.7\% & 3.9\% & 7.0\% \\
\bottomrule
\end{tabular}

\vspace{0.3cm}
(e) MASiVar - Harmonized Data to MTBI\\
\vspace{0.1cm}
\begin{tabular}{lccccccccc}
\toprule
\begin{tabular}[c]{@{}c@{}}CV\\(Std/Mean)\end{tabular}
& ANT & MED & CM & VA & VP & VL & POST & Overall \\
\midrule
CATNUS & {\textbf{3.2\%}} & {\textbf{2.2\%}} &  {\textbf{2.4\%}} & {\textbf{2.0\%}} &  {\textbf{1.6\%}} & {\textbf{2.2\%}} &  {\textbf{1.1\%}} & {\textbf{0.8\%}} \\
HIPS-THOMAS & 6.3\% & 3.0\% & 7.0\% & 3.0\% & 4.4\% & 2.3\% & 2.2\% & 1.6\% \\
FreeSurfer & 6.8\% & 2.7\% & 4.8\% & 2.8\% & 8.1\% & 4.4\% & 3.8\% & 3.9\% \\
\bottomrule
\end{tabular}

\vspace{0.3cm}
(f) MASiVar - Harmonized Data to BLSA\\
\vspace{0.1cm}
\begin{tabular}{lccccccccc}
\toprule
\begin{tabular}[c]{@{}c@{}}CV\\(Std/Mean)\end{tabular}
& ANT & MED & CM & VA & VP & VL & POST & Overall \\
\midrule
CATNUS & {\textbf{3.6\%}} & {\textbf{2.5\%}} & {\textbf{2.2\%}} & {\textbf{2.5\%}} & {\textbf{1.7\%}} & {\textbf{1.9\%}} & {\textbf{1.2\%}} & {\textbf{1.1\%}}\\
HIPS-THOMAS & 5.5\% & 3.3\% & 5.9\% & 3.2\% & 4.8\% & {\textbf{1.9\%}} & 2.1\% & 1.5\% \\
FreeSurfer & 7.4\% & 2.6\% & 3.2\% & 2.7\% & 5.6\% & 3.0\% & 2.8\% & 2.4\% \\
\bottomrule
\end{tabular}
\end{table*}

\begin{figure*}[!ht]
  \centering
  \includegraphics[width= 0.8\linewidth]{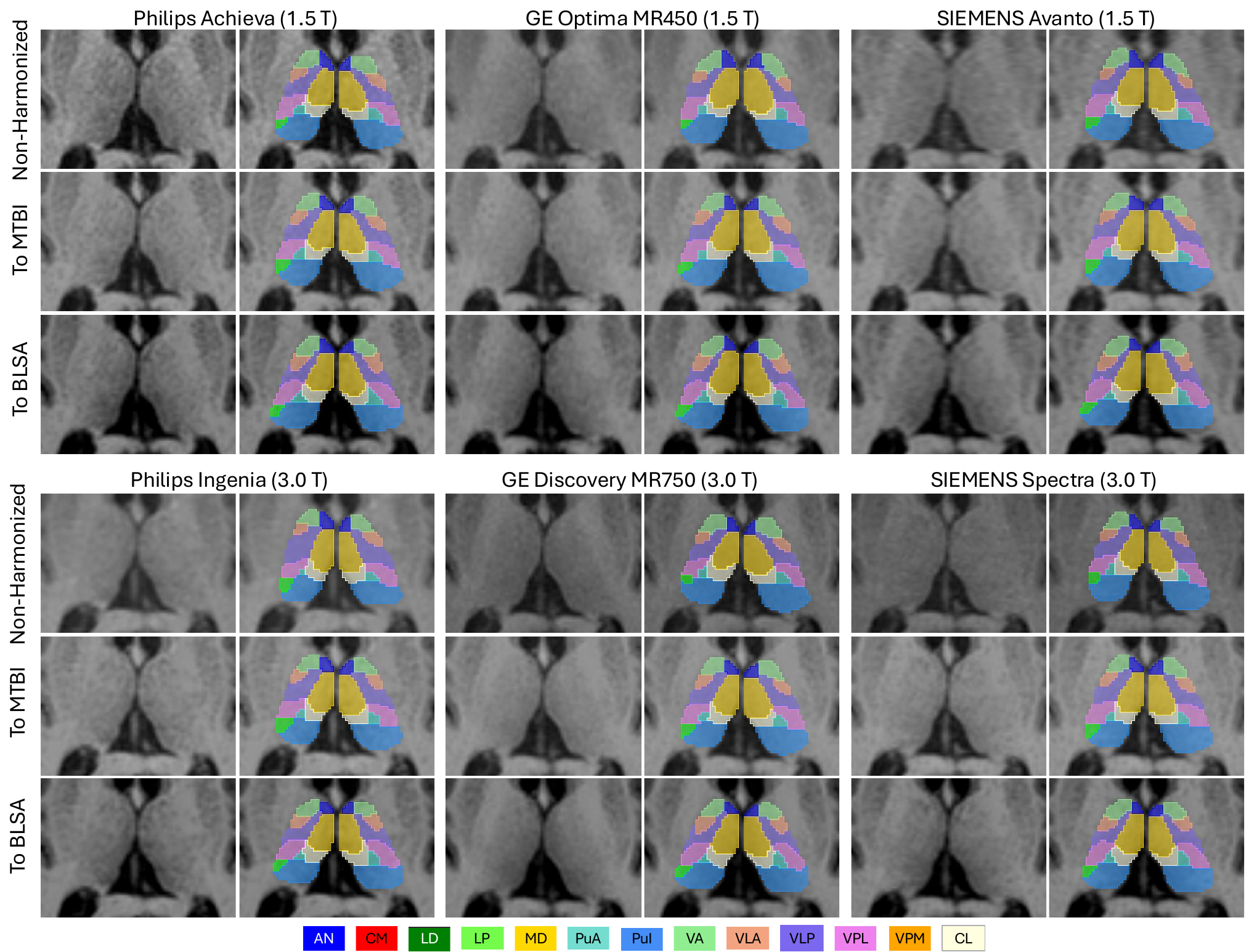}
  \caption{
  Qualitative results of out-of-distribution (OOD) segmentation on the FTHP traveling-subject dataset.
  Results are shown for six scanners spanning 1.5 T and 3.0 T field strengths from three vendors (Philips, GE, SIEMENS). 
  All scans were acquired from the same subject, and the same anatomical slice is shown after registration to the MNI space.
  Rows correspond to three harmonization conditions: non-harmonized, harmonized to MTBI, and harmonized to BLSA.
}
  \label{fig:ood-FTHP}
\end{figure*}

\begin{figure*}[!ht]
  \centering
  \includegraphics[width=0.8\linewidth]{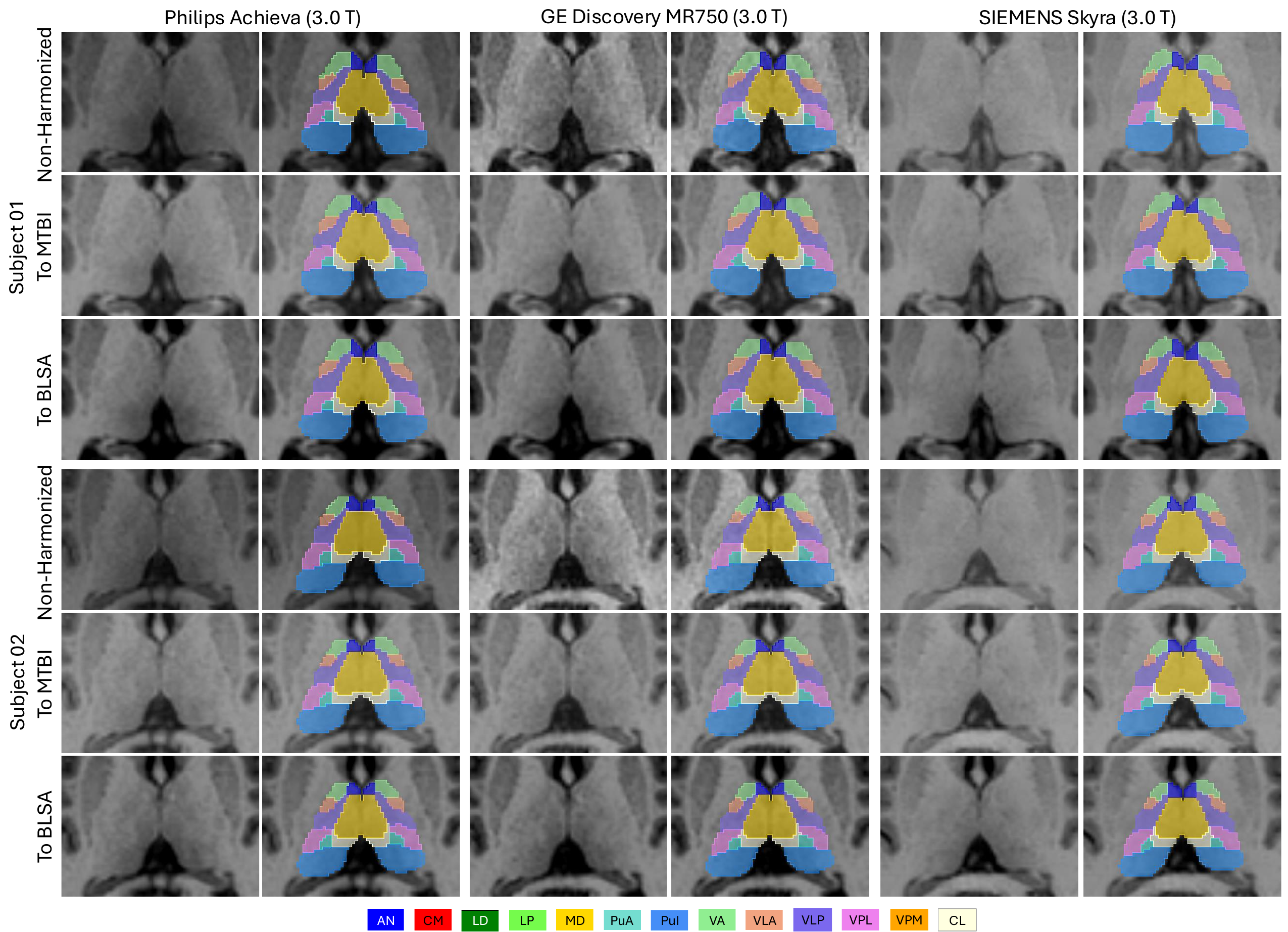}
  \caption{
  Qualitative results of out-of-distribution (OOD) segmentation on the MASiVar traveling-subject dataset.
  Results are shown for two representative subjects scanned on three scanners (Philips Achieva, GE Discovery MR750, and SIEMENS Skyra). 
  For each subject and scanner, the same anatomical slice after registration to the MNI space is shown, with predicted thalamic nuclei overlaid. Rows correspond to three harmonization conditions: non-harmonized, harmonized to MTBI, and harmonized to BLSA T1-weighted data.}
  \label{fig:ood-MASiVar}
\end{figure*}

\textit{Evaluation Strategy:} 
Because these datasets lack manual annotations, evaluation focused on two main aspects: 
(1) quantitative stability, measured by the coefficient of variation (CV = standard deviation/mean volume) across repeated scans of the same subject, and
(2) qualitative consistency, assessed by visual inspection of segmentation plausibility and anatomical coherence across scanners.
For FTHP, CVs were computed across all 80 scans to measure scanner-induced variability.
For MASiVar, subject-level CVs were first computed across scanners and then averaged across five subjects.
We compared CATNUS against FreeSurfer and HIPS-THOMAS. 
HIPS-THOMAS was chosen instead of THOMAS because it is optimized for T1-weighted MRI and provides more accurate and reliable segmentations for conventional MPRAGE scans.
Outputs from all methods were mapped to seven major nuclear groups according to the protocol in Figure~\ref{fig:s2-label_unifying_protocols} for comparison. 

\textit{Quantitative Results:}
Table~\ref{tab:stability} reports the CV across all methods and harmonization conditions.
On the FTHP dataset (Table~\ref{tab:stability}a-c), non-harmonized data showed FreeSurfer achieving the lowest overall CV at 4.0\%, followed by CATNUS at 6.2\% and HIPS-THOMAS at 6.7\%. 
However, harmonization altered this ranking. CATNUS improved nearly six-fold to 1.1\% for both MTBI and BLSA targets, outperforming FreeSurfer (1.9\% MTBI; 1.8\% BLSA) and HIPS-THOMAS (6.5\% for both).
On the MASiVar dataset (Table~\ref{tab:stability}d-f), non-harmonized HIPS-THOMAS achieved the lowest overall CV (1.8\%) compared to CATNUS (5.8\%) and FreeSurfer (7.0\%). After harmonization, CATNUS again achieved the best stability, with overall CVs of 0.8\% (MTBI) and 1.1\% (BLSA), surpassing HIPS-THOMAS (1.6\% MTBI; 1.5\% BLSA) and FreeSurfer (3.9\% MTBI; 2.4\% BLSA). 
Across major nuclear groups, both datasets showed the same pattern: harmonization consistently reduced variability for CATNUS and FreeSurfer, with CATNUS showing consistently low CVs after harmonization, while HIPS-THOMAS remained largely unchanged. 
In addition, performance was nearly identical under MTBI- and BLSA-harmonization (e.g., overall CVs of 1.1\% vs 1.1\% on FTHP, and 0.8\% vs 1.1\% on MASiVar), suggesting stability gains stem from intensity harmonization itself rather than alignment to a specific training domain.

\textit{Qualitative Results:}
Visual inspection showed strong generalization across scanners and sites.
Figure~\ref{fig:ood-FTHP} shows segmentations from the FTHP dataset across six scanners spanning 1.5 T and 3.0 T field strengths and three vendors (Philips, GE, and SIEMENS). 
Despite being trained only on 3.0 T data, our model consistently localized the thalamus and nuclei with smooth, anatomically coherent boundaries across all scanners. 
Figure~\ref{fig:ood-MASiVar} shows segmentations from the MASiVar dataset for two subjects scanned on three different scanners from three vendors (Philips, GE, and SIEMENS). 
Our model accurately captured inter-subject anatomical differences, with thalamic size, nuclear proportions, and shapes varying between individuals. Within each subject, segmentations remained stable across scanners. 
The same anatomical structures maintained consistent boundaries, small nuclei preserved their positions, and spatial relationships between nuclear groups stayed coherent. 
In addition, the model produced anatomically coherent segmentations even on non-harmonized data, indicating strong inherent generalization, with harmonization further improving stability across scanners.

\subsection{Alzheimer's Disease Discrimination Analysis}\label{sec:3.6}

\textit{Experimental Design:}
To further assess the effectiveness of CATNUS in a clinically relevant
downstream setting, we evaluated whether CATNUS-derived thalamic nuclei
volumes could capture anatomical differences associated with Alzheimer's
disease (AD)~\citep{iglesias2018probabilistic, williams2024thalamic}.
We obtained data from the
Alzheimer's Disease Neuroimaging Initiative (ADNI) dataset~\citep{jack2008alzheimer}. The ADNI was launched in 2003 as a
public-private partnership, led by Principal Investigator Michael W. Weiner,
MD. The primary goal of ADNI has been used to test whether serial magnetic
resonance imaging (MRI), positron emission tomography (PET), other biological
markers, and clinical and neuropsychological assessment can be combined to
measure the progression of mild cognitive impairment (MCI) and early
Alzheimer's disease (AD).
For this analysis, we included
one MPRAGE session from each of 430 ADNI participants, consisting of
190 patients with AD and 240 cognitively normal (CN) controls. 
Subjects were restricted to the 70--80 year age range, and the AD and CN
groups showed comparable age distributions ($74.9 \pm 3.0$ and
$74.5 \pm 2.7$ years, respectively). 
The AD
group included 96 female and 94 male subjects, while the CN group
included 139 female and 101 male subjects. The cohort spanned 68 ADNI
site codes and multiple scanner configurations, including 347 3T and
83 1.5T scans acquired on Siemens, GE, and Philips scanners.

\textit{Classification:}
For each subject, MPRAGE images were processed using registration, N4
bias-field correction, and white-matter mean normalization. 
For CATNUS, the preprocessed images were additionally harmonized to the MTBI domain using HACA3. 
For comparison, FreeSurfer and HIPS-THOMAS were applied to the preprocessed MPRAGE images without HACA3 harmonization. 
For each
method, we extracted both whole-thalamus and nuclei-level volumes, and evaluated two classification strategies, one based on the whole-thalamus volume and the other based on all nuclei-level volumes.
Each volume feature was corrected for both age and estimated total
intracranial volume (eTIV), with eTIV obtained from FreeSurfer. Specifically, we fitted an ordinary least-squares model
$y = \beta_0 + \beta_1 \mathrm{age} + \beta_2 \mathrm{eTIV} + \epsilon$
for each feature, and used the residuals added back to the original feature
mean as the corrected volume.
AD/CN classification was
then performed using leave-one-out cross-validation with linear discriminant analysis (LDA).
We generated receiver operating characteristic (ROC) curves and computed the area under the ROC curve (AUC) for all three methods using both whole-thalamus and nuclei-level feature sets. 
For each feature set, we compared the AUC of CATNUS with those of FreeSurfer and HIPS-THOMAS using paired, two-sided DeLong tests \citep{delong1988comparing}. 
Holm--Bonferroni correction was applied across the four comparisons, with p < 0.05 considered statistically significant.

\begin{figure}[ht]
  \centering
  \includegraphics[width=0.7\linewidth]{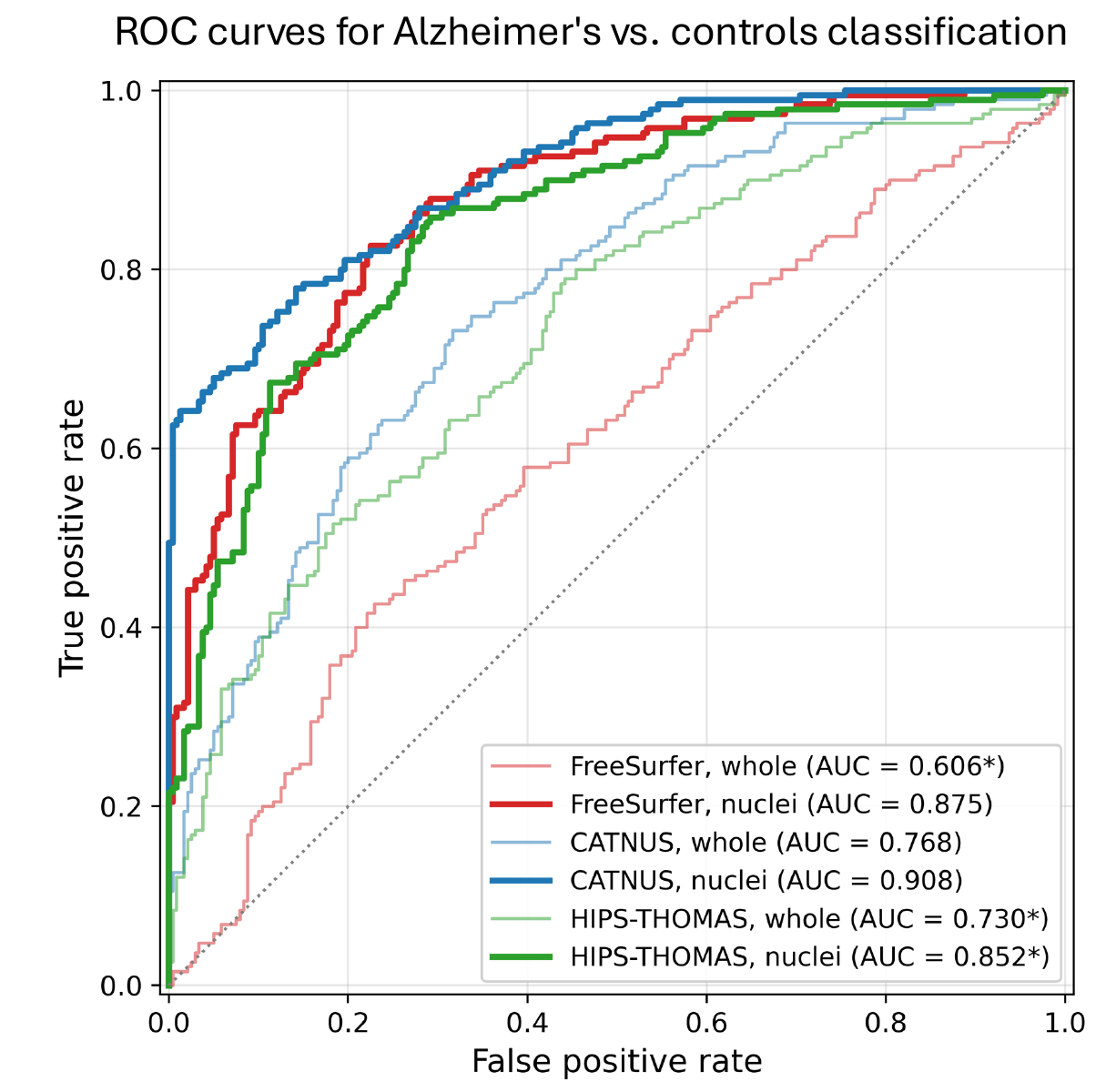}
  \caption{
    ROC curves for Alzheimer's disease (AD) versus cognitively normal controls (CN) classification using whole-thalamus and thalamic nuclei volumes. For each method, lighter curves represent whole-thalamus classification and darker curves represent nuclei-level classification.
     AUCs were computed from LDA decision scores obtained by leave-one-out cross-validation. All volume features were corrected for age and estimated total intracranial volume.
    Asterisks indicate that CATNUS achieved a significantly higher AUC than the marked method, with separate comparisons for whole-thalamus and nuclei-level classification (p < 0.05 after Holm–Bonferroni correction).
}
  \label{fig:adni}
\end{figure}

\textit{Results:}
As shown in Figure~\ref{fig:adni}, nuclei-level volumes yielded higher AUCs than whole-thalamus volume for all three methods.
CATNUS achieved the highest nuclei-level AUC (0.908), with a significant difference from HIPS-THOMAS (0.852; p = 0.0051) but no significant difference from FreeSurfer (0.875; p = 0.0697). 
For whole-thalamus volume, CATNUS achieved an AUC of 0.768, significantly higher than FreeSurfer (0.606; p < 0.001) and HIPS-THOMAS (0.730; p = 0.0412).
These results suggest that CATNUS-derived thalamic nuclei volumes preserve disease-relevant anatomical information useful for downstream neuroimaging analyses.

\section{Discussion}\label{sec:4}
In this study, we developed CATNUS, a coordinate-aware 3D U-Net framework for thalamic nuclei segmentation, providing three single-input models for T1 maps, MPRAGE, and FGATIR. The T1 map model achieved the best overall performance, while the MPRAGE and FGATIR models offer practical alternatives for more widely available clinical scans. CoordConv did not improve all nuclei uniformly, but it markedly reduced recurrent misses and mislocalization of PuA, the smallest nucleus in our label set. We therefore selected the T1 map model with CoordConv as the primary configuration, balancing robust overall segmentation with reliable PuA localization. CATNUS also outperformed established methods in TPR and test–retest reliability, showed consistency with the major thalamic nuclear organization of the Morel atlas, and demonstrated strong generalization across out-of-distribution datasets. In addition, CATNUS-derived volumetric features provided the strongest discrimination between Alzheimer’s disease and cognitively normal groups. Together, these findings support CATNUS as a reliable and practical framework for thalamic nuclei segmentation and downstream neuroimaging analyses.

The stronger overall performance of the T1 map model is consistent with our prior study~\citep{feng2025segmenting}. 
In that study, we compared a broad range of input settings, such as MPRAGE alone, FGATIR alone, T1 map alone, MPRAGE + FGATIR, PD map + T1 map, selected multi-TI subsets, and the full set of 51 multi-TI images. The T1 map alone outperformed or performed comparably to most other input settings.
We therefore used the T1 map as the primary input because it offered strong performance while being simpler and more efficient than stacked multi-TI inputs.
Its strong performance can be explained by MRI signal formation. 
The intensity of a T1-weighted image is governed by two intrinsic tissue properties: the longitudinal relaxation time (T1) and proton density (PD). 
Given these two parameters, one can mathematically synthesize any T1-weighted image, including MPRAGE and FGATIR, by adjusting the inversion and repetition times.
In practice, however, PD maps contribute little unique contrast beyond what is already encoded in the T1 maps~\citep{feng2025segmenting}. 
The important tissue-specific contrast across different T1-weighted acquisitions is inherently captured in the T1 maps.
Although certain features in the T1 maps may not be visually apparent to human raters, deep learning models are capable of leveraging subtle patterns and contextual cues to delineate complex anatomical structures, which may explain why models trained on T1 maps alone can achieve high segmentation accuracy.
That said, should we still compute multi-TI images? 
We argue yes, but for a different purpose. While unnecessary for training segmentation models, multi-TI images offer enhanced visual contrast that helps human raters in identifying thalamic boundaries and selecting optimal TIs for manual annotation. 
Thus, T1 maps are sufficient for model input, while multi-TI images remain valuable for human-guided delineation and quality control.

One might question why we chose single-input models rather than multi-channel inputs. While theoretically we could use MPRAGE, FGATIR, and T1 maps as multi-channel inputs, we chose a single-input strategy for the following reasons: First, not all clinical environments can acquire multiple T1-weighted sequences. Second, when MPRAGE and FGATIR are available, T1 maps can be computed from these acquisitions, and T1 maps inherently encode the core tissue contrast information needed for segmentation. Additional input channels would introduce feature redundancy, potentially dispersing the model's learning capacity and increasing computational complexity. Therefore, we provide three single-input models for different clinical scenarios.

Our finding that the T1 map outperformed FGATIR differs from previous studies favoring WMn or FGATIR-like contrast~\citep{umapathy2022convolutional,ruiz2024deepthalamus}. This difference may partly reflect the source of the labels. \cite{umapathy2022convolutional} used THOMAS-generated labels for both training and evaluation, and the labels used by DeepThalamus were also derived from a THOMAS-based labeling framework. Because THOMAS was developed for WMn-MPRAGE, these labels may be more closely aligned with WMn or FGATIR-like contrast. In contrast, our labels were manually delineated on multi-TI images and verified on quantitative T1 maps. Moreover, FGATIR provides a single T1-weighted contrast governed by tissue T1, whereas the T1 map directly quantifies T1. 

CoordConv was introduced to improve the localization of very small nuclei.
Its benefit was most evident for the anterior pulvinar (PuA), the smallest nucleus in our label set. Without coordinate encoding, PuA was frequently missed or mislocalized, while CoordConv enabled reliable localization.
This benefit reflects the inherent difficulty of segmenting very small structures.
Small nuclei occupy very few voxels, are often underrepresented in training, and are more susceptible to class imbalance. 
In encoder–decoder architectures like U-Net, higher-level features tend to emphasize large, salient regions, making small structures easier to overlook~\citep{lin2017feature}. 
In addition, small nuclei often have complex, irregular boundaries and are more sensitive to spatial perturbations introduced by data augmentation, interpolation, and pooling, leading to unstable predictions and noisy edges.
CoordConv mitigates these issues by explicitly providing the network with normalized spatial coordinates, giving it direct access to positional information~\citep{liu2018intriguing}. 
This reduces the burden of inferring spatial structure from appearance and enables the model to learn location-aware features more efficiently.

CATNUS was benchmarked against three widely used thalamic segmentation tools: FreeSurfer, THOMAS and HIPS-THOMAS. 
Each method was applied to its recommended input modality, and all outputs were mapped to a unified labeling scheme. 
CATNUS outperformed these baselines in segmentation accuracy and test–retest reliability across most thalamic regions.
It is also more efficient, producing a full 3D segmentation in under one minute on a single GPU, compared to approximately 10 minutes for THOMAS and HIPS-THOMAS and several hours for FreeSurfer.
However, this comparison may favor CATNUS because its training and evaluation labels were created using the same annotation protocol, while the benchmark methods were developed using different atlases and labeling conventions. 
Mapping all outputs to a unified scheme reduces but cannot eliminate these differences.

To assess generalization beyond the training domain, we evaluated CATNUS on traveling-subject datasets, which isolate scanner-induced variability while keeping anatomy constant. 
Results showed that CATNUS produced anatomically coherent segmentations even without harmonization, while harmonization further improved cross‐scanner consistency and reduced volumetric variability across repeated scans. 
Therefore, during inference, harmonization is applied as a preprocessing step prior to segmentation to ensure robust performance. 
Different methods benefited from harmonization to varying degrees.
CATNUS and FreeSurfer both showed improved stability, with CATNUS exhibiting a larger gain due to its higher sensitivity to input intensity distribution. 
FreeSurfer, while more robust to contrast variations owing to its probabilistic atlas framework, also achieved better consistency after harmonization. 
In contrast, HIPS‐THOMAS showed little improvement, as its polynomial synthesis model already performs an implicit form of intensity normalization.
We also found that harmonization toward different targets (MTBI vs. BLSA) yielded comparable results, suggesting that the observed improvements stem from consistent intensity statistics rather than matching to the training domain.
However, the current CATNUS framework supports harmonization only for standard T1-weighted images using HACA3. 
We are exploring extensions to T1 maps and FGATIR inputs, which may further enhance the model’s robustness across supported input modalities.

In addition to the limitations discussed above,  several others should be considered. 
First, the labeled dataset included only 24 subjects from a single clinical cohort, limiting anatomical and demographic diversity.
Second, manual delineation was completed by one primary annotator, so inter-rater reliability and agreement in optimal TI selection could not be evaluated.
Third, the manual annotations are sparse, with only high-confidence voxels labeled. 
While this improves label precision, it leaves many thalamic voxels unused during training and reduces the available supervision.
Many unlabeled voxels lie along uncertain nucleus boundaries, limiting direct evaluation of boundary accuracy.
Fourth, our evaluation relied on the true positive rate, which measures sensitivity on labeled voxels but does not penalize false positives in unlabeled regions. 
Conventional metrics such as Dice, Hausdorff distance, and volume difference are also not reliable without complete labels.
Fifth, the current labeling protocol does not include several important thalamic structures, such as the medial and lateral geniculate nuclei (MGN and LGN), habenula (Hb), and mammillothalamic tract (MTT), limiting the applicability of CATNUS in certain studies.
Sixth, our T1 maps were computed from a dual inversion time approach. 
However, T1 
quantification can be performed using various alternative methods, including MP2RAGE~\citep{marques2010mp2rage}, variable flip angle~\citep{fram1987rapid}, and multi-point inversion 
recovery sequence~\citep{messroghli2004modified}, which may produce T1 maps with different contrast characteristics and numerical ranges. 
Model performance on T1 maps derived from these methods remains to be validated.
Seventh, comparison of CATNUS with the Morel atlas evaluates template-level consistency rather than independent validation, and comparison with the Allen Brain Atlas~\citep{ding2016comprehensive} remains to be performed.
Finally, OOD testing was conducted only on standard T1-weighted images because no public datasets currently provide T1 map or FGATIR acquisitions.

Future work will address the limitations described above and further evaluate CATNUS in independent disease cohorts. 
The discrimination of participants with Alzheimer’s disease from cognitively normal participants demonstrates the potential of CATNUS-derived thalamic nucleus volumes for disease-related analyses. Another important direction is deep brain stimulation (DBS). 
The anterior nucleus is an established DBS target for drug-resistant focal epilepsy~\citep{fisher2010electrical,salanova2015long}. 
The CM nucleus is being investigated for Lennox–Gastaut syndrome and other generalized or frontal epilepsies~\citep{ valentin2013deep, dalic2022dbs}. 
The CL nucleus and related central thalamic pathways have been explored to promote recovery in chronic traumatic brain injury and disorders of consciousness and to treat central poststroke pain~\citep{schiff2007behavioural, schiff2023thalamic, nowacki2023deep}. 
The VPL nucleus has been targeted for neuropathic pain after amputation or brachial plexus avulsion~\citep{pereira2013thalamic}. 
The VPM nucleus, together with the periaqueductal gray, has been targeted for refractory facial neuropathic pain~\citep{ben2018deep}, and long-term sensing from these regions has been used to study pain-related neural activity~\citep{ramos2024beta}.
CATNUS could support DBS research by comparing nucleus boundaries with expert-defined targets, electrode locations, and stimulation volumes, and examining their relationships with clinical outcomes. 
However, CATNUS provides structural segmentation only, does not delineate relevant white-matter pathways, and has not been validated for surgical planning. 
Clinical use would require integration with tractography and validation against surgical targets and patient outcomes.

\section{Conclusion}\label{sec:5}
In this work, we developed CATNUS, a coordinate-aware 3D U-Net framework for segmenting 13 thalamic nuclei or nuclear groups. 
To support use across different imaging settings, we provide three pretrained models, one for each input modality: T1 maps, MPRAGE, and FGATIR. 
Overall, CATNUS provides an accurate, robust and generalizable framework for thalamic nuclei segmentation, with strong potential to support both neuroimaging research and clinical practice.

\section*{Ethics Statement}
The traumatic brain injury (TBI) data used in this study were collected under a research protocol approved by the Institutional Review Board (IRB) of the University of Maryland, Baltimore (Protocol No.~HP-00080961, approved on September 4, 2018). All procedures under this protocol were performed in compliance with relevant laws and institutional guidelines. Written informed consent was obtained from all participants before participation in the original study.
The ADNI, BLSA, FTHP, and MASiVar studies were approved by their respective ethics committees, and participants provided informed consent.

\section*{Data and Code Availability Statement}
The MRI data used in this study were obtained from multiple sources, including our institutional MRI scans and publicly available datasets. 
To support validation, one representative subject from our dataset has been released as sample data through the CATNUS GitHub repository 
(\url{https://github.com/ANQIFENG/CATNUS}). This shared dataset was collected and distributed in compliance with Institutional Review Board (IRB) approval and applicable data-sharing regulations.
The external datasets used in this work are publicly accessible, including 
the FTHP dataset (\url{https://www.nitrc.org/projects/fthp/}), the  MASiVar dataset (\url{https://openneuro.org/datasets/ds003416/}), the BLSA dataset (\url{https://www.blsa.nih.gov/}),
and the ADNI dataset (\url{https://adni.loni.usc.edu/}).
The pretrained models and source code are also available in the CATNUS repository and are packaged as a Singularity container, enabling end-to-end execution from preprocessing to thalamic nuclei segmentation.

\section*{Acknowledgment and Disclaimer}
The authors thank all participants of the Baltimore Longitudinal Study of
Aging (BLSA), the Frequently Traveling Human Phantom (FTHP), the Multisite,
Multiscanner, and Multisubject Acquisitions for Studying Variability in
diffusion weighted MRI (MASiVar), and the Alzheimer's Disease Neuroimaging
Initiative (ADNI) datasets. This work was supported in part by the Intramural
Research Program of the National Institutes of Health (NIH), National Institute
on Aging, and in part by the National Institute of Neurological Disorders and
Stroke under grant R01-NS105503 (PIs: Zhuo \& Prince).

Data collection and sharing for the Alzheimer's Disease
Neuroimaging Initiative (ADNI) is funded by the National Institute on Aging
(National Institutes of Health Grant U19 AG024904). The grantee organization is
the Northern California Institute for Research and Education. In the past, ADNI
has also received funding from the National Institute of Biomedical Imaging and
Bioengineering, the Canadian Institutes of Health Research, and private sector
contributions through the Foundation for the National Institutes of Health
(FNIH), including generous contributions from the following: AbbVie,
Alzheimer's Association, Alzheimer's Drug Discovery Foundation, Araclon
Biotech, BioClinica, Inc., Biogen, Bristol-Myers Squibb Company, CereSpir,
Inc., Cogstate, Eisai Inc., Elan Pharmaceuticals, Inc., Eli Lilly and Company,
EuroImmun, F. Hoffmann-La Roche Ltd and its affiliated company Genentech, Inc.,
Fujirebio, GE Healthcare, IXICO Ltd., Janssen Alzheimer Immunotherapy Research
\& Development, LLC., Johnson \& Johnson Pharmaceutical Research \& Development
LLC., Lumosity, Lundbeck, Merck \& Co., Inc., Meso Scale Diagnostics, LLC.,
NeuroRx Research, Neurotrack Technologies, Novartis Pharmaceuticals
Corporation, Pfizer Inc., Piramal Imaging, Servier, Takeda Pharmaceutical
Company, and Transition Therapeutics. As such, the investigators within the
ADNI contributed to the design and implementation of ADNI and/or provided data
but did not participate in analysis or writing of this report. A complete
listing of ADNI investigators can be found at:
\url{http://adni.loni.usc.edu/wp-content/uploads/how_to_apply/ADNI_Acknowledgement_List.pdf}.

The contributions of the NIH authors were made as part of their official duties
as NIH federal employees, are in compliance with agency policy requirements,
and are considered Works of the United States Government. However, the findings
and conclusions presented in this paper are those of the authors and do not
necessarily reflect the views of the NIH or the U.S. Department of Health and
Human Services.

\section*{Declaration of Competing Interest}
The authors declare no conflicts of interest.

\bibliographystyle{elsarticle-harv}
\bibliography{references}

\begin{thebibliography}{80}
\expandafter\ifx\csname natexlab\endcsname\relax\def\natexlab#1{#1}\fi
\providecommand{\url}[1]{\texttt{#1}}
\providecommand{\href}[2]{#2}
\providecommand{\path}[1]{#1}
\providecommand{\DOIprefix}{doi:}
\providecommand{\ArXivprefix}{arXiv:}
\providecommand{\URLprefix}{URL: }
\providecommand{\Pubmedprefix}{pmid:}
\providecommand{\doi}[1]{\href{http://dx.doi.org/#1}{\path{#1}}}
\providecommand{\Pubmed}[1]{\href{pmid:#1}{\path{#1}}}
\providecommand{\bibinfo}[2]{#2}
\ifx\xfnm\relax \def\xfnm[#1]{\unskip,\space#1}\fi
\bibitem[{Alel{\'u}-Paz and Gim{\'e}nez-Amaya(2008)}]{alelu2008mediodorsal}
\bibinfo{author}{Alel{\'u}-Paz, R.}, \bibinfo{author}{Gim{\'e}nez-Amaya, J.M.}, \bibinfo{year}{2008}.
\newblock \bibinfo{title}{The mediodorsal thalamic nucleus and schizophrenia}.
\newblock \bibinfo{journal}{{Journal of Psychiatry and Neuroscience}} \bibinfo{volume}{33}, \bibinfo{pages}{489--498}.
\bibitem[{Avants et~al.(2009)Avants, Tustison, Song et~al.}]{avants2009advanced}
\bibinfo{author}{Avants, B.B.}, \bibinfo{author}{Tustison, N.}, \bibinfo{author}{Song, G.}, et~al., \bibinfo{year}{2009}.
\newblock \bibinfo{title}{{Advanced Normalization Tools (ANTS)}}.
\newblock \bibinfo{journal}{The Insight Journal} \bibinfo{volume}{2}, \bibinfo{pages}{1--35}.
\bibitem[{Battistella et~al.(2017)Battistella, Najdenovska, Maeder, Ghazaleh, Daducci, Thiran, Jacquemont, Tuleasca, Levivier, Bach~Cuadra et~al.}]{battistella2017robust}
\bibinfo{author}{Battistella, G.}, \bibinfo{author}{Najdenovska, E.}, \bibinfo{author}{Maeder, P.}, \bibinfo{author}{Ghazaleh, N.}, \bibinfo{author}{Daducci, A.}, \bibinfo{author}{Thiran, J.P.}, \bibinfo{author}{Jacquemont, S.}, \bibinfo{author}{Tuleasca, C.}, \bibinfo{author}{Levivier, M.}, \bibinfo{author}{Bach~Cuadra, M.}, et~al., \bibinfo{year}{2017}.
\newblock \bibinfo{title}{Robust thalamic nuclei segmentation method based on local diffusion magnetic resonance properties}.
\newblock \bibinfo{journal}{{Brain Structure and Function}} \bibinfo{volume}{222}, \bibinfo{pages}{2203--2216}.
\bibitem[{Behrens et~al.(2003)Behrens, Johansen-Berg, Woolrich, Smith, Wheeler-Kingshott, Boulby, Barker, Sillery, Sheehan, Ciccarelli et~al.}]{behrens2003non}
\bibinfo{author}{Behrens, T.E.}, \bibinfo{author}{Johansen-Berg, H.}, \bibinfo{author}{Woolrich, M.W.}, \bibinfo{author}{Smith, S.M.}, \bibinfo{author}{Wheeler-Kingshott, C.A.}, \bibinfo{author}{Boulby, P.A.}, \bibinfo{author}{Barker, G.J.}, \bibinfo{author}{Sillery, E.}, \bibinfo{author}{Sheehan, K.}, \bibinfo{author}{Ciccarelli, O.}, et~al., \bibinfo{year}{2003}.
\newblock \bibinfo{title}{Non-invasive mapping of connections between human thalamus and cortex using diffusion imaging}.
\newblock \bibinfo{journal}{{Nature Neuroscience}} \bibinfo{volume}{6}, \bibinfo{pages}{750--757}.
\bibitem[{Ben-Haim et~al.(2018)Ben-Haim, Mirzadeh and Rosenberg}]{ben2018deep}
\bibinfo{author}{Ben-Haim, S.}, \bibinfo{author}{Mirzadeh, Z.}, \bibinfo{author}{Rosenberg, W.S.}, \bibinfo{year}{2018}.
\newblock \bibinfo{title}{{Deep brain stimulation for intractable neuropathic facial pain}}.
\newblock \bibinfo{journal}{{Neurosurgical Focus}} \bibinfo{volume}{45}, \bibinfo{pages}{E15}.
\bibitem[{Cai et~al.(2021)Cai, Yang, Kanakaraj, Nath, Newton, Edmonson, Luci, Conrad, Price, Hansen et~al.}]{cai2021masivar}
\bibinfo{author}{Cai, L.Y.}, \bibinfo{author}{Yang, Q.}, \bibinfo{author}{Kanakaraj, P.}, \bibinfo{author}{Nath, V.}, \bibinfo{author}{Newton, A.T.}, \bibinfo{author}{Edmonson, H.A.}, \bibinfo{author}{Luci, J.}, \bibinfo{author}{Conrad, B.N.}, \bibinfo{author}{Price, G.R.}, \bibinfo{author}{Hansen, C.B.}, et~al., \bibinfo{year}{2021}.
\newblock \bibinfo{title}{{MASiVar: Multisite, multiscanner, and multisubject acquisitions for studying variability in diffusion weighted MRI}}.
\newblock \bibinfo{journal}{{Magnetic Resonance in Medicine}} \bibinfo{volume}{86}, \bibinfo{pages}{3304--3320}.
\bibitem[{{\c{C}}i{\c{c}}ek et~al.(2016){\c{C}}i{\c{c}}ek, Abdulkadir, Lienkamp, Brox and Ronneberger}]{cciccek20163d}
\bibinfo{author}{{\c{C}}i{\c{c}}ek, {\"O}.}, \bibinfo{author}{Abdulkadir, A.}, \bibinfo{author}{Lienkamp, S.S.}, \bibinfo{author}{Brox, T.}, \bibinfo{author}{Ronneberger, O.}, \bibinfo{year}{2016}.
\newblock \bibinfo{title}{{3D U-Net: learning dense volumetric segmentation from sparse annotation}}, in: \bibinfo{booktitle}{{Medical Image Computing and Computer-Assisted Intervention--MICCAI 2016: 19th International Conference, Athens, Greece, October 17-21, 2016, Proceedings, Part II 19}}, \bibinfo{organization}{Springer}. pp. \bibinfo{pages}{424--432}.
\bibitem[{Dalic et~al.(2022)Dalic, Warren, Bulluss, Thevathasan, Roten, Churilov and Archer}]{dalic2022dbs}
\bibinfo{author}{Dalic, L.J.}, \bibinfo{author}{Warren, A.E.}, \bibinfo{author}{Bulluss, K.J.}, \bibinfo{author}{Thevathasan, W.}, \bibinfo{author}{Roten, A.}, \bibinfo{author}{Churilov, L.}, \bibinfo{author}{Archer, J.S.}, \bibinfo{year}{2022}.
\newblock \bibinfo{title}{{DBS of thalamic centromedian nucleus for Lennox--Gastaut syndrome (ESTEL trial)}}.
\newblock \bibinfo{journal}{{Annals of Neurology}} \bibinfo{volume}{91}, \bibinfo{pages}{253--267}.
\bibitem[{Datta et~al.(2021)Datta, Bacchus, Kumar, Elliott, Rao, Dolui, Reddy, Banwell and Saranathan}]{datta2021fast}
\bibinfo{author}{Datta, R.}, \bibinfo{author}{Bacchus, M.K.}, \bibinfo{author}{Kumar, D.}, \bibinfo{author}{Elliott, M.A.}, \bibinfo{author}{Rao, A.}, \bibinfo{author}{Dolui, S.}, \bibinfo{author}{Reddy, R.}, \bibinfo{author}{Banwell, B.L.}, \bibinfo{author}{Saranathan, M.}, \bibinfo{year}{2021}.
\newblock \bibinfo{title}{{Fast automatic segmentation of thalamic nuclei from MP2RAGE acquisition at 7 Tesla}}.
\newblock \bibinfo{journal}{{Magnetic Resonance in Medicine}} \bibinfo{volume}{85}, \bibinfo{pages}{2781--2790}.
\bibitem[{DeLong et~al.(1988)DeLong, DeLong and Clarke-Pearson}]{delong1988comparing}
\bibinfo{author}{DeLong, E.R.}, \bibinfo{author}{DeLong, D.M.}, \bibinfo{author}{Clarke-Pearson, D.L.}, \bibinfo{year}{1988}.
\newblock \bibinfo{title}{Comparing the areas under two or more correlated receiver operating characteristic curves: a nonparametric approach}.
\newblock \bibinfo{journal}{Biometrics} , \bibinfo{pages}{837--845}.
\bibitem[{Deoni et~al.(2007)Deoni, Rutt, Parrent and Peters}]{deoni2007segmentation}
\bibinfo{author}{Deoni, S.C.}, \bibinfo{author}{Rutt, B.K.}, \bibinfo{author}{Parrent, A.G.}, \bibinfo{author}{Peters, T.M.}, \bibinfo{year}{2007}.
\newblock \bibinfo{title}{{Segmentation of thalamic nuclei using a modified k-means clustering algorithm and high-resolution quantitative magnetic resonance imaging at 1.5T}}.
\newblock \bibinfo{journal}{NeuroImage} \bibinfo{volume}{34}, \bibinfo{pages}{117--126}.
\bibitem[{Ding et~al.(2016)Ding, Royall, Sunkin, Ng, Facer, Lesnar, Guillozet-Bongaarts, McMurray, Szafer, Dolbeare et~al.}]{ding2016comprehensive}
\bibinfo{author}{Ding, S.L.}, \bibinfo{author}{Royall, J.J.}, \bibinfo{author}{Sunkin, S.M.}, \bibinfo{author}{Ng, L.}, \bibinfo{author}{Facer, B.A.}, \bibinfo{author}{Lesnar, P.}, \bibinfo{author}{Guillozet-Bongaarts, A.}, \bibinfo{author}{McMurray, B.}, \bibinfo{author}{Szafer, A.}, \bibinfo{author}{Dolbeare, T.A.}, et~al., \bibinfo{year}{2016}.
\newblock \bibinfo{title}{{Comprehensive cellular-resolution atlas of the adult human brain}}.
\newblock \bibinfo{journal}{{Journal of Comparative Neurology}} \bibinfo{volume}{524}, \bibinfo{pages}{3127--3481}.
\bibitem[{Fan et~al.(2015)Fan, Nickerson, Li, Ma, Lyu, Miao, Zhuo, Ge, Zou and Gao}]{fan2015functional}
\bibinfo{author}{Fan, Y.}, \bibinfo{author}{Nickerson, L.D.}, \bibinfo{author}{Li, H.}, \bibinfo{author}{Ma, Y.}, \bibinfo{author}{Lyu, B.}, \bibinfo{author}{Miao, X.}, \bibinfo{author}{Zhuo, Y.}, \bibinfo{author}{Ge, J.}, \bibinfo{author}{Zou, Q.}, \bibinfo{author}{Gao, J.H.}, \bibinfo{year}{2015}.
\newblock \bibinfo{title}{{Functional connectivity-based parcellation of the thalamus: an unsupervised clustering method and its validity investigation}}.
\newblock \bibinfo{journal}{{Brain Connectivity}} \bibinfo{volume}{5}, \bibinfo{pages}{620--630}.
\bibitem[{Feng et~al.(2024)Feng, Bian, Dewey, Colinco, Zhuo and Prince}]{feng2024ratnus}
\bibinfo{author}{Feng, A.}, \bibinfo{author}{Bian, Z.}, \bibinfo{author}{Dewey, B.E.}, \bibinfo{author}{Colinco, A.G.}, \bibinfo{author}{Zhuo, J.}, \bibinfo{author}{Prince, J.L.}, \bibinfo{year}{2024}.
\newblock \bibinfo{title}{{RATNUS: Rapid, Automatic Thalamic Nuclei Segmentation Using Multimodal MRI Inputs}}, in: \bibinfo{booktitle}{{International Conference on Medical Image Computing and Computer-Assisted Intervention}}, \bibinfo{organization}{Springer}. pp. \bibinfo{pages}{157--169}.
\bibitem[{Feng et~al.(2025)Feng, Bian, Remedios, Hays, Dewey, Zhuo, Benjamini and Prince}]{feng2025segmenting}
\bibinfo{author}{Feng, A.}, \bibinfo{author}{Bian, Z.}, \bibinfo{author}{Remedios, S.W.}, \bibinfo{author}{Hays, S.P.}, \bibinfo{author}{Dewey, B.E.}, \bibinfo{author}{Zhuo, J.}, \bibinfo{author}{Benjamini, D.}, \bibinfo{author}{Prince, J.L.}, \bibinfo{year}{2025}.
\newblock \bibinfo{title}{{Segmenting Thalamic Nuclei: T1 Maps Provide a Reliable and Efficient Solution}}.
\newblock \bibinfo{journal}{ArXiv Preprint ArXiv:2508.12508} .
\bibitem[{Feng et~al.(2023)Feng, Xue, Wang, Yan, Bian, Shao, Zhuo, Gullapalli, Carass and Prince}]{feng2023label}
\bibinfo{author}{Feng, A.}, \bibinfo{author}{Xue, Y.}, \bibinfo{author}{Wang, Y.}, \bibinfo{author}{Yan, C.}, \bibinfo{author}{Bian, Z.}, \bibinfo{author}{Shao, M.}, \bibinfo{author}{Zhuo, J.}, \bibinfo{author}{Gullapalli, R.P.}, \bibinfo{author}{Carass, A.}, \bibinfo{author}{Prince, J.L.}, \bibinfo{year}{2023}.
\newblock \bibinfo{title}{{Label propagation via random walk for training robust thalamus nuclei parcellation model from noisy annotations}}, in: \bibinfo{booktitle}{{2023 IEEE 20th International Symposium on Biomedical Imaging}}, \bibinfo{organization}{IEEE}. pp. \bibinfo{pages}{1--5}.
\bibitem[{Fisher et~al.(2010)Fisher, Salanova, Witt, Worth, Henry, Gross, Oommen, Osorio, Nazzaro, Labar et~al.}]{fisher2010electrical}
\bibinfo{author}{Fisher, R.}, \bibinfo{author}{Salanova, V.}, \bibinfo{author}{Witt, T.}, \bibinfo{author}{Worth, R.}, \bibinfo{author}{Henry, T.}, \bibinfo{author}{Gross, R.}, \bibinfo{author}{Oommen, K.}, \bibinfo{author}{Osorio, I.}, \bibinfo{author}{Nazzaro, J.}, \bibinfo{author}{Labar, D.}, et~al., \bibinfo{year}{2010}.
\newblock \bibinfo{title}{Electrical stimulation of the anterior nucleus of thalamus for treatment of refractory epilepsy}.
\newblock \bibinfo{journal}{Epilepsia} \bibinfo{volume}{51}, \bibinfo{pages}{899--908}.
\bibitem[{Forno et~al.(2023)Forno, Saranathan, Contador, Guillen, Falg{\`a}s, Tort-Merino, Balasa, Sanchez-Valle, Hornberger and Llad{\'o}}]{forno2023thalamic}
\bibinfo{author}{Forno, G.}, \bibinfo{author}{Saranathan, M.}, \bibinfo{author}{Contador, J.}, \bibinfo{author}{Guillen, N.}, \bibinfo{author}{Falg{\`a}s, N.}, \bibinfo{author}{Tort-Merino, A.}, \bibinfo{author}{Balasa, M.}, \bibinfo{author}{Sanchez-Valle, R.}, \bibinfo{author}{Hornberger, M.}, \bibinfo{author}{Llad{\'o}, A.}, \bibinfo{year}{2023}.
\newblock \bibinfo{title}{{Thalamic nuclei changes in early and late onset Alzheimer's disease}}.
\newblock \bibinfo{journal}{{Current Research in Neurobiology}} \bibinfo{volume}{4}, \bibinfo{pages}{100084}.
\bibitem[{Fram et~al.(1987)Fram, Herfkens, Johnson, Glover, Karis, Shimakawa, Perkins and Pelc}]{fram1987rapid}
\bibinfo{author}{Fram, E.K.}, \bibinfo{author}{Herfkens, R.J.}, \bibinfo{author}{Johnson, G.A.}, \bibinfo{author}{Glover, G.H.}, \bibinfo{author}{Karis, J.P.}, \bibinfo{author}{Shimakawa, A.}, \bibinfo{author}{Perkins, T.G.}, \bibinfo{author}{Pelc, N.J.}, \bibinfo{year}{1987}.
\newblock \bibinfo{title}{{Rapid calculation of T1 using variable flip angle gradient refocused imaging}}.
\newblock \bibinfo{journal}{{Magnetic Resonance Imaging}} \bibinfo{volume}{5}, \bibinfo{pages}{201--208}.
\bibitem[{Gao et~al.(2025)Gao, Wu, Ma, Li, Wang, Guo, Li, Wang, Chu, Madsen et~al.}]{gao2025iterative}
\bibinfo{author}{Gao, C.}, \bibinfo{author}{Wu, X.}, \bibinfo{author}{Ma, L.}, \bibinfo{author}{Li, D.}, \bibinfo{author}{Wang, Y.}, \bibinfo{author}{Guo, C.}, \bibinfo{author}{Li, W.}, \bibinfo{author}{Wang, H.}, \bibinfo{author}{Chu, C.}, \bibinfo{author}{Madsen, K.H.}, et~al., \bibinfo{year}{2025}.
\newblock \bibinfo{title}{{Iterative prior-guided parcellation (iPGP) for capturing inter-subject and inter-nuclei variability in thalamic mapping}}.
\newblock \bibinfo{journal}{NeuroImage} , \bibinfo{pages}{121399}.
\bibitem[{Glaister et~al.(2016)Glaister, Carass, Stough, Calabresi and Prince}]{glaister2016thalamus}
\bibinfo{author}{Glaister, J.}, \bibinfo{author}{Carass, A.}, \bibinfo{author}{Stough, J.V.}, \bibinfo{author}{Calabresi, P.A.}, \bibinfo{author}{Prince, J.L.}, \bibinfo{year}{2016}.
\newblock \bibinfo{title}{Thalamus parcellation using multi-modal feature classification and thalamic nuclei priors}, in: \bibinfo{booktitle}{{Medical Imaging 2016: Image Processing}}, \bibinfo{organization}{SPIE}. pp. \bibinfo{pages}{937--942}.
\bibitem[{Grieve et~al.(2000)Grieve, Acu{\~n}a and Cudeiro}]{grieve2000primate}
\bibinfo{author}{Grieve, K.L.}, \bibinfo{author}{Acu{\~n}a, C.}, \bibinfo{author}{Cudeiro, J.}, \bibinfo{year}{2000}.
\newblock \bibinfo{title}{The primate pulvinar nuclei: vision and action}.
\newblock \bibinfo{journal}{{Trends in Neurosciences}} \bibinfo{volume}{23}, \bibinfo{pages}{35--39}.
\bibitem[{He et~al.(2025)He, Zhu, Zhang, Liu, Xu, Chen, Zekelman, Rushmore, Rathi, Makris et~al.}]{he2025deepnuparc}
\bibinfo{author}{He, H.}, \bibinfo{author}{Zhu, C.}, \bibinfo{author}{Zhang, L.}, \bibinfo{author}{Liu, Y.}, \bibinfo{author}{Xu, X.}, \bibinfo{author}{Chen, Y.}, \bibinfo{author}{Zekelman, L.}, \bibinfo{author}{Rushmore, J.}, \bibinfo{author}{Rathi, Y.}, \bibinfo{author}{Makris, N.}, et~al., \bibinfo{year}{2025}.
\newblock \bibinfo{title}{{DeepNuParc: A novel deep clustering framework for fine-scale parcellation of brain nuclei using diffusion MRI tractography}}.
\newblock \bibinfo{journal}{NeuroImage} , \bibinfo{pages}{121421}.
\bibitem[{Henderson et~al.(2000)Henderson, Carpenter, Cartwright and Halliday}]{henderson2000degeneration}
\bibinfo{author}{Henderson, J.M.}, \bibinfo{author}{Carpenter, K.}, \bibinfo{author}{Cartwright, H.}, \bibinfo{author}{Halliday, G.M.}, \bibinfo{year}{2000}.
\newblock \bibinfo{title}{{Degeneration of the centr{\'e} median--parafascicular complex in Parkinson's disease}}.
\newblock \bibinfo{journal}{Annals of Neurology: Official Journal of the American Neurological Association and the Child Neurology Society} \bibinfo{volume}{47}, \bibinfo{pages}{345--352}.
\bibitem[{Iglesias et~al.(2018)Iglesias, Insausti, Lerma-Usabiaga, Bocchetta, Van~Leemput, Greve, Van~der Kouwe, Fischl, Caballero-Gaudes, Paz-Alonso et~al.}]{iglesias2018probabilistic}
\bibinfo{author}{Iglesias, J.E.}, \bibinfo{author}{Insausti, R.}, \bibinfo{author}{Lerma-Usabiaga, G.}, \bibinfo{author}{Bocchetta, M.}, \bibinfo{author}{Van~Leemput, K.}, \bibinfo{author}{Greve, D.N.}, \bibinfo{author}{Van~der Kouwe, A.}, \bibinfo{author}{Fischl, B.}, \bibinfo{author}{Caballero-Gaudes, C.}, \bibinfo{author}{Paz-Alonso, P.M.}, et~al., \bibinfo{year}{2018}.
\newblock \bibinfo{title}{{A probabilistic atlas of the human thalamic nuclei combining ex vivo MRI and histology}}.
\newblock \bibinfo{journal}{NeuroImage} \bibinfo{volume}{183}, \bibinfo{pages}{314--326}.
\bibitem[{Jack et~al.(2008)Jack, Bernstein, Fox, Thompson, Alexander, Harvey, Borowski, Britson, L.~Whitwell, Ward et~al.}]{jack2008alzheimer}
\bibinfo{author}{Jack, C.R.}, \bibinfo{author}{Bernstein, M.A.}, \bibinfo{author}{Fox, N.C.}, \bibinfo{author}{Thompson, P.}, \bibinfo{author}{Alexander, G.}, \bibinfo{author}{Harvey, D.}, \bibinfo{author}{Borowski, B.}, \bibinfo{author}{Britson, P.J.}, \bibinfo{author}{L.~Whitwell, J.}, \bibinfo{author}{Ward, C.}, et~al., \bibinfo{year}{2008}.
\newblock \bibinfo{title}{{The Alzheimer's disease neuroimaging initiative (ADNI): MRI methods}}.
\newblock \bibinfo{journal}{{Journal of Magnetic Resonance Imaging: An Official Journal of the International Society for Magnetic Resonance in Medicine}} \bibinfo{volume}{27}, \bibinfo{pages}{685--691}.
\bibitem[{Jakab et~al.(2012)Jakab, Blanc, Ber{\'e}nyi and Sz{\'e}kely}]{jakab2012generation}
\bibinfo{author}{Jakab, A.}, \bibinfo{author}{Blanc, R.}, \bibinfo{author}{Ber{\'e}nyi, E.}, \bibinfo{author}{Sz{\'e}kely, G.}, \bibinfo{year}{2012}.
\newblock \bibinfo{title}{{Generation of individualized thalamus target maps by using statistical shape models and thalamocortical tractography}}.
\newblock \bibinfo{journal}{{American Journal of Neuroradiology}} \bibinfo{volume}{33}, \bibinfo{pages}{2110--2116}.
\bibitem[{Johansen-Berg et~al.(2005)Johansen-Berg, Behrens, Sillery, Ciccarelli, Thompson, Smith and Matthews}]{johansen2005functional}
\bibinfo{author}{Johansen-Berg, H.}, \bibinfo{author}{Behrens, T.E.}, \bibinfo{author}{Sillery, E.}, \bibinfo{author}{Ciccarelli, O.}, \bibinfo{author}{Thompson, A.J.}, \bibinfo{author}{Smith, S.M.}, \bibinfo{author}{Matthews, P.M.}, \bibinfo{year}{2005}.
\newblock \bibinfo{title}{Functional--anatomical validation and individual variation of diffusion tractography-based segmentation of the human thalamus}.
\newblock \bibinfo{journal}{{Cerebral Cortex}} \bibinfo{volume}{15}, \bibinfo{pages}{31--39}.
\bibitem[{Jonasson et~al.(2007)Jonasson, Hagmann, Pollo, Bresson, Wilson, Meuli and Thiran}]{jonasson2007level}
\bibinfo{author}{Jonasson, L.}, \bibinfo{author}{Hagmann, P.}, \bibinfo{author}{Pollo, C.}, \bibinfo{author}{Bresson, X.}, \bibinfo{author}{Wilson, C.R.}, \bibinfo{author}{Meuli, R.}, \bibinfo{author}{Thiran, J.P.}, \bibinfo{year}{2007}.
\newblock \bibinfo{title}{{A level set method for segmentation of the thalamus and its nuclei in DT-MRI}}.
\newblock \bibinfo{journal}{Signal Processing} \bibinfo{volume}{87}, \bibinfo{pages}{309--321}.
\bibitem[{Jones(2012)}]{jones2012thalamus}
\bibinfo{author}{Jones, E.G.}, \bibinfo{year}{2012}.
\newblock \bibinfo{title}{{The Thalamus}}.
\newblock \bibinfo{publisher}{Springer Science \& Business Media}.
\bibitem[{Keun et~al.(2021)Keun, van Heese, Laansma, Weeland, de~Joode, van~den Heuvel, Gool, Kasprzak, Bright, Vriend et~al.}]{keun2021structural}
\bibinfo{author}{Keun, J.T.B.}, \bibinfo{author}{van Heese, E.M.}, \bibinfo{author}{Laansma, M.A.}, \bibinfo{author}{Weeland, C.J.}, \bibinfo{author}{de~Joode, N.T.}, \bibinfo{author}{van~den Heuvel, O.A.}, \bibinfo{author}{Gool, J.K.}, \bibinfo{author}{Kasprzak, S.}, \bibinfo{author}{Bright, J.K.}, \bibinfo{author}{Vriend, C.}, et~al., \bibinfo{year}{2021}.
\newblock \bibinfo{title}{{Structural assessment of thalamus morphology in brain disorders: A review and recommendation of thalamic nucleus segmentation and shape analysis}}.
\newblock \bibinfo{journal}{{Neuroscience \& Biobehavioral Reviews}} \bibinfo{volume}{131}, \bibinfo{pages}{466--478}.
\bibitem[{Krauth et~al.(2010)Krauth, Blanc, Poveda, Jeanmonod, Morel and Sz{\'e}kely}]{krauth2010mean}
\bibinfo{author}{Krauth, A.}, \bibinfo{author}{Blanc, R.}, \bibinfo{author}{Poveda, A.}, \bibinfo{author}{Jeanmonod, D.}, \bibinfo{author}{Morel, A.}, \bibinfo{author}{Sz{\'e}kely, G.}, \bibinfo{year}{2010}.
\newblock \bibinfo{title}{A mean three-dimensional atlas of the human thalamus: generation from multiple histological data}.
\newblock \bibinfo{journal}{{NeuroImage}} \bibinfo{volume}{49}, \bibinfo{pages}{2053--2062}.
\bibitem[{Kumar et~al.(2017)Kumar, Van~Oort, Scheffler, Beckmann and Grodd}]{kumar2017functional}
\bibinfo{author}{Kumar, V.J.}, \bibinfo{author}{Van~Oort, E.}, \bibinfo{author}{Scheffler, K.}, \bibinfo{author}{Beckmann, C.F.}, \bibinfo{author}{Grodd, W.}, \bibinfo{year}{2017}.
\newblock \bibinfo{title}{{Functional anatomy of the human thalamus at rest}}.
\newblock \bibinfo{journal}{{NeuroImage}} \bibinfo{volume}{147}, \bibinfo{pages}{678--691}.
\bibitem[{Lin et~al.(2017)Lin, Doll{\'a}r, Girshick, He, Hariharan and Belongie}]{lin2017feature}
\bibinfo{author}{Lin, T.Y.}, \bibinfo{author}{Doll{\'a}r, P.}, \bibinfo{author}{Girshick, R.}, \bibinfo{author}{He, K.}, \bibinfo{author}{Hariharan, B.}, \bibinfo{author}{Belongie, S.}, \bibinfo{year}{2017}.
\newblock \bibinfo{title}{Feature pyramid networks for object detection}, in: \bibinfo{booktitle}{{Proceedings of the IEEE Conference on Computer Vision and Pattern Recognition}}, pp. \bibinfo{pages}{2117--2125}.
\bibitem[{Liu et~al.(2018)Liu, Lehman, Molino, Petroski~Such, Frank, Sergeev and Yosinski}]{liu2018intriguing}
\bibinfo{author}{Liu, R.}, \bibinfo{author}{Lehman, J.}, \bibinfo{author}{Molino, P.}, \bibinfo{author}{Petroski~Such, F.}, \bibinfo{author}{Frank, E.}, \bibinfo{author}{Sergeev, A.}, \bibinfo{author}{Yosinski, J.}, \bibinfo{year}{2018}.
\newblock \bibinfo{title}{{An intriguing failing of convolutional neural networks and the coordconv solution}}.
\newblock \bibinfo{journal}{{Advances in Neural Information Processing Systems}} \bibinfo{volume}{31}.
\bibitem[{Liu et~al.(2020)Liu, D'Haese, Newton and Dawant}]{liu2020generation}
\bibinfo{author}{Liu, Y.}, \bibinfo{author}{D'Haese, P.F.}, \bibinfo{author}{Newton, A.T.}, \bibinfo{author}{Dawant, B.M.}, \bibinfo{year}{2020}.
\newblock \bibinfo{title}{{Generation of human thalamus atlases from 7T data and application to intrathalamic nuclei segmentation in clinical 3T T1-weighted images}}.
\newblock \bibinfo{journal}{{Magnetic Resonance Imaging}} \bibinfo{volume}{65}, \bibinfo{pages}{114--128}.
\bibitem[{{Lopez Ramos} et~al.(2024){Lopez Ramos}, Rockhill, Shahin, Gragg, Tan, Yamamoto, Fecker, Ismail, Cleary and Raslan}]{ramos2024beta}
\bibinfo{author}{{Lopez Ramos}, C.G.}, \bibinfo{author}{Rockhill, A.P.}, \bibinfo{author}{Shahin, M.N.}, \bibinfo{author}{Gragg, A.}, \bibinfo{author}{Tan, H.}, \bibinfo{author}{Yamamoto, E.A.}, \bibinfo{author}{Fecker, A.L.}, \bibinfo{author}{Ismail, M.}, \bibinfo{author}{Cleary, D.R.}, \bibinfo{author}{Raslan, A.M.}, \bibinfo{year}{2024}.
\newblock \bibinfo{title}{{Beta oscillations in the sensory thalamus during severe facial neuropathic pain using novel sensing deep brain stimulation}}.
\newblock \bibinfo{journal}{{Neuromodulation: Technology at the Neural Interface}} \bibinfo{volume}{27}, \bibinfo{pages}{1419--1427}.
\bibitem[{Majdi et~al.(2020)Majdi, Keerthivasan, Rutt, Zahr, Rodriguez and Saranathan}]{majdi2020automated}
\bibinfo{author}{Majdi, M.S.}, \bibinfo{author}{Keerthivasan, M.B.}, \bibinfo{author}{Rutt, B.K.}, \bibinfo{author}{Zahr, N.M.}, \bibinfo{author}{Rodriguez, J.J.}, \bibinfo{author}{Saranathan, M.}, \bibinfo{year}{2020}.
\newblock \bibinfo{title}{Automated thalamic nuclei segmentation using multi-planar cascaded convolutional neural networks}.
\newblock \bibinfo{journal}{{Magnetic Resonance Imaging}} \bibinfo{volume}{73}, \bibinfo{pages}{45--54}.
\bibitem[{Mang et~al.(2012)Mang, Busza, Reiterer, Grodd and Klose}]{mang2012thalamus}
\bibinfo{author}{Mang, S.C.}, \bibinfo{author}{Busza, A.}, \bibinfo{author}{Reiterer, S.}, \bibinfo{author}{Grodd, W.}, \bibinfo{author}{Klose, U.}, \bibinfo{year}{2012}.
\newblock \bibinfo{title}{{Thalamus segmentation based on the local diffusion direction: A group study}}.
\newblock \bibinfo{journal}{{Magnetic Resonance Medicine}} \bibinfo{volume}{67}, \bibinfo{pages}{118--126}.
\bibitem[{Marques et~al.(2010)Marques, Kober, Krueger, van~der Zwaag, Van~de Moortele and Gruetter}]{marques2010mp2rage}
\bibinfo{author}{Marques, J.P.}, \bibinfo{author}{Kober, T.}, \bibinfo{author}{Krueger, G.}, \bibinfo{author}{van~der Zwaag, W.}, \bibinfo{author}{Van~de Moortele, P.F.}, \bibinfo{author}{Gruetter, R.}, \bibinfo{year}{2010}.
\newblock \bibinfo{title}{{MP2RAGE, a self bias-field corrected sequence for improved segmentation and T1-mapping at high field}}.
\newblock \bibinfo{journal}{{NeuroImage}} \bibinfo{volume}{49}, \bibinfo{pages}{1271--1281}.
\bibitem[{Messroghli et~al.(2004)Messroghli, Radjenovic, Kozerke, Higgins, Sivananthan and Ridgway}]{messroghli2004modified}
\bibinfo{author}{Messroghli, D.R.}, \bibinfo{author}{Radjenovic, A.}, \bibinfo{author}{Kozerke, S.}, \bibinfo{author}{Higgins, D.M.}, \bibinfo{author}{Sivananthan, M.U.}, \bibinfo{author}{Ridgway, J.P.}, \bibinfo{year}{2004}.
\newblock \bibinfo{title}{{Modified Look-Locker inversion recovery (MOLLI) for high-resolution T1 mapping of the heart}}.
\newblock \bibinfo{journal}{{Magnetic Resonance in Medicine: An Official Journal of the International Society for Magnetic Resonance in Medicine}} \bibinfo{volume}{52}, \bibinfo{pages}{141--146}.
\bibitem[{Minagar et~al.(2013)Minagar, Barnett, Benedict, Pelletier, Pirko, Sahraian, Frohman and Zivadinov}]{minagar2013thalamus}
\bibinfo{author}{Minagar, A.}, \bibinfo{author}{Barnett, M.H.}, \bibinfo{author}{Benedict, R.H.}, \bibinfo{author}{Pelletier, D.}, \bibinfo{author}{Pirko, I.}, \bibinfo{author}{Sahraian, M.A.}, \bibinfo{author}{Frohman, E.}, \bibinfo{author}{Zivadinov, R.}, \bibinfo{year}{2013}.
\newblock \bibinfo{title}{{The thalamus and multiple sclerosis: Modern views on pathologic, imaging, and clinical aspects}}.
\newblock \bibinfo{journal}{Neurology} \bibinfo{volume}{80}, \bibinfo{pages}{210--219}.
\bibitem[{Mitchell(2015)}]{mitchell2015mediodorsal}
\bibinfo{author}{Mitchell, A.S.}, \bibinfo{year}{2015}.
\newblock \bibinfo{title}{{The mediodorsal thalamus as a higher order thalamic relay nucleus important for learning and decision-making}}.
\newblock \bibinfo{journal}{{Neuroscience \& Biobehavioral Reviews}} \bibinfo{volume}{54}, \bibinfo{pages}{76--88}.
\bibitem[{Morel et~al.(1997)Morel, Magnin and Jeanmonod}]{morel1997multiarchitectonic}
\bibinfo{author}{Morel, A.}, \bibinfo{author}{Magnin, M.}, \bibinfo{author}{Jeanmonod, D.}, \bibinfo{year}{1997}.
\newblock \bibinfo{title}{{Multiarchitectonic and stereotactic atlas of the human thalamus}}.
\newblock \bibinfo{journal}{{Journal of Comparative Neurology}} \bibinfo{volume}{387}, \bibinfo{pages}{588--630}.
\bibitem[{Nowacki et~al.(2023)Nowacki, Zhang, Barlatey, Ai-Schl{\"a}ppi, Rosner, Arnold and Pollo}]{nowacki2023deep}
\bibinfo{author}{Nowacki, A.}, \bibinfo{author}{Zhang, D.}, \bibinfo{author}{Barlatey, S.}, \bibinfo{author}{Ai-Schl{\"a}ppi, J.}, \bibinfo{author}{Rosner, J.}, \bibinfo{author}{Arnold, M.}, \bibinfo{author}{Pollo, C.}, \bibinfo{year}{2023}.
\newblock \bibinfo{title}{{Deep brain stimulation of the central lateral and ventral posterior thalamus for central poststroke pain syndrome: preliminary experience}}.
\newblock \bibinfo{journal}{{Neuromodulation: Technology at the Neural Interface}} \bibinfo{volume}{26}, \bibinfo{pages}{1747--1756}.
\bibitem[{O'Muircheartaigh et~al.(2011)O'Muircheartaigh, Vollmar, Traynor, Barker, Kumari, Symms, Thompson, Duncan, Koepp and Richardson}]{o2011clustering}
\bibinfo{author}{O'Muircheartaigh, J.}, \bibinfo{author}{Vollmar, C.}, \bibinfo{author}{Traynor, C.}, \bibinfo{author}{Barker, G.J.}, \bibinfo{author}{Kumari, V.}, \bibinfo{author}{Symms, M.R.}, \bibinfo{author}{Thompson, P.}, \bibinfo{author}{Duncan, J.S.}, \bibinfo{author}{Koepp, M.J.}, \bibinfo{author}{Richardson, M.P.}, \bibinfo{year}{2011}.
\newblock \bibinfo{title}{Clustering probabilistic tractograms using independent component analysis applied to the thalamus}.
\newblock \bibinfo{journal}{NeuroImage} \bibinfo{volume}{54}, \bibinfo{pages}{2020--2032}.
\bibitem[{Opfer et~al.(2023)Opfer, Kr{\"u}ger, Spies, Ostwaldt, Kitzler, Schippling and Buchert}]{opfer2023automatic}
\bibinfo{author}{Opfer, R.}, \bibinfo{author}{Kr{\"u}ger, J.}, \bibinfo{author}{Spies, L.}, \bibinfo{author}{Ostwaldt, A.C.}, \bibinfo{author}{Kitzler, H.H.}, \bibinfo{author}{Schippling, S.}, \bibinfo{author}{Buchert, R.}, \bibinfo{year}{2023}.
\newblock \bibinfo{title}{{Automatic segmentation of the thalamus using a massively trained 3D convolutional neural network: higher sensitivity for the detection of reduced thalamus volume by improved inter-scanner stability}}.
\newblock \bibinfo{journal}{{European Radiology}} \bibinfo{volume}{33}, \bibinfo{pages}{1852--1861}.
\bibitem[{Pereira et~al.(2013)Pereira, Boccard, Linhares, Chamadoira, Rosas, Abreu, Rebelo, Vaz and Aziz}]{pereira2013thalamic}
\bibinfo{author}{Pereira, E.A.}, \bibinfo{author}{Boccard, S.G.}, \bibinfo{author}{Linhares, P.}, \bibinfo{author}{Chamadoira, C.}, \bibinfo{author}{Rosas, M.J.}, \bibinfo{author}{Abreu, P.}, \bibinfo{author}{Rebelo, V.}, \bibinfo{author}{Vaz, R.}, \bibinfo{author}{Aziz, T.Z.}, \bibinfo{year}{2013}.
\newblock \bibinfo{title}{{Thalamic deep brain stimulation for neuropathic pain after amputation or brachial plexus avulsion}}.
\newblock \bibinfo{journal}{{Neurosurgical Focus}} \bibinfo{volume}{35}, \bibinfo{pages}{E7}.
\bibitem[{Price(2002)}]{price2002central}
\bibinfo{author}{Price, D.D.}, \bibinfo{year}{2002}.
\newblock \bibinfo{title}{Central neural mechanisms that interrelate sensory and affective dimensions of pain}.
\newblock \bibinfo{journal}{{Molecular Interventions}} \bibinfo{volume}{2}, \bibinfo{pages}{392}.
\bibitem[{Reinhold et~al.(2019)Reinhold, Dewey, Carass and Prince}]{reinhold2019evaluating}
\bibinfo{author}{Reinhold, J.C.}, \bibinfo{author}{Dewey, B.E.}, \bibinfo{author}{Carass, A.}, \bibinfo{author}{Prince, J.L.}, \bibinfo{year}{2019}.
\newblock \bibinfo{title}{{Evaluating the impact of intensity normalization on MR image synthesis}}, in: \bibinfo{booktitle}{{Proceedings of SPIE--the International Society for Optical Engineering}}, p. \bibinfo{pages}{109493H}.
\bibitem[{Resnick et~al.(2000)Resnick, Goldszal, Davatzikos, Golski, Kraut, Metter, Bryan and Zonderman}]{resnick2000one}
\bibinfo{author}{Resnick, S.M.}, \bibinfo{author}{Goldszal, A.F.}, \bibinfo{author}{Davatzikos, C.}, \bibinfo{author}{Golski, S.}, \bibinfo{author}{Kraut, M.A.}, \bibinfo{author}{Metter, E.J.}, \bibinfo{author}{Bryan, R.N.}, \bibinfo{author}{Zonderman, A.B.}, \bibinfo{year}{2000}.
\newblock \bibinfo{title}{{One-year age changes in MRI brain volumes in older adults}}.
\newblock \bibinfo{journal}{{Cerebral Cortex}} \bibinfo{volume}{10}, \bibinfo{pages}{464--472}.
\bibitem[{Rittner et~al.(2010)Rittner, Lotufo, Campbell and Pike}]{rittner2010segmentation}
\bibinfo{author}{Rittner, L.}, \bibinfo{author}{Lotufo, R.A.}, \bibinfo{author}{Campbell, J.}, \bibinfo{author}{Pike, G.B.}, \bibinfo{year}{2010}.
\newblock \bibinfo{title}{{Segmentation of thalamic nuclei based on tensorial morphological gradient of diffusion tensor fields}}, in: \bibinfo{booktitle}{{2010 IEEE International Symposium on Biomedical Imaging: From Nano to Macro}}, \bibinfo{organization}{IEEE}. pp. \bibinfo{pages}{1173--1176}.
\bibitem[{Ruiz-Perez et~al.(2024)Ruiz-Perez, Morell-Ortega, Gadea, Vivo-Hernando, Rubio, Aparici, de~la Iglesia-Vaya, Tourdias, Coup{\'e} and Manj{\'o}n}]{ruiz2024deepthalamus}
\bibinfo{author}{Ruiz-Perez, M.}, \bibinfo{author}{Morell-Ortega, S.}, \bibinfo{author}{Gadea, M.}, \bibinfo{author}{Vivo-Hernando, R.}, \bibinfo{author}{Rubio, G.}, \bibinfo{author}{Aparici, F.}, \bibinfo{author}{de~la Iglesia-Vaya, M.}, \bibinfo{author}{Tourdias, T.}, \bibinfo{author}{Coup{\'e}, P.}, \bibinfo{author}{Manj{\'o}n, J.V.}, \bibinfo{year}{2024}.
\newblock \bibinfo{title}{{DeepThalamus: A novel deep learning method for automatic segmentation of brain thalamic nuclei from multimodal ultra-high resolution MRI}}.
\newblock \bibinfo{journal}{arXiv preprint arXiv:2401.07751} .
\bibitem[{Salanova et~al.(2015)Salanova, Witt, Worth, Henry, Gross, Nazzaro, Labar, Sperling, Sharan, Sandok et~al.}]{salanova2015long}
\bibinfo{author}{Salanova, V.}, \bibinfo{author}{Witt, T.}, \bibinfo{author}{Worth, R.}, \bibinfo{author}{Henry, T.R.}, \bibinfo{author}{Gross, R.E.}, \bibinfo{author}{Nazzaro, J.M.}, \bibinfo{author}{Labar, D.}, \bibinfo{author}{Sperling, M.R.}, \bibinfo{author}{Sharan, A.}, \bibinfo{author}{Sandok, E.}, et~al., \bibinfo{year}{2015}.
\newblock \bibinfo{title}{{Long-term efficacy and safety of thalamic stimulation for drug-resistant partial epilepsy}}.
\newblock \bibinfo{journal}{Neurology} \bibinfo{volume}{84}, \bibinfo{pages}{1017--1025}.
\bibitem[{Schiff et~al.(2023)Schiff, Giacino, Butson, Choi, Baker, O’Sullivan, Janson, Bergin, Bronte-Stewart, Chua et~al.}]{schiff2023thalamic}
\bibinfo{author}{Schiff, N.D.}, \bibinfo{author}{Giacino, J.T.}, \bibinfo{author}{Butson, C.R.}, \bibinfo{author}{Choi, E.Y.}, \bibinfo{author}{Baker, J.L.}, \bibinfo{author}{O’Sullivan, K.P.}, \bibinfo{author}{Janson, A.P.}, \bibinfo{author}{Bergin, M.}, \bibinfo{author}{Bronte-Stewart, H.M.}, \bibinfo{author}{Chua, J.}, et~al., \bibinfo{year}{2023}.
\newblock \bibinfo{title}{{Thalamic deep brain stimulation in traumatic brain injury: a phase 1, randomized feasibility study}}.
\newblock \bibinfo{journal}{{Nature Medicine}} \bibinfo{volume}{29}, \bibinfo{pages}{3162--3174}.
\bibitem[{Schiff et~al.(2007)Schiff, Giacino, Kalmar, Victor, Baker, Gerber, Fritz, Eisenberg, O’connor, Kobylarz et~al.}]{schiff2007behavioural}
\bibinfo{author}{Schiff, N.D.}, \bibinfo{author}{Giacino, J.T.}, \bibinfo{author}{Kalmar, K.}, \bibinfo{author}{Victor, J.D.}, \bibinfo{author}{Baker, K.}, \bibinfo{author}{Gerber, M.}, \bibinfo{author}{Fritz, B.}, \bibinfo{author}{Eisenberg, B.}, \bibinfo{author}{O’connor, J.}, \bibinfo{author}{Kobylarz, E.}, et~al., \bibinfo{year}{2007}.
\newblock \bibinfo{title}{{Behavioural improvements with thalamic stimulation after severe traumatic brain injury}}.
\newblock \bibinfo{journal}{{Nature}} \bibinfo{volume}{448}, \bibinfo{pages}{600--603}.
\bibitem[{Segobin et~al.(2024)Segobin, Haast, Kumar, Lella, Alkemade, Bach~Cuadra, Barbeau, Felician, Pergola, Pitel et~al.}]{segobin2024roadmap}
\bibinfo{author}{Segobin, S.}, \bibinfo{author}{Haast, R.A.}, \bibinfo{author}{Kumar, V.J.}, \bibinfo{author}{Lella, A.}, \bibinfo{author}{Alkemade, A.}, \bibinfo{author}{Bach~Cuadra, M.}, \bibinfo{author}{Barbeau, E.J.}, \bibinfo{author}{Felician, O.}, \bibinfo{author}{Pergola, G.}, \bibinfo{author}{Pitel, A.L.}, et~al., \bibinfo{year}{2024}.
\newblock \bibinfo{title}{{A roadmap towards standardized neuroimaging approaches for human thalamic nuclei}}.
\newblock \bibinfo{journal}{{Nature Reviews Neuroscience}} \bibinfo{volume}{25}, \bibinfo{pages}{792--808}.
\bibitem[{Sherman(2007)}]{sherman2007thalamus}
\bibinfo{author}{Sherman, S.M.}, \bibinfo{year}{2007}.
\newblock \bibinfo{title}{{The thalamus is more than just a relay}}.
\newblock \bibinfo{journal}{{Current Opinion in Neurobiology}} \bibinfo{volume}{17}, \bibinfo{pages}{417--422}.
\bibitem[{Sherman(2016)}]{sherman2016thalamus}
\bibinfo{author}{Sherman, S.M.}, \bibinfo{year}{2016}.
\newblock \bibinfo{title}{{Thalamus plays a central role in ongoing cortical functioning}}.
\newblock \bibinfo{journal}{{Nature Neuroscience}} \bibinfo{volume}{19}, \bibinfo{pages}{533--541}.
\bibitem[{Sherman and Guillery(2006)}]{sherman2006exploring}
\bibinfo{author}{Sherman, S.M.}, \bibinfo{author}{Guillery, R.W.}, \bibinfo{year}{2006}.
\newblock \bibinfo{title}{{Exploring the thalamus and its role in cortical function}}.
\newblock \bibinfo{publisher}{{MIT Press}}.
\bibitem[{Sommer(2003)}]{sommer2003role}
\bibinfo{author}{Sommer, M.A.}, \bibinfo{year}{2003}.
\newblock \bibinfo{title}{The role of the thalamus in motor control}.
\newblock \bibinfo{journal}{{Current Opinion in Neurobiology}} \bibinfo{volume}{13}, \bibinfo{pages}{663--670}.
\bibitem[{Stehling et~al.(1991)Stehling, Turner and Mansfield}]{stehling1991echo}
\bibinfo{author}{Stehling, M.K.}, \bibinfo{author}{Turner, R.}, \bibinfo{author}{Mansfield, P.}, \bibinfo{year}{1991}.
\newblock \bibinfo{title}{{Echo-planar imaging: magnetic resonance imaging in a fraction of a second}}.
\newblock \bibinfo{journal}{Science} \bibinfo{volume}{254}, \bibinfo{pages}{43--50}.
\bibitem[{Stough et~al.(2014)Stough, Glaister, Ye, Ying, Prince and Carass}]{stough2014automatic}
\bibinfo{author}{Stough, J.V.}, \bibinfo{author}{Glaister, J.}, \bibinfo{author}{Ye, C.}, \bibinfo{author}{Ying, S.H.}, \bibinfo{author}{Prince, J.L.}, \bibinfo{author}{Carass, A.}, \bibinfo{year}{2014}.
\newblock \bibinfo{title}{Automatic method for thalamus parcellation using multi-modal feature classification}, in: \bibinfo{booktitle}{{Medical Image Computing and Computer-Assisted Intervention--MICCAI 2014: 17th International Conference, Boston, MA, USA, September 14-18, 2014, Proceedings, Part III 17}}, \bibinfo{organization}{Springer}. pp. \bibinfo{pages}{169--176}.
\bibitem[{Su et~al.(2019)Su, Thomas, Kasoff, Tourdias, Choi, Rutt and Saranathan}]{su2019thalamus}
\bibinfo{author}{Su, J.H.}, \bibinfo{author}{Thomas, F.T.}, \bibinfo{author}{Kasoff, W.S.}, \bibinfo{author}{Tourdias, T.}, \bibinfo{author}{Choi, E.Y.}, \bibinfo{author}{Rutt, B.K.}, \bibinfo{author}{Saranathan, M.}, \bibinfo{year}{2019}.
\newblock \bibinfo{title}{{Thalamus Optimized Multi Atlas Segmentation (THOMAS): fast, fully automated segmentation of thalamic nuclei from structural MRI}}.
\newblock \bibinfo{journal}{NeuroImage} \bibinfo{volume}{194}, \bibinfo{pages}{272--282}.
\bibitem[{Torrico and Munakomi(2023)}]{torrico2019neuroanatomy}
\bibinfo{author}{Torrico, T.J.}, \bibinfo{author}{Munakomi, S.}, \bibinfo{year}{2023}.
\newblock \bibinfo{title}{Neuroanatomy, thalamus}.
\newblock \bibinfo{howpublished}{In: \emph{StatPearls} [Internet]. Treasure Island (FL): StatPearls Publishing}.
\newblock \URLprefix \url{https://www.ncbi.nlm.nih.gov/books/NBK542184/}.
\bibitem[{Tourdias et~al.(2014)Tourdias, Saranathan, Levesque, Su and Rutt}]{tourdias2014visualization}
\bibinfo{author}{Tourdias, T.}, \bibinfo{author}{Saranathan, M.}, \bibinfo{author}{Levesque, I.R.}, \bibinfo{author}{Su, J.}, \bibinfo{author}{Rutt, B.K.}, \bibinfo{year}{2014}.
\newblock \bibinfo{title}{{Visualization of intra-thalamic nuclei with optimized white-matter-nulled MPRAGE at 7T}}.
\newblock \bibinfo{journal}{NeuroImage} \bibinfo{volume}{84}, \bibinfo{pages}{534--545}.
\bibitem[{Traynor et~al.(2011)Traynor, Barker, Crum, Williams and Richardson}]{traynor2011segmentation}
\bibinfo{author}{Traynor, C.R.}, \bibinfo{author}{Barker, G.J.}, \bibinfo{author}{Crum, W.R.}, \bibinfo{author}{Williams, S.C.}, \bibinfo{author}{Richardson, M.P.}, \bibinfo{year}{2011}.
\newblock \bibinfo{title}{{Segmentation of the thalamus in MRI based on T1 and T2}}.
\newblock \bibinfo{journal}{NeuroImage} \bibinfo{volume}{56}, \bibinfo{pages}{939--950}.
\bibitem[{Tregidgo et~al.(2023)Tregidgo, Soskic, Althonayan, Maffei, Van~Leemput, Golland, Insausti, Lerma-Usabiaga, Caballero-Gaudes, Paz-Alonso et~al.}]{tregidgo2023accurate}
\bibinfo{author}{Tregidgo, H.F.}, \bibinfo{author}{Soskic, S.}, \bibinfo{author}{Althonayan, J.}, \bibinfo{author}{Maffei, C.}, \bibinfo{author}{Van~Leemput, K.}, \bibinfo{author}{Golland, P.}, \bibinfo{author}{Insausti, R.}, \bibinfo{author}{Lerma-Usabiaga, G.}, \bibinfo{author}{Caballero-Gaudes, C.}, \bibinfo{author}{Paz-Alonso, P.M.}, et~al., \bibinfo{year}{2023}.
\newblock \bibinfo{title}{{Accurate Bayesian segmentation of thalamic nuclei using diffusion MRI and an improved histological atlas}}.
\newblock \bibinfo{journal}{NeuroImage} \bibinfo{volume}{274}, \bibinfo{pages}{120129}.
\bibitem[{Tustison et~al.(2010)Tustison, Avants, Cook, Zheng, Egan, Yushkevich and Gee}]{tustison2010n4itk}
\bibinfo{author}{Tustison, N.J.}, \bibinfo{author}{Avants, B.B.}, \bibinfo{author}{Cook, P.A.}, \bibinfo{author}{Zheng, Y.}, \bibinfo{author}{Egan, A.}, \bibinfo{author}{Yushkevich, P.A.}, \bibinfo{author}{Gee, J.C.}, \bibinfo{year}{2010}.
\newblock \bibinfo{title}{{N4ITK: improved N3 bias correction}}.
\newblock \bibinfo{journal}{{IEEE Transactions on Medical Imaging}} \bibinfo{volume}{29}, \bibinfo{pages}{1310--1320}.
\bibitem[{Umapathy et~al.(2022)Umapathy, Keerthivasan, Zahr, Bilgin and Saranathan}]{umapathy2022convolutional}
\bibinfo{author}{Umapathy, L.}, \bibinfo{author}{Keerthivasan, M.B.}, \bibinfo{author}{Zahr, N.M.}, \bibinfo{author}{Bilgin, A.}, \bibinfo{author}{Saranathan, M.}, \bibinfo{year}{2022}.
\newblock \bibinfo{title}{{Convolutional neural network based frameworks for fast automatic segmentation of thalamic nuclei from native and synthesized contrast structural MRI}}.
\newblock \bibinfo{journal}{Neuroinformatics} \bibinfo{volume}{20}, \bibinfo{pages}{651--664}.
\bibitem[{Valent{\'\i}n et~al.(2013)Valent{\'\i}n, Garc{\'\i}a~Navarrete, Chelvarajah, Torres, Navas, Vico, Torres, Pastor, Selway, Sola et~al.}]{valentin2013deep}
\bibinfo{author}{Valent{\'\i}n, A.}, \bibinfo{author}{Garc{\'\i}a~Navarrete, E.}, \bibinfo{author}{Chelvarajah, R.}, \bibinfo{author}{Torres, C.}, \bibinfo{author}{Navas, M.}, \bibinfo{author}{Vico, L.}, \bibinfo{author}{Torres, N.}, \bibinfo{author}{Pastor, J.}, \bibinfo{author}{Selway, R.}, \bibinfo{author}{Sola, R.G.}, et~al., \bibinfo{year}{2013}.
\newblock \bibinfo{title}{{Deep brain stimulation of the centromedian thalamic nucleus for the treatment of generalized and frontal epilepsies}}.
\newblock \bibinfo{journal}{{Epilepsia}} \bibinfo{volume}{54}, \bibinfo{pages}{1823--1833}.
\bibitem[{Van~Oort et~al.(2018)Van~Oort, Mennes, Schr{\"o}der, Kumar, Jimenez, Grodd, Doeller and Beckmann}]{van2018functional}
\bibinfo{author}{Van~Oort, E.S.}, \bibinfo{author}{Mennes, M.}, \bibinfo{author}{Schr{\"o}der, T.N.}, \bibinfo{author}{Kumar, V.J.}, \bibinfo{author}{Jimenez, N.I.Z.}, \bibinfo{author}{Grodd, W.}, \bibinfo{author}{Doeller, C.F.}, \bibinfo{author}{Beckmann, C.F.}, \bibinfo{year}{2018}.
\newblock \bibinfo{title}{{Functional parcellation using time courses of instantaneous connectivity}}.
\newblock \bibinfo{journal}{{NeuroImage}} \bibinfo{volume}{170}, \bibinfo{pages}{31--40}.
\bibitem[{Vidal et~al.(2024)Vidal, Danet, P{\'e}ran, Pariente, Bach~Cuadra, Zahr, Barbeau and Saranathan}]{vidal2024robust}
\bibinfo{author}{Vidal, J.P.}, \bibinfo{author}{Danet, L.}, \bibinfo{author}{P{\'e}ran, P.}, \bibinfo{author}{Pariente, J.}, \bibinfo{author}{Bach~Cuadra, M.}, \bibinfo{author}{Zahr, N.M.}, \bibinfo{author}{Barbeau, E.J.}, \bibinfo{author}{Saranathan, M.}, \bibinfo{year}{2024}.
\newblock \bibinfo{title}{{Robust thalamic nuclei segmentation from T1-weighted MRI using polynomial intensity transformation}}.
\newblock \bibinfo{journal}{Brain Structure and Function} \bibinfo{volume}{229}, \bibinfo{pages}{1087--1101}.
\bibitem[{Wiegell et~al.(2003)Wiegell, Tuch, Larsson and Wedeen}]{wiegell2003automatic}
\bibinfo{author}{Wiegell, M.R.}, \bibinfo{author}{Tuch, D.S.}, \bibinfo{author}{Larsson, H.B.}, \bibinfo{author}{Wedeen, V.J.}, \bibinfo{year}{2003}.
\newblock \bibinfo{title}{Automatic segmentation of thalamic nuclei from diffusion tensor magnetic resonance imaging}.
\newblock \bibinfo{journal}{NeuroImage} \bibinfo{volume}{19}, \bibinfo{pages}{391--401}.
\bibitem[{Williams et~al.(2024)Williams, Nguyen, Vidal and Saranathan}]{williams2024thalamic}
\bibinfo{author}{Williams, B.}, \bibinfo{author}{Nguyen, D.}, \bibinfo{author}{Vidal, J.P.}, \bibinfo{author}{Saranathan, M.}, \bibinfo{year}{2024}.
\newblock \bibinfo{title}{{Thalamic nuclei segmentation from T1-weighted MRI: Unifying and benchmarking state-of-the-art methods}}.
\newblock \bibinfo{journal}{{Imaging Neuroscience}} \bibinfo{volume}{2}, \bibinfo{pages}{imag--2}.
\bibitem[{Yan et~al.(2023)Yan, Shao, Bian, Feng, Xue, Zhuo, Gullapalli, Carass and Prince}]{yan2023segmenting}
\bibinfo{author}{Yan, C.}, \bibinfo{author}{Shao, M.}, \bibinfo{author}{Bian, Z.}, \bibinfo{author}{Feng, A.}, \bibinfo{author}{Xue, Y.}, \bibinfo{author}{Zhuo, J.}, \bibinfo{author}{Gullapalli, R.P.}, \bibinfo{author}{Carass, A.}, \bibinfo{author}{Prince, J.L.}, \bibinfo{year}{2023}.
\newblock \bibinfo{title}{{Segmenting thalamic nuclei from manifold projections of multi-contrast MRI}}, in: \bibinfo{booktitle}{{Medical Imaging 2023: Image Processing}}, \bibinfo{organization}{SPIE}. pp. \bibinfo{pages}{727--734}.
\bibitem[{Ye et~al.(2013)Ye, Bogovic, Ying and Prince}]{ye2013parcellation}
\bibinfo{author}{Ye, C.}, \bibinfo{author}{Bogovic, J.A.}, \bibinfo{author}{Ying, S.H.}, \bibinfo{author}{Prince, J.L.}, \bibinfo{year}{2013}.
\newblock \bibinfo{title}{{Parcellation of the thalamus using diffusion tensor images and a multi-object geometric deformable model}}, in: \bibinfo{booktitle}{{Proceedings of SPIE--the International Society for Optical Engineering}}, pp. \bibinfo{pages}{10--1117}.
\bibitem[{Zhang et~al.(2008)Zhang, Snyder, Fox, Sansbury, Shimony and Raichle}]{zhang2008intrinsic}
\bibinfo{author}{Zhang, D.}, \bibinfo{author}{Snyder, A.Z.}, \bibinfo{author}{Fox, M.D.}, \bibinfo{author}{Sansbury, M.W.}, \bibinfo{author}{Shimony, J.S.}, \bibinfo{author}{Raichle, M.E.}, \bibinfo{year}{2008}.
\newblock \bibinfo{title}{{Intrinsic functional relations between human cerebral cortex and thalamus}}.
\newblock \bibinfo{journal}{{Journal of Neurophysiology}} \bibinfo{volume}{100}, \bibinfo{pages}{1740--1748}.
\bibitem[{Ziyan et~al.(2006)Ziyan, Tuch and Westin}]{ziyan2006segmentation}
\bibinfo{author}{Ziyan, U.}, \bibinfo{author}{Tuch, D.}, \bibinfo{author}{Westin, C.F.}, \bibinfo{year}{2006}.
\newblock \bibinfo{title}{{Segmentation of thalamic nuclei from DTI using spectral clustering}}, in: \bibinfo{booktitle}{{Medical Image Computing and Computer-Assisted Intervention--MICCAI 2006: 9th International Conference, Copenhagen, Denmark, October 1-6, 2006. Proceedings, Part II 9}}, \bibinfo{organization}{Springer}. pp. \bibinfo{pages}{807--814}.
\bibitem[{Zuo et~al.(2023)Zuo, Liu, Xue, Dewey, Remedios, Hays, Bilgel, Mowry, Newsome, Calabresi et~al.}]{zuo2023haca3}
\bibinfo{author}{Zuo, L.}, \bibinfo{author}{Liu, Y.}, \bibinfo{author}{Xue, Y.}, \bibinfo{author}{Dewey, B.E.}, \bibinfo{author}{Remedios, S.W.}, \bibinfo{author}{Hays, S.P.}, \bibinfo{author}{Bilgel, M.}, \bibinfo{author}{Mowry, E.M.}, \bibinfo{author}{Newsome, S.D.}, \bibinfo{author}{Calabresi, P.A.}, et~al., \bibinfo{year}{2023}.
\newblock \bibinfo{title}{{HACA3: A unified approach for multi-site MR image harmonization}}.
\newblock \bibinfo{journal}{{Computerized Medical Imaging and Graphics}} \bibinfo{volume}{109}, \bibinfo{pages}{102285}.

\end{thebibliography}

\end{document}


\section*{Supplementary Note S1. Multi-TI Reference Selection} 

To support the selection of TI images for manual annotation, we first computed the local contrast of each registered Morel nucleus across TI images. For each nucleus, the surrounding region was defined as a 4-voxel dilation shell around the registered Morel label, excluding the original nucleus mask. For each subject and TI image, local contrast was defined as:
\[
C(\mathrm{TI}) =
\frac{
\left|\mu_{\mathrm{nuc}}(\mathrm{TI}) - \mu_{\mathrm{sur}}(\mathrm{TI})\right|
}{
\sigma_{\mathrm{local}}(\mathrm{TI})
},
\]
where
\[
\sigma_{\mathrm{local}}(\mathrm{TI}) =
\sqrt{
\frac{
\sigma_{\mathrm{nuc}}^2(\mathrm{TI}) +
\sigma_{\mathrm{sur}}^2(\mathrm{TI})
}{2}
}.
\]
Here, \(\mu_{\mathrm{nuc}}\) and \(\mu_{\mathrm{sur}}\) are the mean intensities in the registered Morel nucleus and surrounding shell, and \(\sigma_{\mathrm{nuc}}\) and \(\sigma_{\mathrm{sur}}\) are the corresponding standard deviations. For each nucleus and TI image, \(C(\mathrm{TI})\) was computed separately for each subject and then averaged across the 24 annotated subjects.

\begin{figure}[H]
\centering\includegraphics[width=\textwidth]{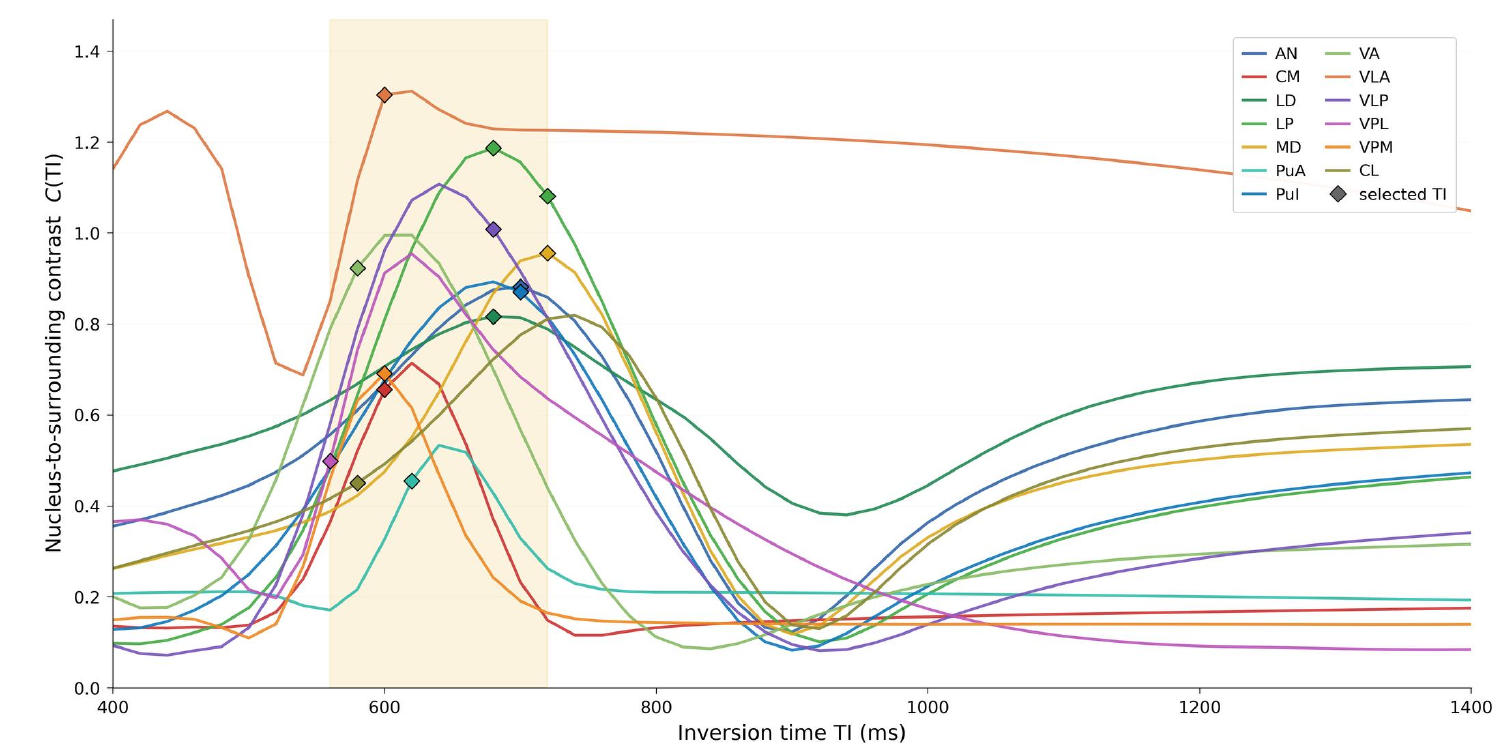}
\caption{Local contrast across TI images. For each nucleus and TI image, contrast was computed between the registered Morel nucleus and a 4-voxel surrounding shell separately for each subject, and then averaged across the 24 annotated subjects. Diamond markers indicate the TI images selected for manual annotation. The shaded region indicates the range of selected TI values across nuclei.}
    \label{fig:s3_ti_selection}
\end{figure}

\medskip\noindent
Supplementary Figure~\ref{fig:s3_ti_selection} shows that local contrast varied across TI images and nuclei. For most nuclei, the highest or near-peak contrast occurred within approximately 560--720 ms. Because the contrast peaks occurred at different TI values across nuclei, a single TI image was not sufficient for delineating all nuclei. Based on these contrast profiles, the manual annotation protocol selected TI images within this range, generally close to the contrast peak for each nucleus. The final selection also considered image quality, boundary visibility, subject-specific anatomy, and multi-view landmarks, so the selected TI values did not exactly match the contrast peak for every nucleus.

\newpage
\section*{Supplementary Note S2. Manual Annotation Protocol}
This section summarizes the thalamic nuclei delineation protocol developed by A.C. and J.Z. to guide manual annotation.
A.C. performed the manual annotations according to this protocol under the supervision of J.Z.
Supplementary Table~\ref{tab:annotation_protocol} lists the corresponding Morel atlas nuclei, the selected TI image used as the primary reference, the preferred starting point, and the landmarks used in coronal, axial, and sagittal views.

\begin{table}[H]
\centering
\scriptsize
\renewcommand{\arraystretch}{1.25}
\setlength{\tabcolsep}{3.2pt}

\caption{Manual annotation protocol for the 13 thalamic nuclei. For each label, the table lists the corresponding Morel atlas nuclei, the selected TI image used as the primary reference, the preferred starting point, and the view-specific landmarks used during manual refinement.}
\label{tab:annotation_protocol}

\begin{adjustbox}{width=\textwidth}
\begin{tabular}{|C{1.1cm}|C{1.8cm}|C{1.1cm}|C{3.0cm}|C{3.8cm}|C{3.8cm}|C{3.8cm}|}
\hline
\rowcolor{gray!20}
\textbf{Label} &
\textbf{Morel label} &
\textbf{TI} &
\textbf{Starting point} &
\textbf{Coronal view} &
\textbf{Axial view} &
\textbf{Sagittal view} \\
\hline
\rowcolor{ANcolor}
\wt{\textbf{AN}} &
\wt{AD, AM, AV} &
\wt{700 ms} &
\wt{Half-arch region in the coronal view} &
\wt{Thin half-arch adjacent to the ventricles.} &
\wt{Thicker segment adjacent to the thin ventricular border and medial to a darker lateral band.} &
\wt{Half-arch along the inferior ventricular border, superior to the thalamus.} \\
\hline

\rowcolor{CMcolor}
\wt{\textbf{CM}} &
\wt{CM} &
\wt{600 ms} &
\wt{Small axial region} &
\wt{Small dark region inferolateral to a large bright region near the medial ventricle.} &
\wt{Dark region immediately lateral to a medium-sized bright region adjacent to the medial ventricle.} &
\wt{Small dark region located medially within the thalamus, at the base of an elongated dark band extending superiorly.} \\
\hline

\rowcolor{LDcolor}
\wt{\textbf{LD}} &
\wt{LD} &
\wt{680 ms} &
\wt{Linear region along the medial edge of the thalamus in the coronal view} &
\wt{Triangular dark region inferior to the lateral ventricle, between a large bright medial region and a smaller bright lateral region.} &
\wt{Slightly darker elongated region along the inferior thalamus, superior to an elongated bright region.} &
\wt{Small, slightly darker region near the superior aspect of the thalamus.} \\
\hline

\rowcolor{LPcolor}
\bt{\textbf{LP}} &
\bt{LP} &
\bt{680 ms, 720 ms} &
\bt{Oval-shaped region extending from the medial ventricle toward the lateral thalamus in the axial view} &
\bt{Light region near the superior-lateral portion of the thalamus.} &
\bt{Medium-sized light region near the inferior-lateral tip of the thalamus.} &
\bt{Light region near the superior aspect of the thalamus. TI = 680 ms was used near the VPM/VLP boundary, and TI = 720 ms was used near the pulvinar boundary.} \\
\hline

\rowcolor{MDcolor}
\bt{\textbf{MD}} &
\bt{MDmc, MDpc} &
\bt{720 ms} &
\bt{Medial region adjacent to the ventricle in the coronal view} &
\bt{Large region adjacent to the medial ventricle.} &
\bt{Large region adjacent to the medial ventricle.} &
\bt{Dark region in the central thalamus.} \\
\hline

\rowcolor{PuAcolor}
\bt{\textbf{PuA}} &
\bt{PuA} &
\bt{620 ms} &
\bt{Small region in the axial view} &
\bt{Lateral dark-gray region.} &
\bt{Very small region, refined primarily using coronal and sagittal views.} &
\bt{Small light region near the posterior thalamus, surrounded by dark bands.} \\
\hline

\rowcolor{PuIcolor}
\wt{\textbf{PuI}} &
\wt{PuI, PuL, PuM} &
\wt{700 ms} &
\wt{Coronal view, where the medial portion is well defined} &
\wt{Large light region adjacent to the medial ventricle, occupying much of the thalamus in this view.} &
\wt{Large light region along the inferior thalamus.} &
\wt{Large light region in the posterior thalamus, extending from the inferior to superior thalamus.} \\
\hline

\rowcolor{VAcolor}
\bt{\textbf{VA}} &
\bt{VAmc, VApc} &
\bt{580 ms} &
\bt{Elongated region in the coronal view or half-circle region in the sagittal view} &
\bt{Medium dark region surrounded by two dark linear regions.} &
\bt{Small dark region near the superior tip of the thalamus.} &
\bt{Dark region near the anterior thalamus, surrounded by dark regions on three sides.} \\
\hline

\rowcolor{VLAcolor}
\bt{\textbf{VLA}} &
\bt{VLA} &
\bt{600 ms} &
\bt{Medium-sized region in the lateral thalamus in the coronal view} &
\bt{Dark region within a lateral dark band along the side of the thalamus.} &
\bt{Lateral dark region located approximately midway along the thalamus.} &
\bt{Medium-dark region curving around a larger light region.} \\
\hline

\rowcolor{VLPcolor}
\wt{\textbf{VLP}} &
\wt{VLP} &
\wt{680 ms} &
\wt{Tadpole-shaped region in the coronal view} &
\wt{Lateral light half-arch in the superior-lateral thalamus.} &
\wt{Larger lateral light region near the anterior thalamus.} &
\wt{Medium-dark region adjacent to a dark band.} \\
\hline

\rowcolor{VPLcolor}
\bt{\textbf{VPL}} &
\bt{VPLa, VPLp} &
\bt{560 ms} &
\bt{Medium oblong region in the lateral thalamus in the coronal view} &
\bt{Lateral region containing two light regions separated by a slightly darker band.} &
\bt{Horizontal, slightly dark lateral region.} &
\bt{Light region adjacent to a dark band.} \\
\hline

\rowcolor{VPMcolor}
\bt{\textbf{VPM}} &
\bt{VPM} &
\bt{600 ms} &
\bt{Dark region in axial or coronal view, slightly below the mid-thalamic level near the third ventricle} &
\bt{Dark region adjacent to the third ventricle.} &
\bt{Dark region immediately adjacent to the third ventricle.} &
\bt{Dark region in the anterior-middle portion of the thalamus.} \\
\hline

\rowcolor{CLcolor}
\bt{\textbf{CL}} &
\bt{CL} &
\bt{580 ms} &
\bt{Round region adjacent to the medial ventricle in the coronal view} &
\bt{Dark region between a light medial region near the ventricle and a more lateral light region.} &
\bt{Elongated arch surrounding a light region adjacent to the medial ventricle.} &
\bt{Darker arch along the inferior ventricular border, surrounding a lighter central region.} \\
\hline

\end{tabular}
\end{adjustbox}
\end{table}

\clearpage
\section*{Supplementary Note S3. Manual Correction and Consistency Check}
\begin{figure}[H]
    \centering
    \includegraphics[width=0.89\textwidth]{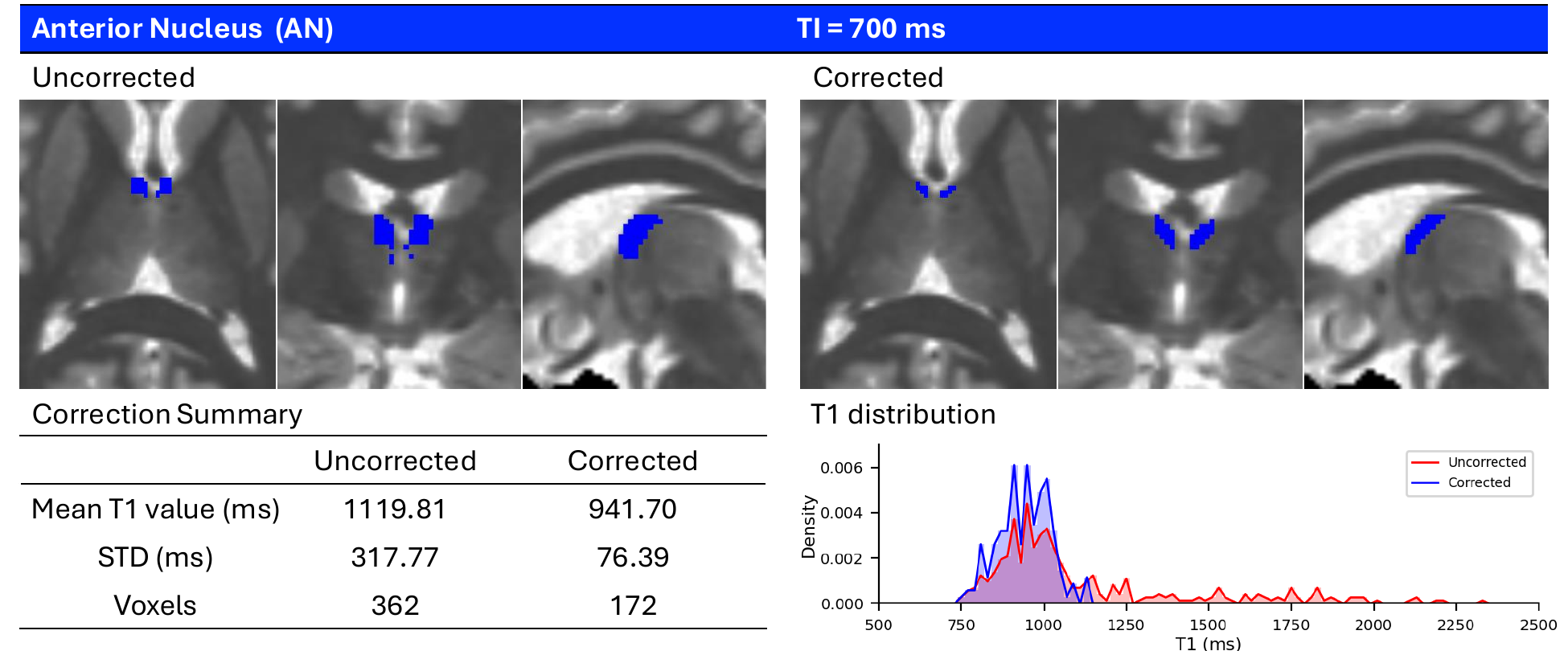}
    \caption{Manual correction example for the anterior nucleus (AN), with uncorrected and corrected labels, the correction summary, and the T1 distribution before and after correction.}
    \label{fig:s2_an}
\end{figure}

\begin{figure}[H]
    \centering
    \includegraphics[width=0.89\textwidth]{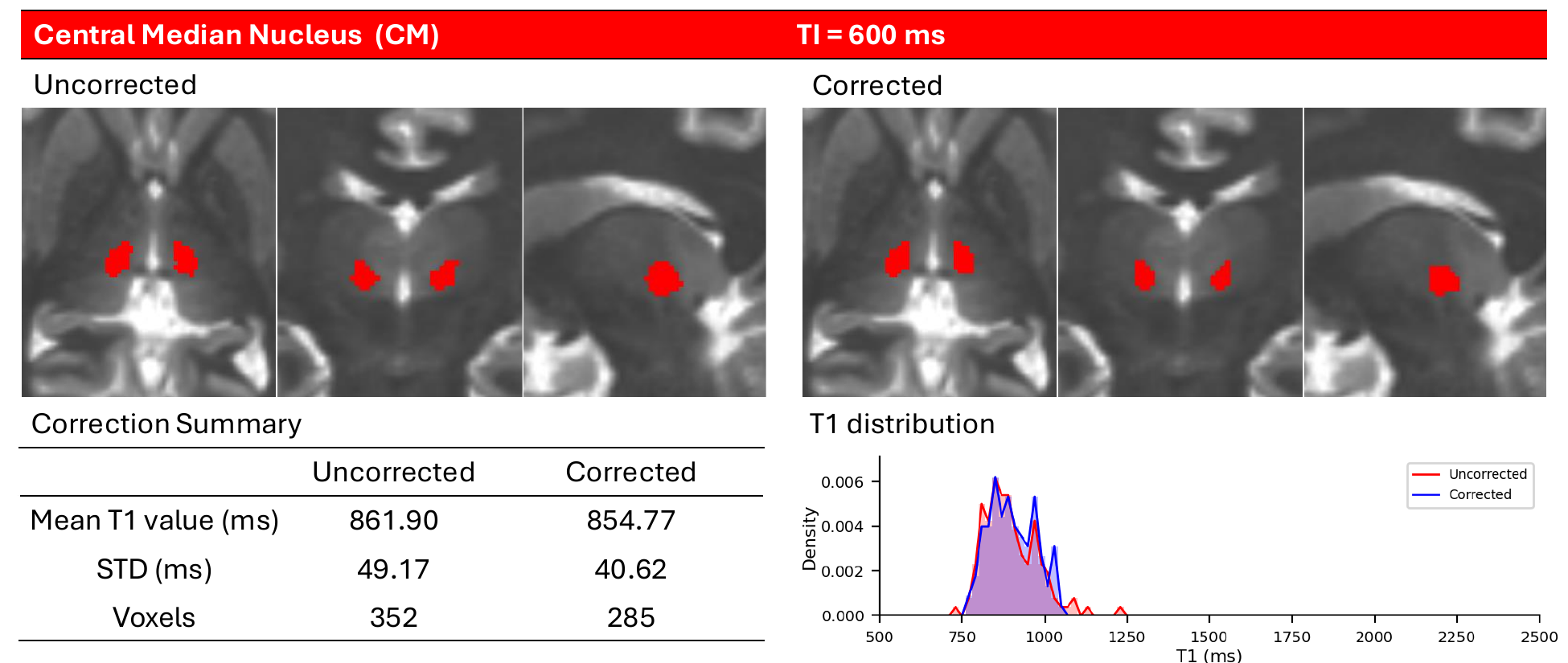}
    \caption{Manual correction example for the central median nucleus (CM), with uncorrected and corrected labels, the correction summary, and the T1 distribution before and after correction.}
    \label{fig:s2_cm}
\end{figure}

\begin{figure}[H]
    \centering
    \includegraphics[width=0.9\textwidth]{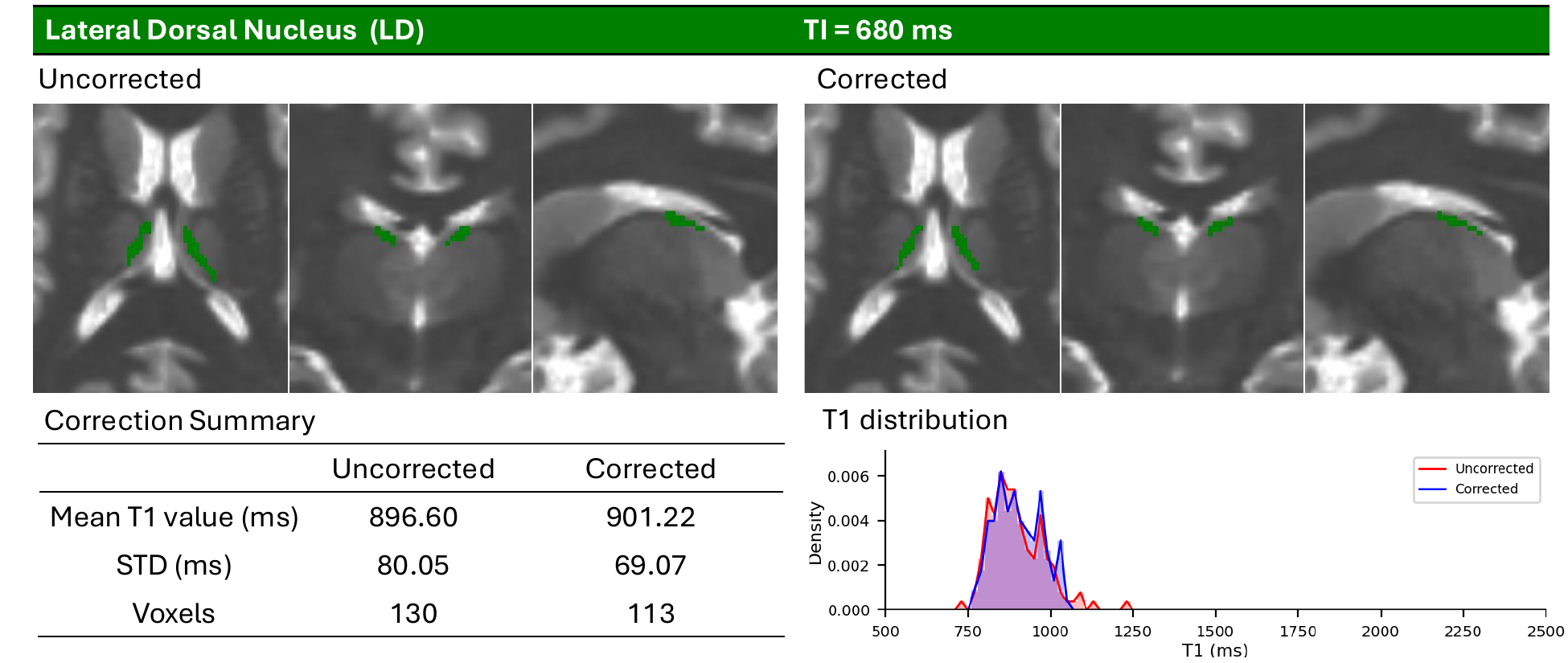}
    \caption{Manual correction example for the lateral dorsal nucleus (LD), with uncorrected and corrected labels, the correction summary, and the T1 distribution before and after correction.}
    \label{fig:s2_ld}
\end{figure}

\begin{figure}[H]
    \centering
    \includegraphics[width=0.9\textwidth]{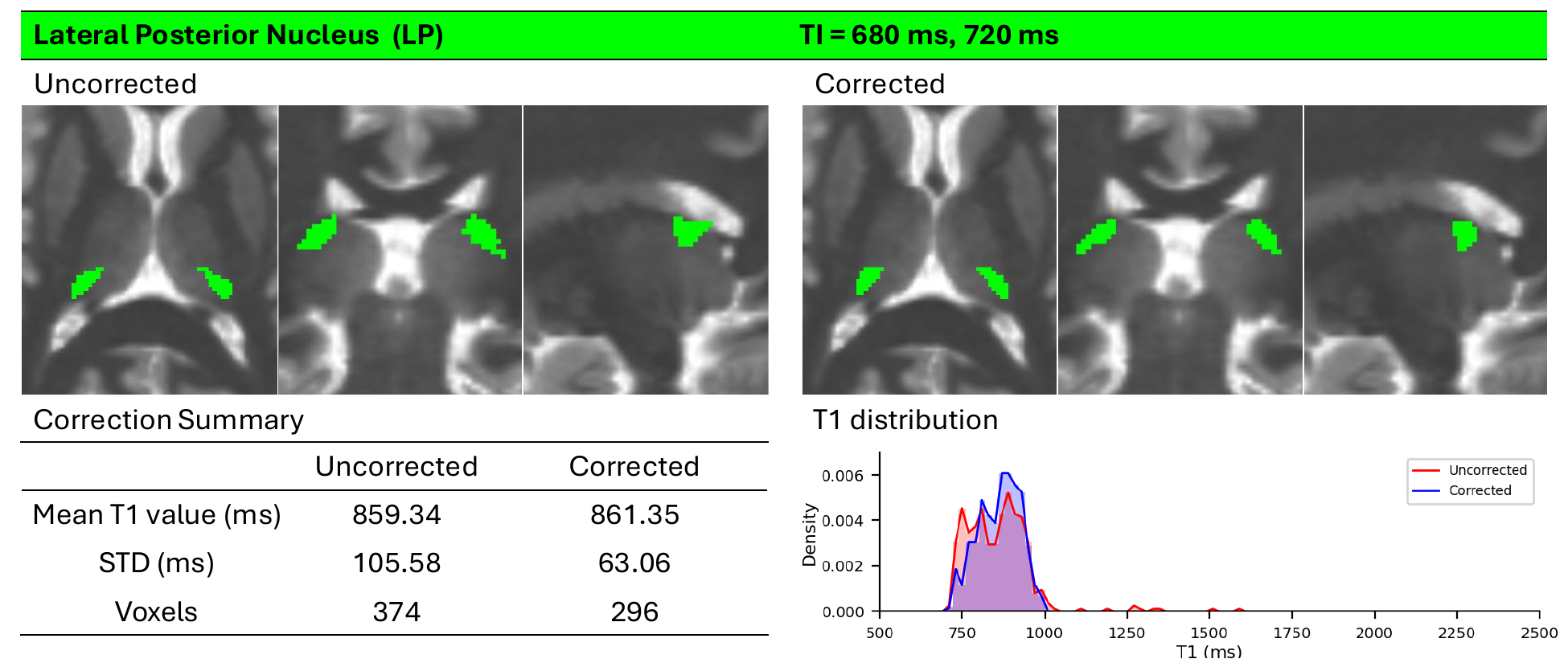}
    \caption{Manual correction example for the lateral posterior nucleus (LP), with uncorrected and corrected labels, the correction summary, and the T1 distribution before and after correction.}
    \label{fig:s2_lp}
\end{figure}

\begin{figure}[H]
    \centering
    \includegraphics[width=0.9\textwidth]{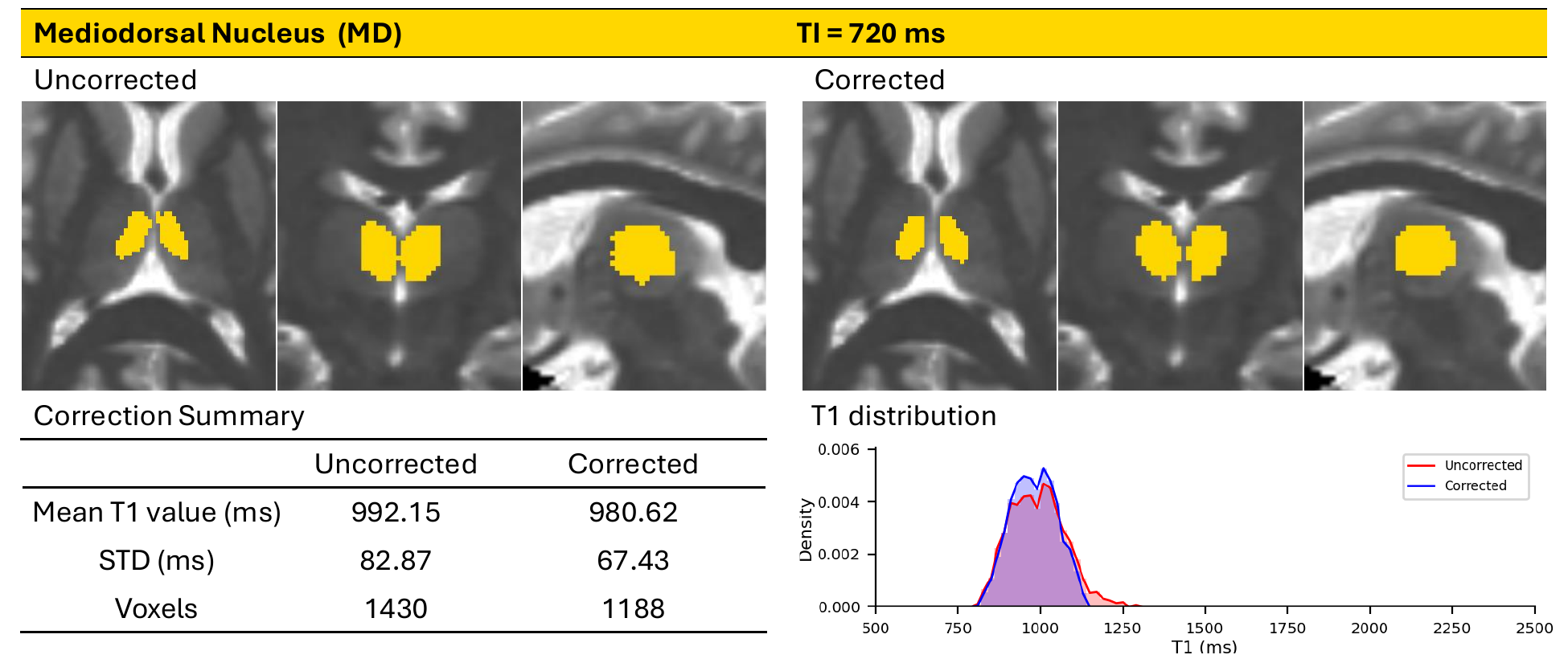}
    \caption{Manual correction example for the mediodorsal nucleus (MD), with uncorrected and corrected labels, the correction summary, and the T1 distribution before and after correction.}
    \label{fig:s2_md}
\end{figure}

\begin{figure}[H]
    \centering
    \includegraphics[width=0.9\textwidth]{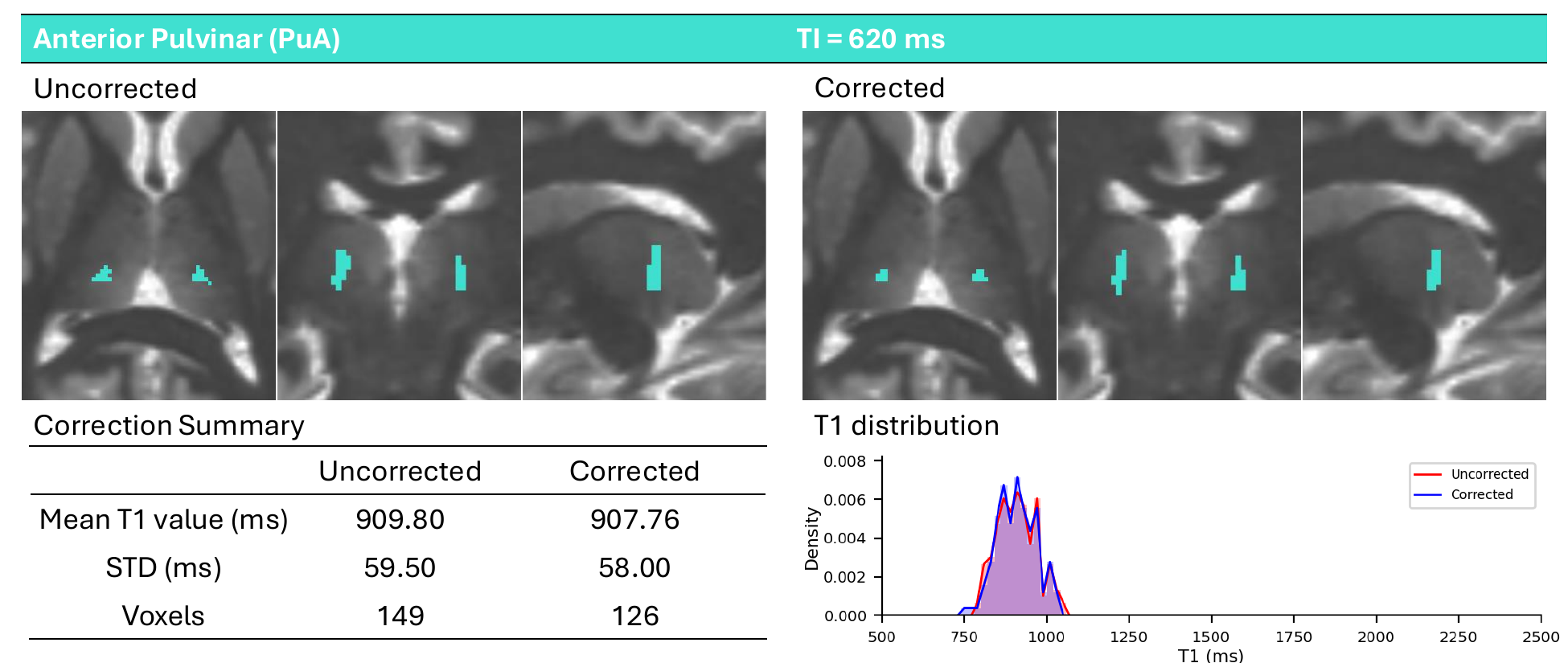}
    \caption{Manual correction example for the anterior pulvinar (PuA), with uncorrected and corrected labels, the correction summary, and the T1 distribution before and after correction.}
    \label{fig:s2_pua}
\end{figure}

\begin{figure}[H]
    \centering
    \includegraphics[width=0.9\textwidth]{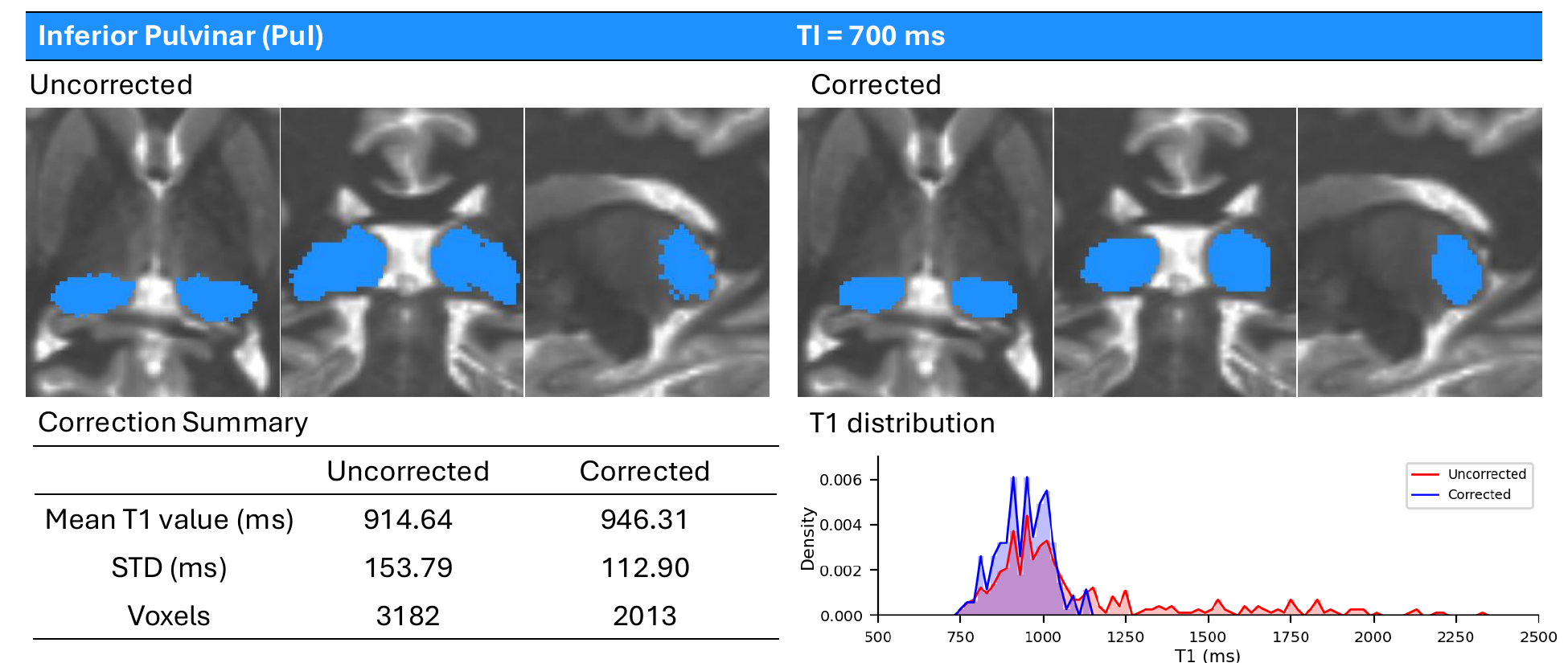}
    \caption{Manual correction example for the inferior pulvinar (PuI), with uncorrected and corrected labels, the correction summary, and the T1 distribution before and after correction.}
    \label{fig:s2_pui}
\end{figure}

\begin{figure}[H]
    \centering
    \includegraphics[width=0.9\textwidth]{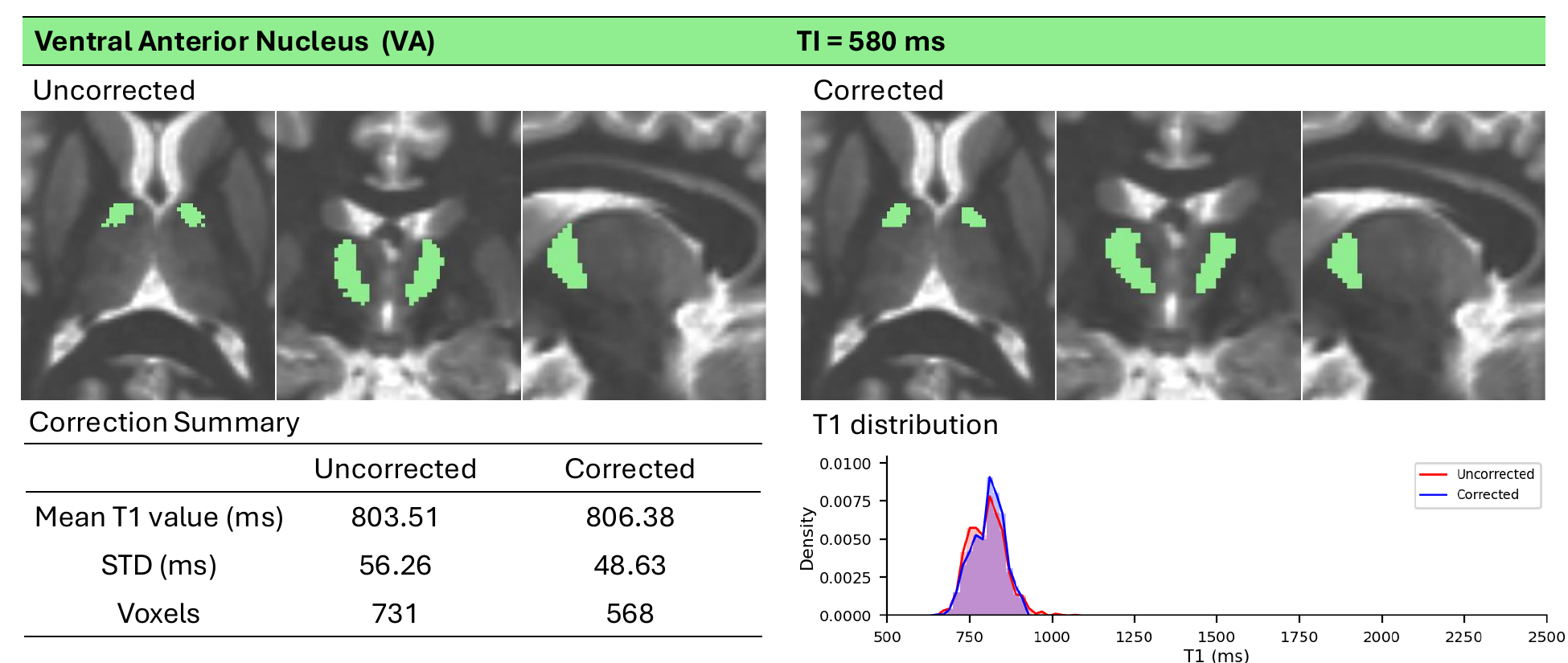}
    \caption{Manual correction example for the ventral anterior nucleus (VA), with uncorrected and corrected labels, the correction summary, and the T1 distribution before and after correction.}
    \label{fig:s2_va}
\end{figure}

\begin{figure}[H]
    \centering
    \includegraphics[width=0.9\textwidth]{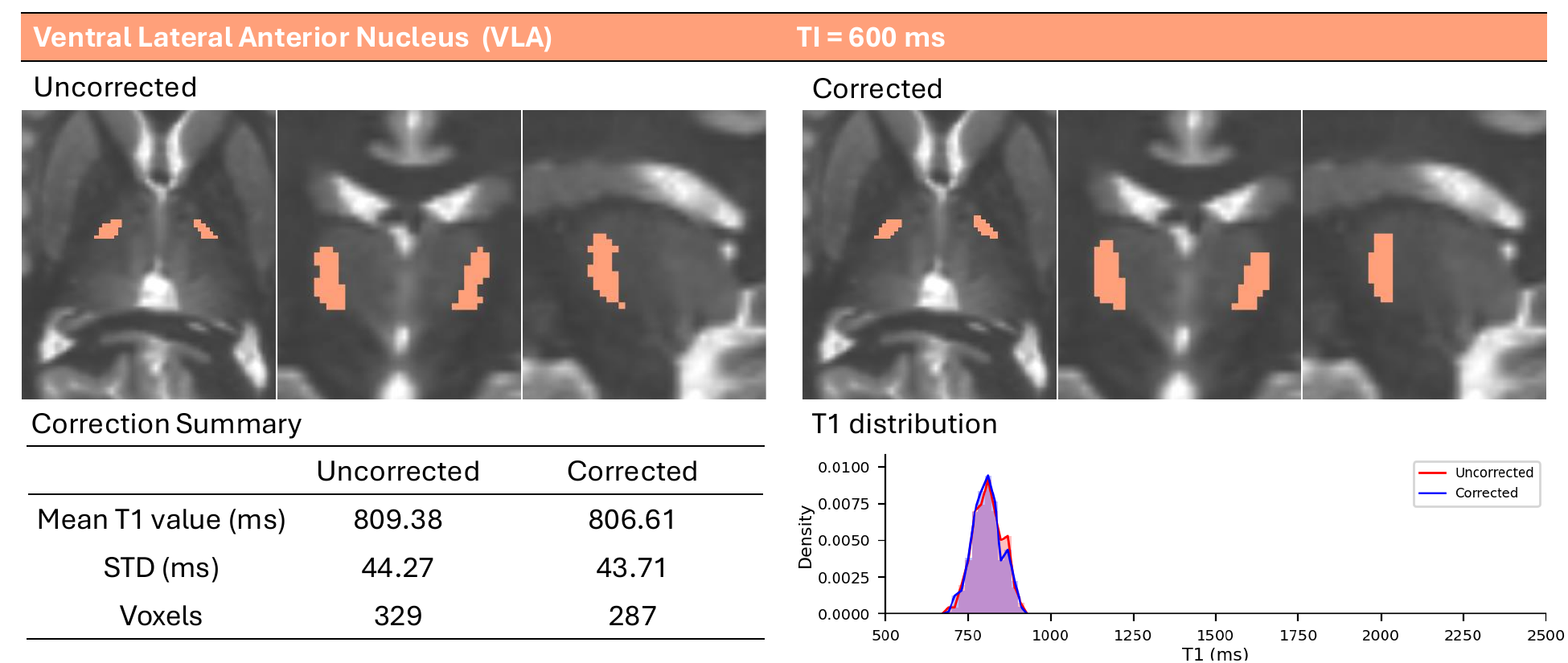}
    \caption{Manual correction example for the ventral lateral anterior nucleus (VLA), with uncorrected and corrected labels, the correction summary, and the T1 distribution before and after correction.}
    \label{fig:s2_vla}
\end{figure}

\begin{figure}[H]
    \centering
    \includegraphics[width=0.9\textwidth]{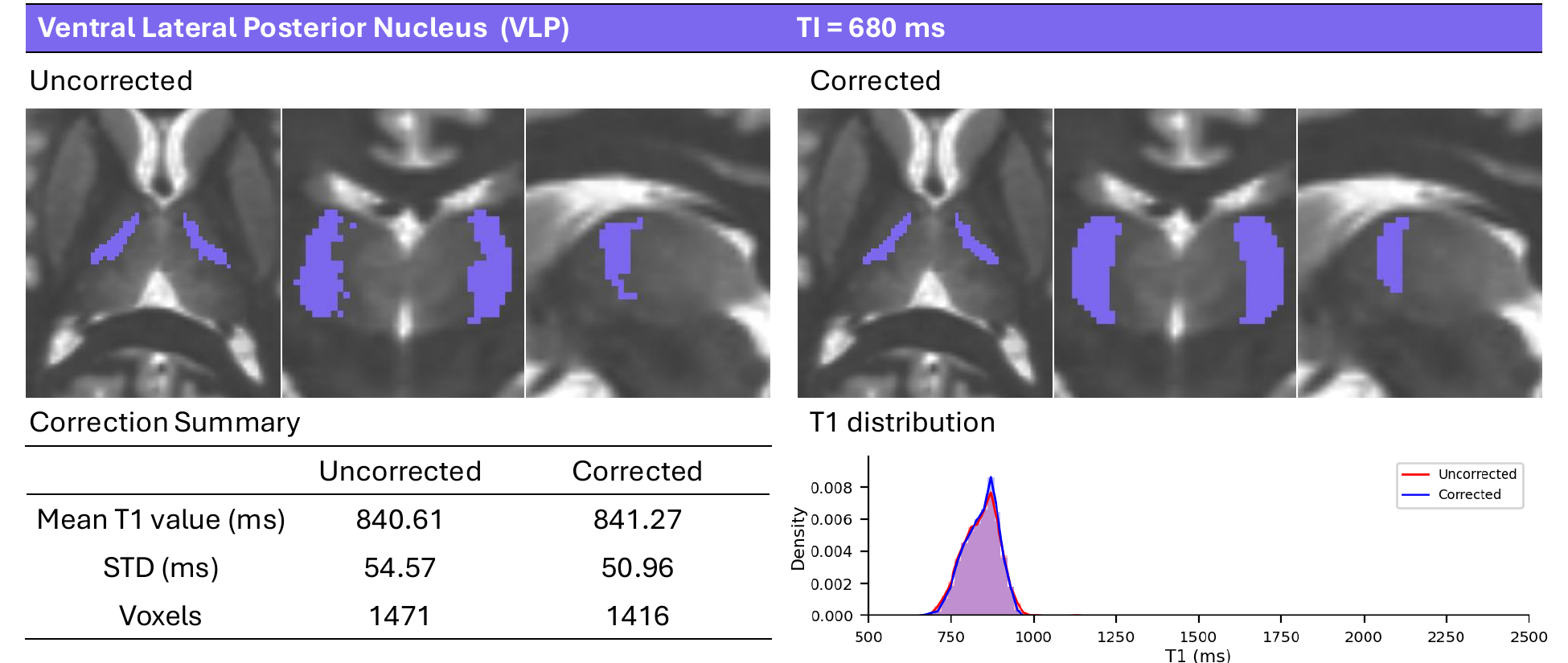}
    \caption{Manual correction example for the ventral lateral posterior nucleus (VLP), with uncorrected and corrected labels, the correction summary, and the T1 distribution before and after correction.}
    \label{fig:s2_vlp}
\end{figure}

\begin{figure}[H]
    \centering
    \includegraphics[width=0.9\textwidth]{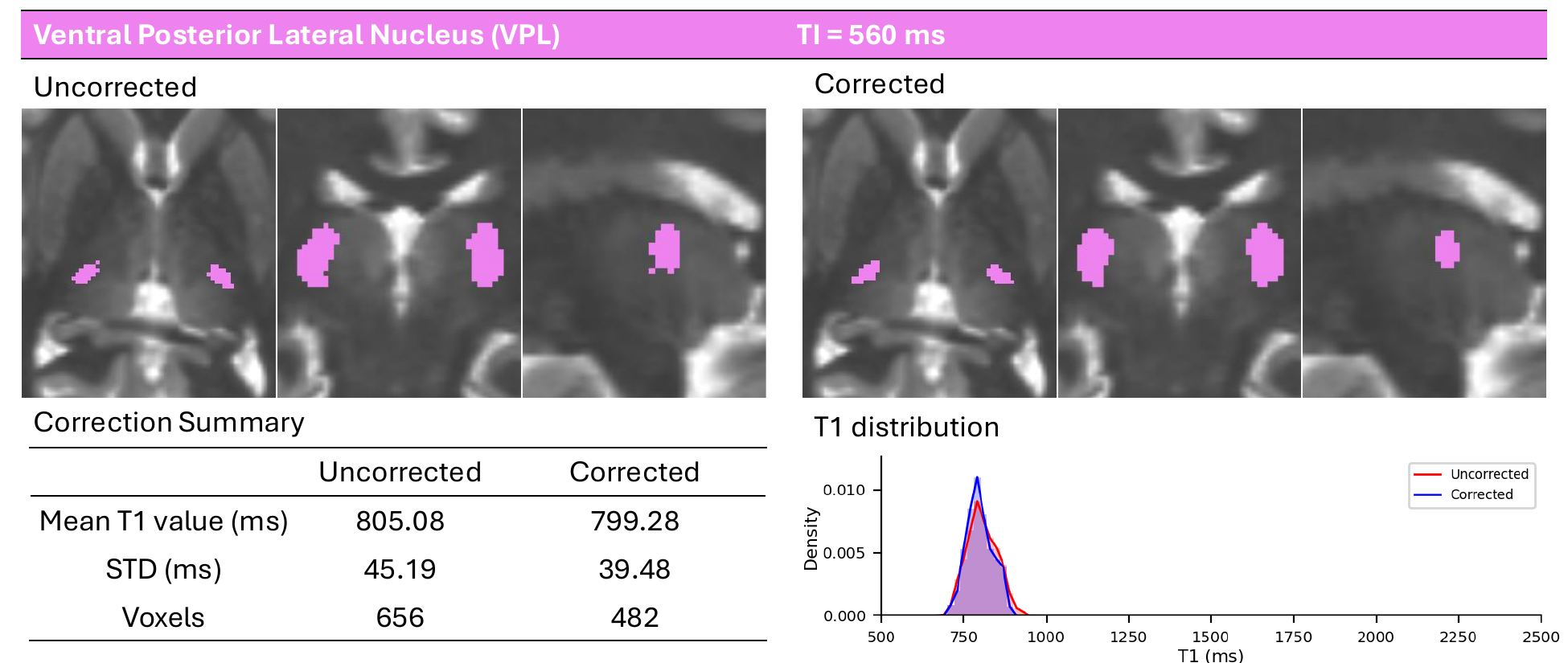}
    \caption{Manual correction example for the ventral posterior lateral nucleus (VPL), with uncorrected and corrected labels, the correction summary, and the T1 distribution before and after correction.}
    \label{fig:s2_vpl}
\end{figure}

\begin{figure}[H]
    \centering
    \includegraphics[width=0.9\textwidth]{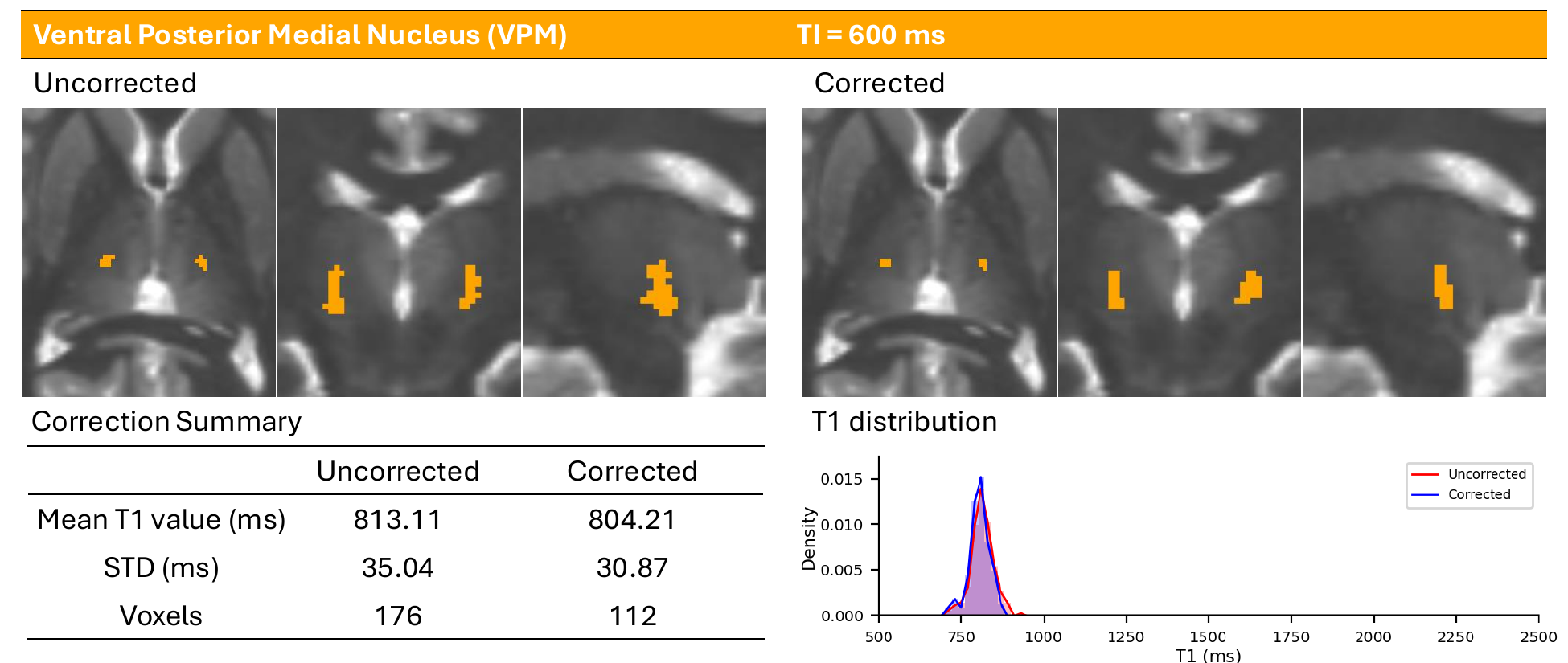}
    \caption{Manual correction example for the ventral posterior medial nucleus (VPM), with uncorrected and corrected labels, the correction summary, and the T1 distribution before and after correction.}
    \label{fig:s2_vpm}
\end{figure}

\begin{figure}[H]
    \centering
    \includegraphics[width=0.9\textwidth]{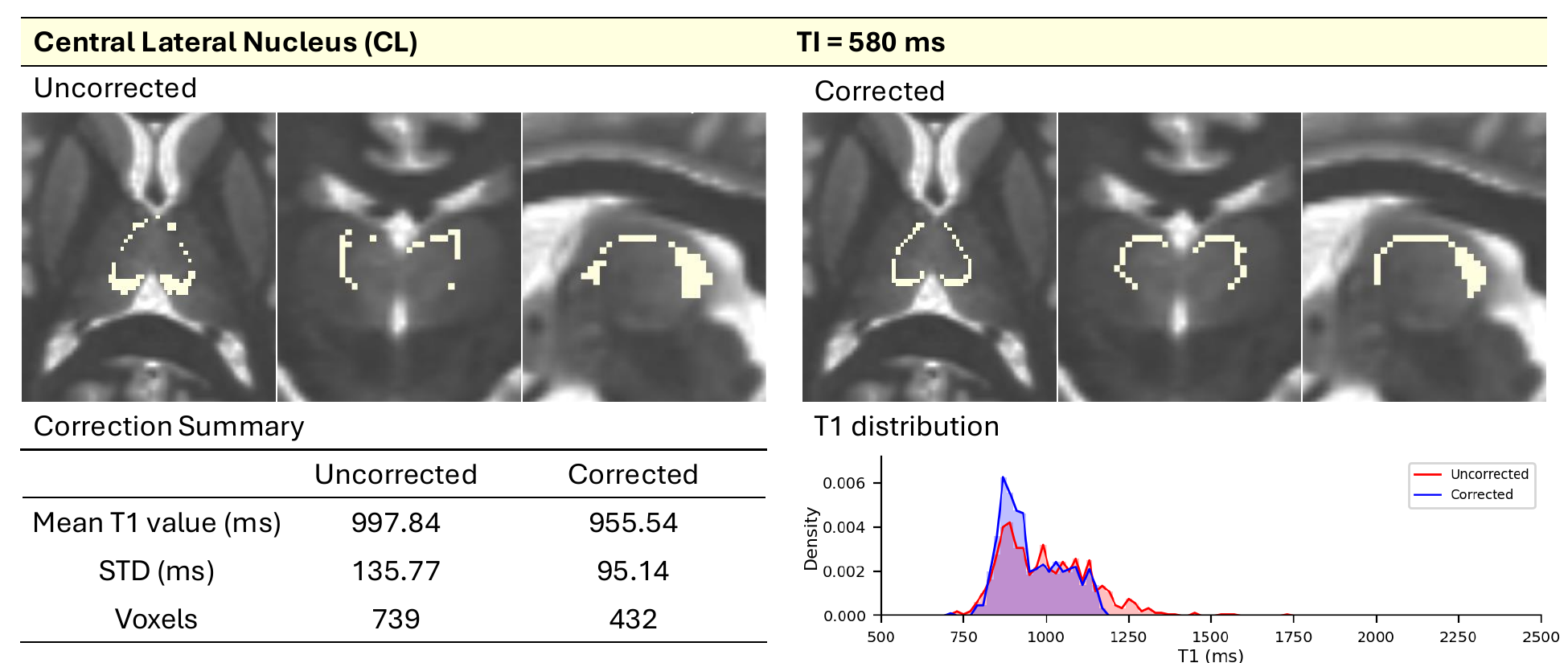}
    \caption{Manual correction example for the central lateral nucleus (CL), with uncorrected and corrected labels, the correction summary, and the T1 distribution before and after correction.}
    \label{fig:s2_cl}
\end{figure}

\begin{figure}[H]
    \centering\includegraphics[width=\textwidth]{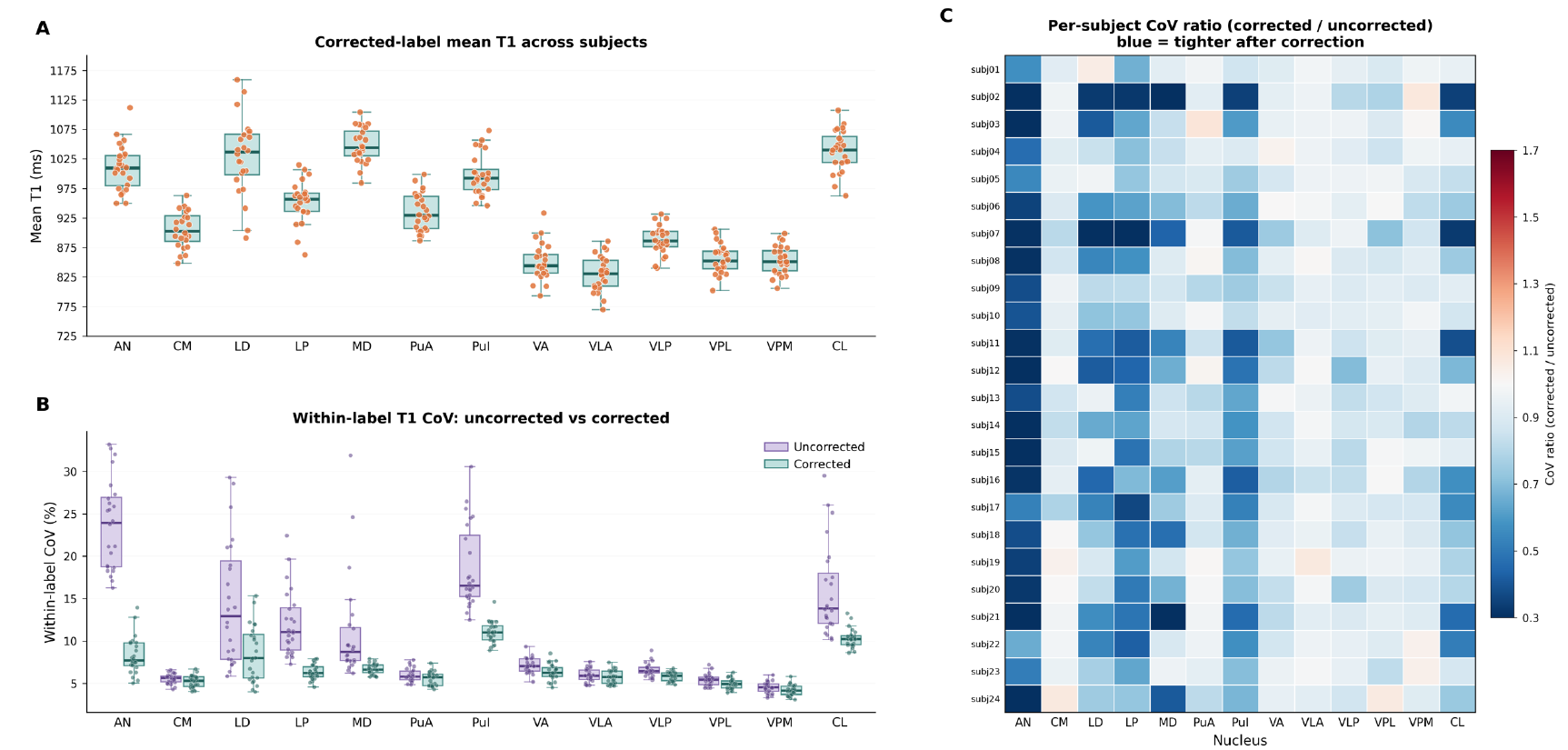}
    \caption{
    Group-level consistency check of corrected manual labels.
    (A) Boxplots of subject-level mean T1 values within the corrected labels for each thalamic nucleus.
    (B) Boxplots of within-label T1 coefficient of variation (CoV) before and after correction.
    (C) Heatmap of the per-subject CoV ratio, defined as corrected CoV divided by uncorrected CoV, for each nucleus.
    }
    \label{fig:s2_qc}
\end{figure}

Supplementary Figures~\ref{fig:s2_an}--\ref{fig:s2_cl} show manual correction examples for the 13 thalamic nuclei from one representative subject. For each nucleus, we show the uncorrected and corrected labels, a correction summary, and the T1 distribution before and after correction. These examples illustrate common types of manual edits, including removal of visually inconsistent or low-confidence voxels and refinement based on the multi-TI contrasts. After correction, the labels better matched the intended anatomical regions, and the corresponding T1 distributions were generally more concentrated.
Supplementary Figure~\ref{fig:s2_qc} shows a group-level consistency check of the corrected labels. 
Panel A shows the subject-level mean T1 values within each thalamic nucleus after correction. 
Panel B shows the T1 coefficient of variation (CoV) of each thalamic nucleus before and after correction, where CoV was defined as the standard deviation of T1 values divided by the mean T1 value. 
Panel C shows the per-subject, per-nucleus CoV ratio, defined as corrected CoV divided by uncorrected CoV, where values below one indicate lower within-label T1 variability after correction. 
Together, these results show consistent nucleus-wise T1 patterns across subjects and more compact within-label T1 distributions after correction.

\clearpage
\section*{Supplementary Note S4. Sensitivity to Reduced Training Labels}

Across the 24 subjects, manually labeled voxels covered
$68.3\% \pm 4.0\%$ of the manually localized thalamic region.
The median coverage was 69.6\%, with an IQR of 65.9\% to 70.6\%
and a range of 57.1\% to 74.1\%.

\medskip\noindent
We generated four reduced training label sets containing 80\%, 60\%, 40\%, and 20\% of the original labeled voxels. 
The original labels served as the 100\% reference. 
Label reduction was performed separately for each subject, nucleus, and hemisphere. For each labeled region, we calculated the Euclidean distance from each voxel to the region boundary. 
Voxels closest to the boundary were changed first from their nucleus label to the ignore label 100 to simulate progressively reduced supervision near nucleus boundaries. 
Voxels with the same distance were randomly ordered using a fixed seed of 1234. 
The same voxel ranking was used at all four levels, so every voxel included at a lower percentage was also included at all higher percentages.
The corresponding annotation coverage at each level is summarized in Supplementary Table~\ref{tab:label_coverage}.

\begin{table}[htbp]
\centering
\caption{Annotation coverage at different training label levels across 24 subjects.}
\label{tab:label_coverage}
\begin{tabular}{lccccc}
\toprule
Original labels used
& 100\% & 80\% & 60\% & 40\% & 20\% \\
\midrule
Average labeled voxels per subject
& 7,746.5
& 6,197.1
& 4,647.9
& 3,098.6
& 1,549.4 \\
Average thalamic region coverage
& $68.3 \pm 4.0\%$
& $54.6 \pm 3.2\%$
& $41.0 \pm 2.4\%$
& $27.3 \pm 1.6\%$
& $13.7 \pm 0.8\%$ \\
\bottomrule
\end{tabular}
\end{table}

\medskip\noindent
Reduced labels were used only for training. Validation and test labels remained unchanged. The six model configurations, comprising three modalities with and without CoordConv, were retrained at the 80\%, 60\%, 40\%, and 20\% label levels, following the same 8-fold cross-validation procedure across the 24 subjects. The original models served as the 100\% reference. Performance was evaluated on the unchanged test labels using per-nucleus TPRs and volume-weighted average (VWA) TPR without postprocessing.

\begin{figure}[h]
\centering
\includegraphics[width=\textwidth]{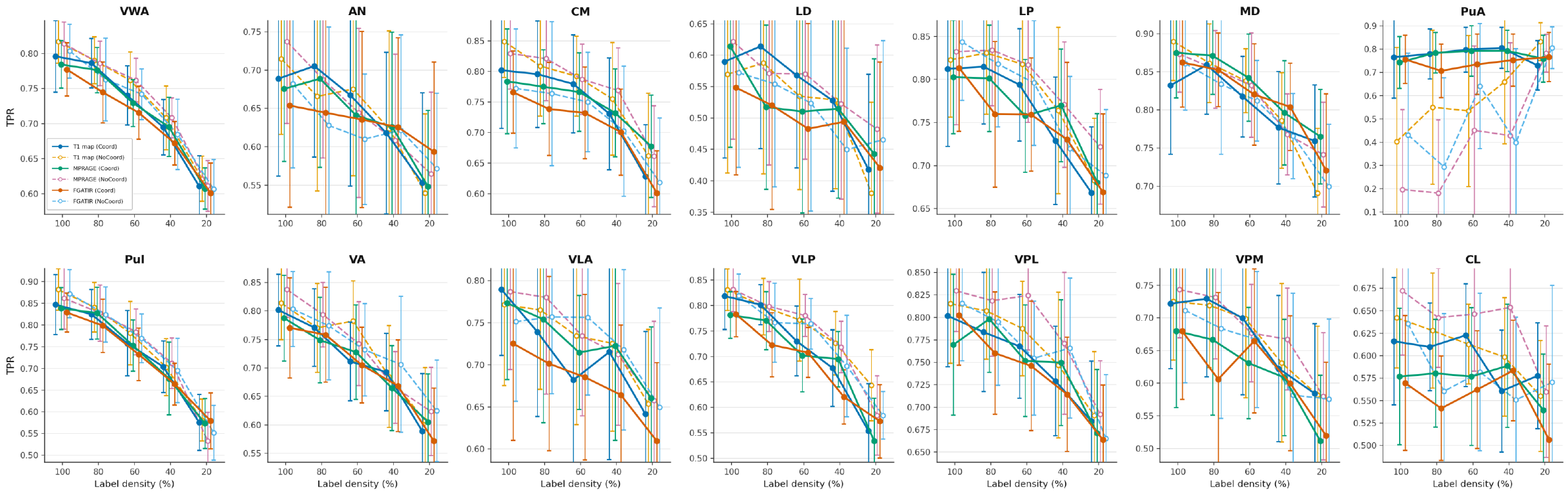}
\caption{Sensitivity to reduced training labels. Volume-weighted average (VWA) TPR and per-nucleus TPRs are shown for the six model configurations trained using 100\%, 80\%, 60\%, 40\%, and 20\% of the original labeled voxels. Points and error bars indicate mean $\pm$ SD across 24 subjects.}
\label{fig:label_sensitivity}
\end{figure}

\noindent
As shown in Supplementary Figure~\ref{fig:label_sensitivity}, both VWA and per-nucleus TPRs generally decreased across the six model configurations as fewer training labels were used. At 80\% of the original labels, VWA TPR ranged from 0.745 to 0.790, slightly lower than the 0.777–0.817 range at 100\%. With further reductions to 60\%, 40\%, and 20\%, the corresponding ranges decreased to 0.716--0.762, 0.672--0.708, and 0.601--0.621. Most nuclei followed the same overall trend. However, individual trajectories were not always monotonic. PuA was the clearest exception. Models without CoordConv showed larger fluctuations, while models with CoordConv remained relatively stable. 
Overall, the models showed some robustness to reduced training labels, but both VWA and per-nucleus performance declined as fewer labels were used.